\documentclass[twocolumn]{aa}
\usepackage{graphicx}
\usepackage{rotating}
\usepackage{subfigure}
\usepackage{longtable}
\usepackage{lscape}
\usepackage{txfonts}
\begin{document}
   \title{{\it XMM-Newton} observations of the Lockman Hole V: time variability of the brightest AGN
   } \titlerunning{X-ray spectra LH}

   \subtitle{}
   \author{
     S. Mateos  \inst{1}
    \and
     X. Barcons \inst{2}
    \and
     F. J. Carrera  \inst{2}
    \and
     M. J. Page \inst{3}
    \and
     M. T. Ceballos  \inst{2}
    \and
     G. Hasinger \inst{4}
    \and
     A. C. Fabian  \inst{5}
    }

           \authorrunning{S. Mateos et al.}

   \offprints{S. Mateos, \email{sm279@star.le.ac.uk}}
   \institute{X-ray Astronomy Group, Department of Physics and Astronomy, Leicester University, Leicester LE1 7RH, UK 
     \and Instituto de F\'\i sica de Cantabria (CSIC-UC), 39005 Santander, Spain
     \and MSSL, University College London, Holmbury St. Mary, Dorking, Surrey RH5 6NT, UK
     \and Max-Planck-Institut f\"{u}r Extraterrestrische Physik, Giessenbachstrasse, Garching D-85748, Germany.
      \and Institute of Astronomy, University of Cambridge, Madingley Road, Cambridge CB3 OHA, UK. }
 \date{10 July 2007}

   \abstract{This paper presents the results of a study of X-ray spectral and flux variability on time 
scales from months to years, of the 123 brightest objects (including 46 type-1 AGN and 28 type-2 AGN) 
detected with {\it XMM-Newton} in the {\it Lockman Hole} field. 
We detected flux variability with a significance $\ge$3$\sigma$ in $\sim$50\% of the objects, including 
68$\pm$11\% and 48$\pm$15\% among our samples of type-1 and type-2 AGN. However we found that the fraction of sources 
with best quality light curves that exhibit flux variability on the time scales sampled by our data is $\ge80\%$, i.e the great majority of 
the AGN population may actually vary in flux on long time scales. The mean relative intrinsic amplitude of flux variability 
was found to be $\sim$0.15 although with a large dispersion in measured values, from $\sim$0.1 to $\sim$0.65. 
The flux variability properties of our samples of AGN (fraction of variable objects and 
amplitude of variability) do not significantly depend on the redshift or X-ray luminosity of the objects and seem 
to be similar for the two AGN types.
Using a broad band X-ray colour we found that the fraction of sources showing spectral variability with a 
significance $\ge$3$\sigma$ is $\sim$40\% i.e. less common than flux variability. Spectral variability was found 
to be more common in type-2 AGN than in type-1 AGN with a 
significance of more than 99\%. This result is consistent with the fact that part of the soft emission in type-2 AGN
comes from scattered radiation, and this component is expected to be much less variable than the hard component.
The observed flux and spectral variability properties of our objects and especially the 
lack of correlation between flux and spectral variability in most of them
cannot be explained as being produced by variability of one spectral component alone, for example changes in $\Gamma$
associated with changes in the mass accretion rate, or variability in the amount of X-ray absorption. At least 
two spectral components must vary in order to explain the X-ray variability of our objects.
    } \maketitle
%

\section{Introduction} 
\label{Introduction}
Active Galactic Nuclei (AGN) are strongly variable sources in all wave bands 
(Edelson et al.~\cite{Edelson96}; Nandra et al.~\cite{Nandra98}) and on different time scales.
The largest amplitude and fastest variability is usually observed in X-rays. 
This is taken as evidence for X-ray emission in AGN originating
in a small region close to the central object, 
which is thought to be a supermassive ($10^6-10^9 {\rm M_\odot}$) black hole.

It has been found that AGN are generally more variable 
on long time scales than on short time scales (see Barr \& Mushotzky~\cite{Barr86}; Nandra et al.~\cite{Nandra97}; Markowitz \& Edelson~\cite{Markowitz01}). In addition, it is well known that variability amplitude on 
short (Barr \& Mushotzky~\cite{Barr86}; Nandra et al.~\cite{Nandra97}; 
Turner et al.~\cite{Turner99}; Lawrence \& Papadakis~\cite{Lawrence93}) 
and long (Markowitz \& Edelson~\cite{Markowitz01}) time scales, when measured 
over a fixed temporal frequency, is anti-correlated with the X-ray luminosity, i.e. lower variability 
amplitudes are seen in more luminous sources. 
All these results are consistent with recent results based on 
power spectrum analysis of light curves of AGN (Markowitz et al.~\cite{Markowitz03a}; McHardy et al.~\cite{McHardy06}).
There are also indications that 
the strength of the anti-correlation might decrease towards longer time scales (Markowitz et al.~\cite{Markowitz04}).

All these findings are consistent with a scenario where 
more luminous sources, hosting more massive black holes, have larger X-ray emitting regions,
and therefore the variability needs more time to propagate through the X-ray emitting region.

X-ray spectral variability studies of nearby objects
have shown that most AGN tend to become softer when they brighten 
(see e.g. Markowitz et al.~\cite{Markowitz03b}; Nandra et al.~\cite{Nandra97}) similar to the 
different flux-spectral states seen in black hole X-ray binaries (BHXRB) (McClintock et al.~\cite{McClintock03}). 
In addition, nearly all AGN exhibit stronger variability at soft X-rays on all time 
scales (Markowitz et al.~\cite{Markowitz04}). 

Two phenomenological model parameterisations have been proposed to explain the variability patterns observed 
in AGN (see e.g. Taylor et al.~\cite{Taylor03}): in the {\it two-component spectral model} (${\rm M^c}$Hardy et al.~\cite{McHardy98}, Shih et al.~\cite{Shih02})
a softer continuum emission power law component of constant slope but variable flux 
(probably associated with the emission from the hot corona) is superimposed on a harder 
spectral component with constant flux (likely associated with the Compton reflection hump). 
This model does not require the form of any of the spectral 
components to vary, what varies is the relative contribution 
of the soft and hard spectral components with time. When the source becomes brighter the soft component 
dominates the spectrum and the source becomes softer. In the {\it two-component spectral model} it is 
expected that as the sources become brighter their X-ray spectral slope will saturate to the 
slope of the soft component. However, some bright objects show evidence for variations in their 
underlying continuum spectrum (e.g. NGC 4151 see Perola et al.~\cite{Perola86}; 
Yaqoob \& Warwick~\cite{Yaqoob91}; Yaqoob et al.~\cite{Yaqoob93}). For these objects 
the {\it two component spectral model} cannot explain their variability properties. 

An alternative parameterisation is the {\it spectral pivoting model}: 
in this model flux-correlated changes in the continuum shape are due to changes 
in a single variable component, that becomes softer as the source brightens (see e.g. Zdziarski et al.~\cite{Zdziarski03}).

Although X-ray spectral variability in a significant number of AGN can be explained as 
due to changes in the primary continuum shape, 
changes in the column density of the cold gas responsible for the absorption in X-rays have been reported on 
time-scales of months-years in both type-1 and type-2 Seyfert galaxies 
(Risaliti et al.~\cite{Risaliti02}; Malizia et al.~\cite{Malizia97}; 
Puccetti et al.~\cite{Puccetti04}). 
The absorbing material in these objects 
must be clumpy and due to the time scales involved, it must 
be close to the central black hole, much nearer than the standard torus of 
AGN unification models (Antonucci et al.~\cite{Antonucci93}).

\begin{table*}
\caption{Summary of {\sl XMM-Newton} EPIC-pn observations in the {\it Lockman Hole}.}
\begin{center}
{\large  
\begin{tabular}{ccccccccccc}
\hline
{\rm Rev/ObsId} & obs phase  & R.A. &Dec & obs. date & ${\rm Filter}$ & ${\rm GTI}$ \\
(1) & (2) & (3) & (4) & (5) &  (6) & (7)\\
\hline
\hline
 070\,/\,0123700101  & PV & 10\,\,52\,\,43.0 & +57\,\,28\,\,48 & 2000-04-27 & Th & 34 \\[0.5ex]
 073\,/\,0123700401  & PV & 10\,\,52\,\,43.0 & +57\,\,28\,\,48 & 2000-05-02 & Th & 14 \\[0.5ex]
 074\,/\,0123700901  & PV & 10\,\,52\,\,41.8 & +57\,\,28\,\,59 & 2000-05-05 & Th & 5 \\[0.5ex]
 081\,/\,0123701001  & PV & 10\,\,52\,\,41.8 & +57\,\,28\,\,59 & 2000-05-19 & Th & 27 \\[0.5ex]
 345\,/\,0022740201  & AO1 & 10\,\,52\,43.0 & +57\,\,28\,\,48 & 2001-10-27  & M & 40 \\[0.5ex]
 349\,/\,0022740301  & AO1 & 10\,\,52\,43.0 & +57\,\,28\,\,48 & 2001-11-04  & M & 35 \\[0.5ex]
 522\,/\,0147510101  & AO2 & 10\,\,51\,\,03.4 & +57\,\,27\,\,50 & 2002-10-15 &M & 79 \\[0.5ex]
 523\,/\,0147510801  & AO2 & 10\,\,51\,\,27.7 & +57\,\,28\,\,07 & 2002-10-17 &M & 55 \\[0.5ex]
 524\,/\,0147510901  & AO2 & 10\,\,52\,\,43.0 & +57\,\,28\,\,48 & 2002-10-19 &M & 55 \\[0.5ex]
 525\,/\,0147511001  & AO2 & 10\,\,52\,\,08.1 & +57\,\,28\,\,29 & 2002-10-21 &M & 78 \\[0.5ex]
 526\,/\,0147511101  & AO2 & 10\,\,53\,\,17.9 & +57\,\,29\,\,07 & 2002-10-23 &M & 45 \\[0.5ex]
 527\,/\,0147511201  & AO2 & 10\,\,53\,\,58.3 & +57\,\,29\,\,29 & 2002-10-25 &M & 30 \\[0.5ex]
 528\,/\,0147511301  & AO2 & 10\,\,54\,\,29.5 & +57\,\,29\,\,46 & 2002-10-27 &M & 28 \\[0.5ex]
 544\,/\,0147511601  & AO2 & 10\,\,52\,\,43.0 & +57\,\,28\,\,48 & 2002-11-27 &M & 104\\[0.5ex]
 547\,/\,0147511701  & AO2 & 10\,\,52\,\,40.6 & +57\,\,28\,\,29 & 2002-12-04 &M & 98 \\[0.5ex]
 548\,/\,0147511801  & AO2 & 10\,\,52\,\,45.3 & +57\,\,29\,\,07 & 2002-12-06 &M & 86 \\[0.5ex]
\hline
\end{tabular}
\label{tab_observations}
}  
\end{center}
Columns are as follows: (1) {\sl XMM-Newton} revolution and observation identifier; 
(2) Observation phase: Verification Phase (PV), first and second announcements of 
opportunity (AO1 and AO2); (3) and (4) pointing coordinates in right ascension and declination (J2000); 
(5) Observation date; (6) and (7) EPIC-pn blocking filter$^{\mathrm{a}}$ of the observation and good time interval (in ksec) 
after subtracting periods of the observation affected by high background flares.\\  
$^{\mathrm{a}}$ Blocking filters: Th: Thin at 40nm A1; M: Medium at 80nm A1.
\end{table*}

In this paper we present the results of a study of the flux and spectral variability properties 
of the 123 brightest objects detected with {\sl XMM-Newton} in the {\it Lockman Hole} field, 
for which their time-averaged X-ray spectral properties are known, and 
were presented in a previous paper (Mateos et al.~\cite{Mateos05b}). 
The aim of this study is to provide further insight into the 
origin of the X-ray emission in AGN, by analysing the variability properties of the objects on long time scales, from 
months to years.
It is important to note that the variability analysis presented here  
does not allow us to study, for example, fluctuations in the accretion disc, as 
the dynamical time scales (light-crossing time) are much shorter than the ones sampled in our 
analysis. Intra-orbit variability studies (i.e. on time scales $<$2 days) cannot be performed 
on the {\it Lockman Hole} sources, as they are too faint.
With our analysis we 
can detect, for example, variability related to changes in the global accretion rate, such as 
variations in the accretion rate propagating 
inwards (Lyubarskii~\cite{Lyubarskii97}).
In addition, we can investigate whether the detected long-term X-ray variability 
in our objects can be explained as being due to variations in the obscuration of the nuclear engine. 

This paper is organised as follows: Sec.~\ref{observations}
describes the X-ray data that was used for the analysis 
and explains how we built our sample of objects; in Sec.~\ref{var_analysis} we describe the approach that 
we followed to carry out our variability analysis; the results of the study of flux and spectral 
variability in our objects are shown in Sec.~\ref{flux_var} and Sec.~\ref{sp_var};
Sec.~\ref{fracc_var} explores the true fraction of sources where either flux or spectral 
variability are present; a discussion of the results and possible interpretations for the origin of 
the long-term variability are presented in Sec.~\ref{origin_var}; Sec.~\ref{sp_var_unabs_type2} presents the variability properties 
of the type-2 AGN in our sample without any sign of X-ray absorption in their co-added X-ray spectra; finally the results of our analysis are summarised in Sec.~\ref{conclusions}. 

 Throughout this paper we have adopted the {\it WMAP} concordance cosmology model parameters (Spergel et al.~\cite{Spergel03}) 
with ${\rm H_0=70\,km\,s}^{-1}\,{\rm Mpc}^{-1}$, $\Omega_M=0.3$ and $\Omega_{\Lambda}=0.7$.  

\section{{\it XMM-Newton} deep survey in the {\it Lockman Hole}}
\label{observations}
The {\it XMM-Newton} observatory has carried out its deepest observation in 
the direction of the {\it Lockman Hole} field, centred at R.A.:10:52:43 and 
Dec:+57:28:48 (J2000). 
The {\sl XMM-Newton} deep survey in the {\it Lockman Hole} is composed of 16 
observations carried out from 2000 to 2002, which allow us to study the X-ray variability properties of our sources 
on long time scales, from months to years. 
The {\it Lockman Hole} was also observed during revolutions 071 ($\sim$61 ksec) 
and 344 ($\sim$80 ksec duration). However at the time of this analysis there was no Observation Data File (ODF) available for the observation in revolution 071, and hence, we could not use that data.
We could not use the data from revolution 344 because most of the observation was affected by high and flaring background.
We report the summary of the {\sl XMM-Newton} observations in the {\it Lockman Hole} used in our study  
in Table~\ref{tab_observations}. 
The first column shows the revolution
number and observation identifier. The second column shows the name of the 
phase of observation (PV for observations during the Payload
Verification Phase, and AO1 and AO2 for observations during the first and
second Announcements of Opportunity). The third and fourth columns list the pointing coordinates 
of the observation. Column five lists the observation dates, while 
the last two columns show the filters that were used during each observation for 
the EPIC-pn X-ray detector, together with the exposure times after removal of periods of high 
background. The 16 {\sl XMM-Newton} observations gave a total exposure time (after removal of periods of
high background) of $\sim$650 ksec for the EPIC-pn data. 

\subsection{Sample of sources}
\label{sample}
A detailed description of how the list of {\it XMM-Newton} detected sources 
in the total observation of the {\it Lockman Hole} was obtained can be found in 
Mateos et al.(~\cite{Mateos05b}). 
In brief, the {\it XMM-Newton} Science Analysis Software 
(SAS, Gabriel et al.~\cite{Gabriel2004}) was used to analyse the X-ray data. 
The event files of each observation were filtered to remove periods of time affected by high 
background. Images, exposure maps and background maps were obtained for each 
individual observation and for the EPIC-pn detector 
on the five standard {\it XMM-Newton} energy bands (0.5-2, 2-4.5, 4.5-7.5, 
7.5-12 keV). 
The data were summed to obtain the total observation of the field for each 
energy band. The SAS source detection algorithm {\tt eboxdetect-emldetect} was 
run on the five {\sl XMM-Newton} energy bands simultaneously\footnote{This increases 
the sensitivity of detection of sources on each individual energy band, and 
allows to obtain source parameters on each individual energy band.} 
to produce an X-ray source list for the total observation. 

From the final list of sources the 123 brightest objects (with more than 500 source-background subtracted 
0.2-12 keV counts) were selected for analysis. 
Because the main goal was to study the X-ray properties of AGN,
objects identified as clusters of galaxies or stars were excluded from the sample.
At the time of the analysis, 74 ($\sim$ 60\%) of the selected objects 
had optical spectroscopic identifications available. Of these, 46 were 
optically classified as type-1 AGN and 28 as type-2 AGN.

\section{X-ray variability analysis}
\label{var_analysis}
The three {\sl XMM-Newton} EPIC cameras (M1, M2 and pn) have different geometries, therefore for 
a given observation, it is common to find that a significant number of serendipitously detected 
objects fall near or inside CCD gaps in at least one of the cameras. 
This means that, for the same object, we will have light curves from different data sets 
(i.e. data from different revolutions) for each camera and therefore the sampling is different. 
In addition, because of their different instrumental responses we cannot 
combine the count rates from the three EPIC cameras. 
The fact that light curves from each EPIC camera 
are sampled differently in most sources, makes
comparing the results of our variability analysis between detectors very difficult. 
Due to the geometry of the pn detector, the number of sources 
falling in the pn CCD gaps will be larger than in MOS detectors; however the former 
provides the deepest observation of the field for each 
exposure\footnote{Note that in general pn observations receive $\sim$twice 
the number of photons than M1 and M2, and hence pn observations are 
in general deeper, by a factor of $\sim$2, than M1 and M2 observations. 
Although the total exposure time of the EPIC-pn data is lower than the exposure 
times of M1 and M2 data, the pn total observation still collects more counts.}, which allows us to detect 
lower variability amplitudes, and to measure better the variability amplitude. 
Therefore, we have used only pn data in the analysis presented in this paper.
The pn source lists obtained for each revolution were visually screened to 
remove spurious detections of hot pixels still present in the X-ray images. 

Source detection was carried out on the EPIC-pn data from each individual 
revolution to obtain all the relevant parameters for the selected sources 
from all observations where they were detected. 
We have used measured 0.2-12 keV count rates\footnote{The SAS source detection task {\tt emldetect} 
provides count rates corrected for vignetting and Point Spread Function (PSF) losses on each energy band.}
from each individual observation where 
sources were detected, to build light curves.

Source parameters become very uncertain for objects detected close to CCD gaps.
To avoid this problem, 
we did not use data from observations where the objects were detected near CCD gaps, bad columns, or near the edge of the Field of View (FOV).
To remove these cases we created detector masks, one for each observation,
increasing the size of the CCD gaps up to the radius that contains 80\% of the telescope PSF
on each detector point. We used these masks to remove automatically non desired data.
 
Each point on the light curves corresponds to the average count rate of the sources 
in that revolution. The exposure times of the EPIC-pn observations varied significantly 
between revolutions and therefore also the uncertainty in the measured count rate values. In addition, the 
time interval between points is not constant, 
with some data points in the light curves separated by days, and others by years.
Therefore we are not studying variability properties on a well defined 
temporal scale. In addition we do not have 
the same number of points in the light curves for all sources. 
All this needs to be taken into account when comparing the variability properties 
of different sources. 

\section{Flux variability}
\label{flux_var}

\begin{figure}
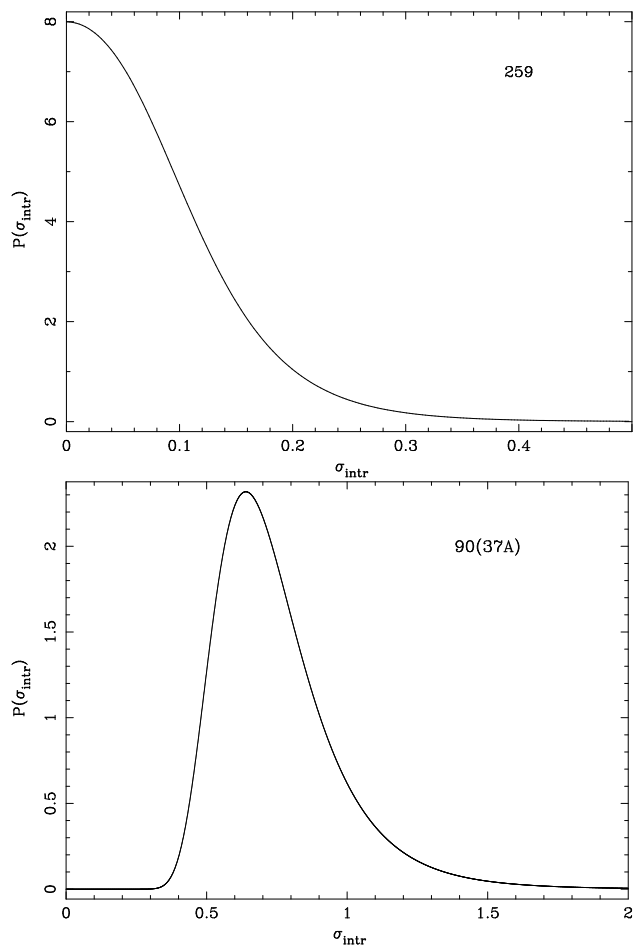

    \hbox{
    \includegraphics[angle=-90,width=0.45\textwidth]{fig1.ps}\hspace{0.3cm}}
    \hbox{
    \includegraphics[angle=-90,width=0.45\textwidth]{fig2.ps}}
    \caption{
    Typical probability density distributions of the excess variance, $\sigma_{{\rm intr}}$, for sources
    without detected variability (top) and detected variability (bottom). 
    }
    \label{psi_examples}
\end{figure}

In order to search for flux variability, we created a light curve for each source using
the count rates in the observed 0.2-12 keV energy band in each revolution. We took this energy band 
because it was used to study the time averaged spectral emission properties of the sources
and therefore we can refer our variability studies to the time averaged spectral properties.
In addition, using 0.2-12 keV count rates we have more than 10 background subtracted counts 
on each point of the light curves, and therefore we 
can assume Gaussian statistics during the analysis.
Using EPIC-pn 0.2-12 keV count rates we obtained light curves with at least two 
data points for 120 out of the 123 sources in our sample, including 
45 type-1 AGN and 27 type-2 AGN. 

\begin{table}[!tb]
\caption{Summary of detection of flux and spectral variability in the {\it Lockman Hole} sources.}
\begin{center}
\begin{tabular}{ccccccccc}
\hline
Group & ${\rm N_{tot}}$ & ${\rm n_{flux}}$ & ${\rm fracc_{flux}(\%)}$ & ${\rm n_{sp}}$ & ${\rm fracc_{sp}(\%)}$ \\
(1) & (2) & (3) & (4) & (5) & (6)\\
\hline
\hline
 All        & 120 & 62 & 51$\pm$7   & 24 & 20$\pm$6\\
 type-1 AGN &  45 & 31 & 68$\pm$11  &  6 & 14$\pm$8 \\
 type-2 AGN &  27 & 13 & 48$\pm$15  &  9 & 34$\pm$14\\
 Unidentified &48 & 18 & 37$\pm$11 &   9 & 19$\pm$9\\
\hline 
\end{tabular}
\label{tab_fracc_det_var}
\begin{list}{}{}
Columns are as follows: (1) group of sources; (2) total number of objects in the group; 
(3) number of sources with detected flux variability (confidence $\ge3\sigma$); 
(4) fraction (corrected for spurious detections; see Mateos et al.~\cite{Mateos05a}) 
of sources in group with detected flux variability; 
(5) number of sources with detected spectral variability (confidence $\ge3\sigma$); 
(6) fraction (corrected for spurious detections) of sources in group with detected spectral variability. 
Errors correspond to the 1$\sigma$ confidence interval.
\end{list}
\end{center}
\end{table}

Not all EPIC-pn observations were carried out with the same blocking filter 
(see Table~\ref{tab_observations}). We cannot directly compare the observed count rates, 
because blocking filters affect in a different way the low energy photons 
and hence measured 0.2-12 keV count rates will be different 
even for a non varying source. Moreover, changes in 
background modelling/calibration during the lifetime of the mission can introduce systematic 
differences between measured count rates. 
The observations we are using were carried out in a time interval spanning two years, 
and therefore we need to study whether any instrumental drift is present in our data, 
and whether it is affecting the measured variability properties of our sources. In order to 
correct for both different filters and instrumental drifts, we have calculated the mean (averaged over all sources) 
deviation of the count rates measured on each observation from the 
mean count rates of the light curves. We then corrected the count rates from these mean 
deviations. This is explained in detail in Appendix~\ref{appendix_A}. 
All count rates used for the study presented in this paper are corrected for this otherwise small 
effect ($\lesssim$10\%).

To search for deviations from the null hypothesis (that 
the 0.2-12 keV mean count rates of the objects have remained constant 
during all observations) we used $\chi^2$ 
\begin{equation}\chi^2={\sum_{i=1}^{N}{(x_i-\langle x\rangle)^2 \over \sigma_i^2}}\end{equation}
where {\it $x_i$} are the 0.2-12 keV count rates of each source on each bin and $\sigma_i$ 
the corresponding 1$\sigma$ statistical errors, {\it N} is the 
number of points on each light curve and $\langle x \rangle$ is the unweighted 
mean count rate for that source.
We accepted a source as variable if the significance
of $\chi^2$ being higher than the obtained value just by chance is
lower than $2.7 \times 10^{-3}$ i.e. a 3$\sigma$ detection confidence.

The results of detection of flux variability are summarised 
in columns 3 and 4 of Table~\ref{tab_fracc_det_var}. 
We corrected the fractions of sources with detected flux variability for 
spurious detections for the given confidence level (3$\sigma$) 
using the bayesian method described in Mateos et al.~(\cite{Mateos05a}) and in Stevens et al.~(\cite{Stevens05}): 
the probability distribution of the ``true'' fraction of objects with a detected property is calculated as a 
function of the selected confidence level, the sample size and the measured number of detections. The latter 
has two different contributions, the spurious detections allowed by the selected confidence level and 
the ``true'' detections.

Overall we detect flux variability in $\sim$50\% of the sources with a significance of more than 
3$\sigma$. Flux variability was detected in 31 out of 45 (68$\pm$11\%) type-1 AGN and 13 out of 
27 (48$\pm$15\%) type-2 AGN. 

The fraction of AGN with detected flux variability was not found to 
vary with redshift, i.e., the different sampling of rest-frame energies for sources at different redshifts 
does not seem to affect the detection of flux variability.
We compared the observed fractions of varying sources for different samples of objects using the method described 
in Mateos et al. (\cite{Mateos05a}). We found the difference in the fractions of flux 
variable objects among our samples of type-1 and type-2 AGN to be significant at only the 
$\sim$75\% confidence level. 
Therefore there is no evidence that the fraction of sources showing flux variability 
in the 0.2-12 keV band on long time scales is significantly different for type-1 and type-2 AGN. We see that the 
fraction of variable sources among unidentified objects is lower than the fractions obtained 
for type-1 and type-2 AGN, and for the whole sample of objects. This could be 
due to the fact that most unidentified objects are among the faintest sources  
in the sample, and therefore they will tend to have light curves with 
lower quality, where variations are more difficult to detect. In addition, some of these unidentified sources might 
not be variable. For example, unidentified heavily obscured AGN 
(i.e. much more obscured than the type-2 AGN identified in our sample) 
may show only reprocessed emission which will also have lower variability amplitude.

We found our sources to exhibit a whole range of flux variability patterns. Some objects became fainter with 
time, while others became brighter. However for a significant fraction of sources, we found
irregular flux variations with respect to their mean flux level. 
Some examples of light curves used in our analysis are shown in Fig.~\ref{flux_var_unabsorbed_agn}. 

\begin{figure}[!tb]
    \hbox{
    \includegraphics[angle=90,width=0.5\textwidth]{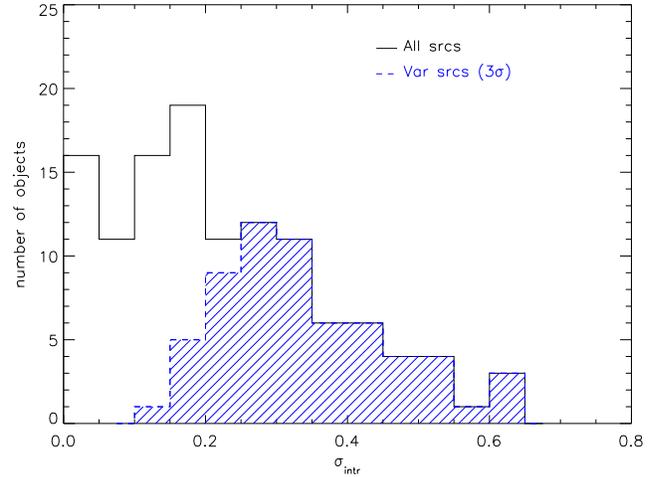}}
    \caption{Distribution of measured amplitude of variability for all objects (solid line) and 
      for the sources where flux variability was detected with a confidence level $\ge3\sigma$ (filled histogram).}
    \label{hist_excess_var}
\end{figure}

\subsection{Amplitude of flux variability} 
\label{var_amplitude}

The method we have used to calculate the 
amplitude of flux variability (noise-subtracted) in our light curves is fully described 
in Almaini et al. (\cite{Almaini00}). This method is the most appropriate in the regime of Gaussian 
statistics, an assumption that, as we said before, is satisfied by our data. In addition, this method 
is appropriate 
for light curves with points having significantly different measurement errors. 

The method assumes that the measured dispersion in the light curves has two different contributions:
\begin{equation}\sigma_{{\rm total}}^2=\sigma_{{\rm noise}}^2+\sigma_Q^2 \end{equation}
The first contribution, $\sigma_{noise}$, is the statistical error
in the data points and the second, $\sigma_Q$, 
the true fluctuation in the source flux. We are 
interested in the second quantity, $\sigma_Q$, as it measures the intrinsic variability in flux 
of the objects. In order to calculate $\sigma_Q$, a maximum likelihood method is used.

\begin{table*}[!htb]
\caption{Measured values of 0.2-12 keV flux variability amplitudes in {\it Lockman Hole} sources.}
\begin{center}
\begin{tabular}{ccccccccc}
\hline
Sample & ${\rm \langle\sigma_{{\rm intr}}\rangle_{all}}$ 
& ${\rm \langle\sigma_{{\rm intr}}\rangle_{all}^{w}}$ & ${\rm \langle\sigma_{{\rm intr}}\rangle_{var}}$ 
& ${\rm \langle\sigma_{{\rm intr}}\rangle_{var}^{w}}$ & ${\rm \sigma_{{\rm intr}}}$\\
(1) & (2) & (3) & (4) & (5) & (6) \\
\hline
\hline
 All          &  0.22$\pm$0.01 & $0.15\pm0.01$ & 0.34$\pm$0.01  & 0.28$\pm$0.01 & $\le$0.36\\ 
 type-1 AGN   &  0.27$\pm$0.02 & $0.18\pm0.02$ & 0.34$\pm$0.02  & 0.27$\pm$0.02 & $0.27_{-0.08}^{+0.47}$\\ 
 type-2 AGN   &  0.21$\pm$0.03 & $0.15\pm0.02$ & 0.34$\pm$0.04  & 0.28$\pm$0.02 & $\le0.35$\\ 
 Unidentified &  0.19$\pm$0.02& $0.11\pm0.02$ & 0.35$\pm$0.03  & 0.30$\pm$0.02 & $\le0.33$\\ 
\hline 
\end{tabular}
\label{tab_sigmaq}
\begin{list}{}{}
Columns are as follows: (1) group of sources; 
(2) arithmetic means and corresponding 1$\sigma$ errors of the measured variability amplitudes for different types of sources; 
(3) weighted means and corresponding 1$\sigma$ errors of the measured variability amplitudes for different types of sources; 
(4) and (5) arithmetic and weighted means of the measured variability amplitudes considering only objects with detected variability 
from the $\chi^2$ test; (6) mode and 1$\sigma$ confidence intervals from 
the average ${\rm P(\sigma_{{\rm intr}})}$ distributions. 
For the cases where the integral of the ${\rm P(\sigma_{{\rm intr}})}$ distribution to the left reached zero 
68\% confidence upper limits are given. 
\end{list}
\end{center}
\end{table*}

For each object, a probability distribution for $\sigma_Q$ is obtained. The maximum of this function gives the most probable 
value of $\sigma_Q$ for the given set of data points, while the uncertainty in $\sigma_Q$ is obtained by integrating the 
function from the maximum value to both sides until the desired probability is 
encompassed. We have calculated 1$\sigma$ errors for the measured variability amplitudes. 
In the cases where the lower error bound of the integrals reached 
zero we calculated 1$\sigma$ upper limits for $\sigma_Q$.    
 
To allow comparison of variability amplitudes for sources with different 
mean count rates we have calculated the value of the normalised excess variance, $\sigma_{{\rm intr}}$, 
for each source, 
\begin{equation} \sigma_{{\rm intr}}={\sigma_Q \over \langle {\it CR}\rangle} \label{eq:sintr} \end{equation}
where $\langle {\it CR}\rangle$ is the unweighted mean 0.2-12 keV count rate of that source.
This parameter gives the fraction of the total flux that is variable, and therefore 
can be used to compare amplitudes of flux variability for sources with different fluxes.
Fig.~\ref{psi_examples} shows typical probability distributions of $\sigma_{{\rm intr}}$
of sources without detected (top) and detected (bottom) variability.

The distribution of $\sigma_{{\rm intr}}$ values obtained from the light curves of our sources 
is shown in Fig.~\ref{hist_excess_var} for the whole sample of objects (solid line) 
and for objects where flux variability was detected in terms of the $\chi^2$ test with a 
confidence $\ge$3$\sigma$ (filled histogram).

We see that our variable sources have a broad range of values of the measured excess variance from 
$\sim$10\% to $\sim$65\%, with the maximum of the distribution (for objects with detected 
variability\footnote{Note that the values outside the filled area correspond to sources with undetected 
variability ($<$3$\sigma$) and therefore they cannot be considered significant detections.})  
being at a value of $\sim$30\%. 
Values of the excess variance above 
$\sim$50-60\% are not common in our sources. 

We expect the efficiency of detection of variability to be a strong function 
of the quality of the light curves, but also of the amplitude of variability. 
Therefore the fraction of sources with detected 
variability from the $\chi^2$ test should decrease for lower variability 
amplitudes. This effect is evident from Fig.~\ref{hist_excess_var}, where we see that, 
for variability amplitudes lower than 20\%, 
the values of the amplitude formally obtained are very rarely significant at 3$\sigma$.

Therefore our distribution of observed flux variability amplitudes for variable objects should 
not be interpreted as a true distribution of variability amplitudes 
in our sample of sources. Indeed we might be missing sources with low variability 
amplitudes or faint sources with strong variability due to the large statistical errors in the data. 
We will return to this point in Sec.~\ref{sims_sigma}.

\subsection{Mean variability amplitude}
\label{mean_sigma} 

\begin{figure}[!tb]
    \hbox{
    \includegraphics[angle=0,width=0.46\textwidth]{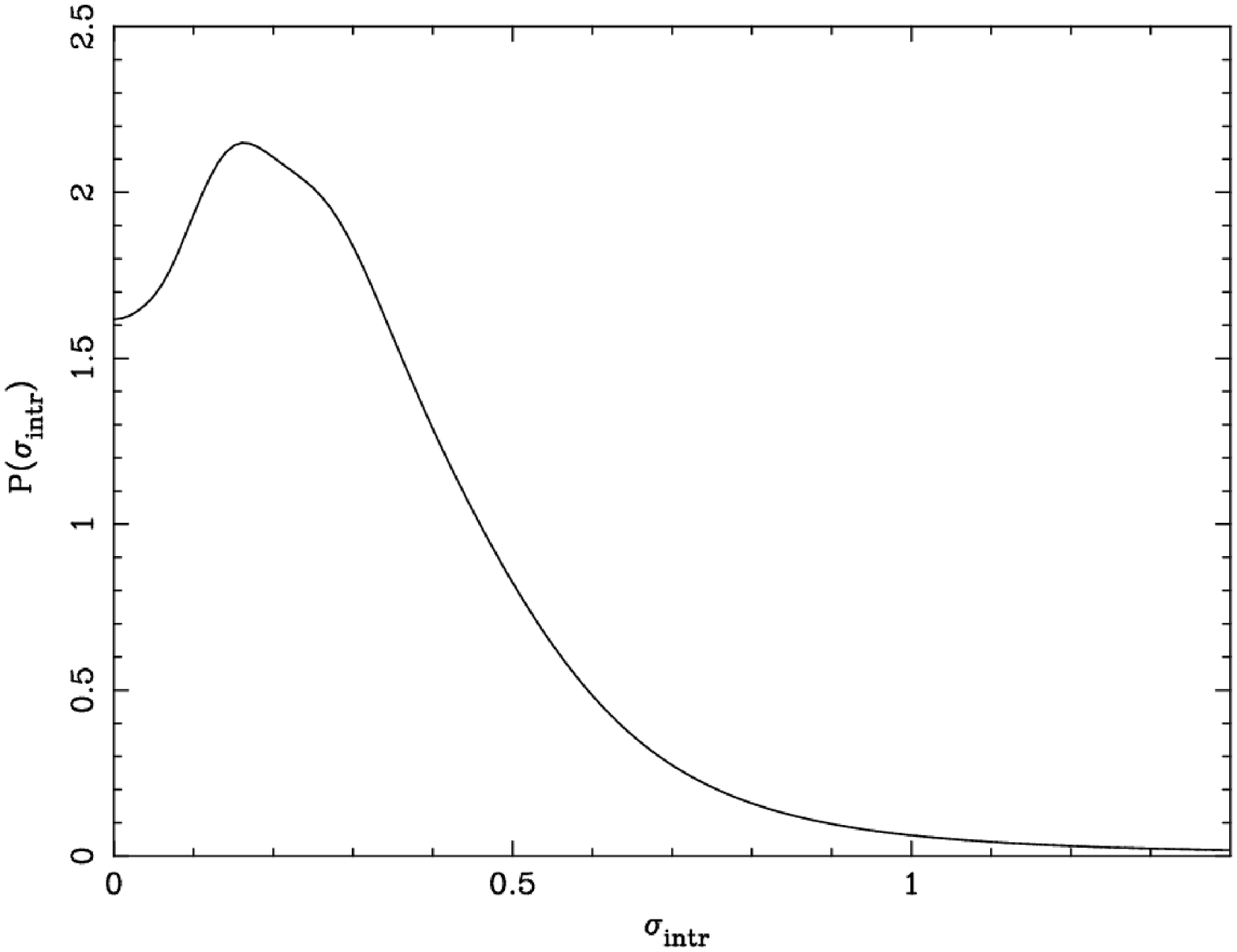}}
    \hbox{
     \hspace{0.0125\textwidth}\includegraphics[angle=90,width=0.51\textwidth]{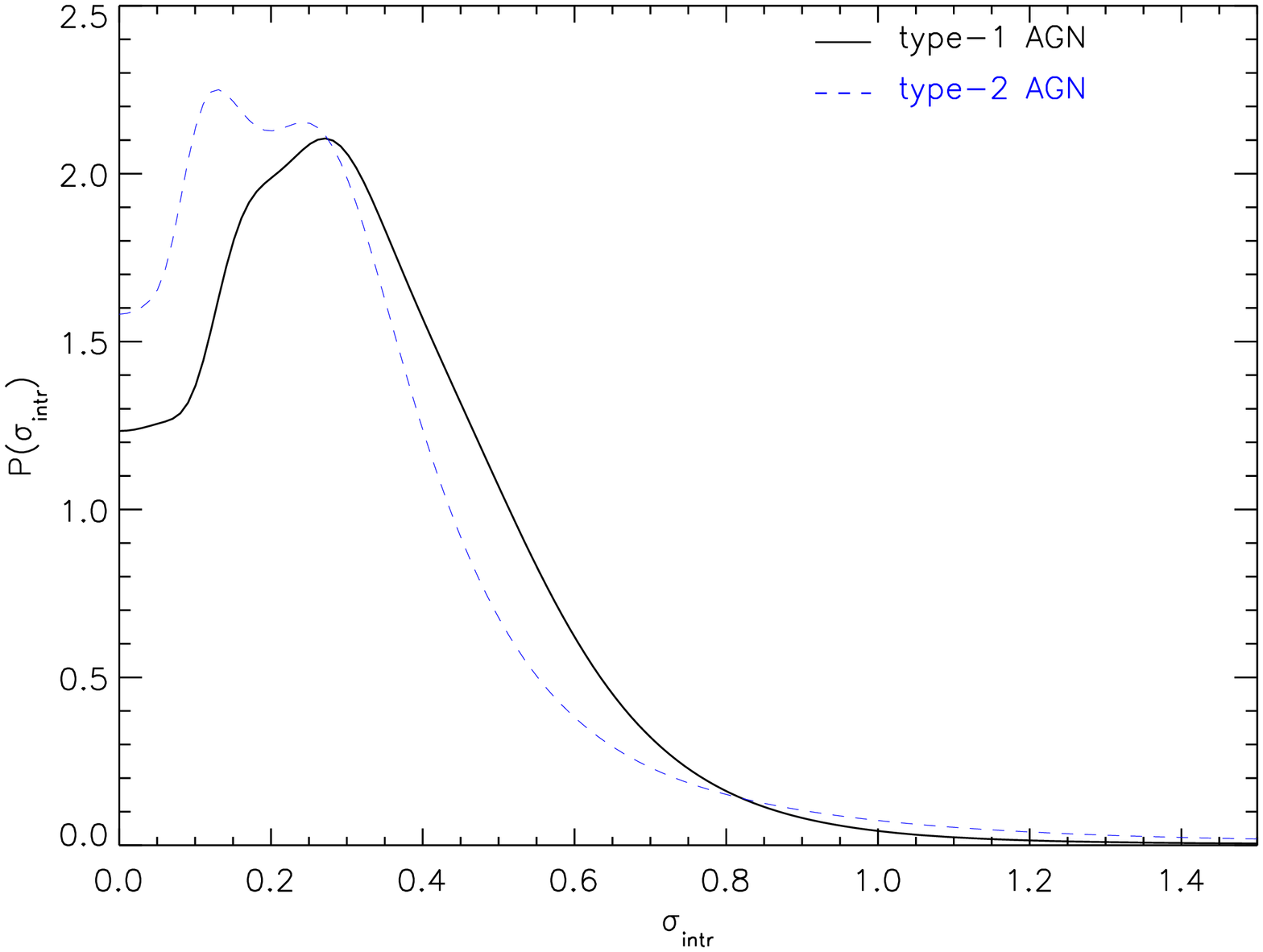}}
    \caption{
    Mean probability density functions of the excess variance, $P(\sigma_{{\rm intr}})$, 
    for all objects (top) and for type-1 and type-2 AGN (bottom). The distributions were obtained from the mean of the 
    individual probability distributions of $\sigma_{{\rm intr}}$. 
    }
    \label{dist_qdp_all}
\end{figure}

Mean values of $\sigma_{{\rm intr}}$ for the whole sample of sources and for type-1 and type-2 AGN and 
unidentified subsamples are listed in Table~\ref{tab_sigmaq}. We show values obtained 
using both the arithmetic and weighted mean for comparison. We see that average values 
from the weighted mean are significantly lower that those obtained with the 
arithmetic mean in all cases. This is because, as we want to study the variability properties of 
our sample, we included in the calculations all measured values of ${\rm \sigma_{intr}}$ 
independently on whether we detected or not significant variability in the light curves. Because of that,
we have a number of sources with measured $\sigma_{{\rm intr}}$ values of zero but with errors of the same order as the ones for 
$\sigma_{{\rm intr}}$$>$0. These values shift the weighted means to lower values.

We see in Table~\ref{tab_sigmaq} that the mean values for the amplitude of flux variability 
for type-1 and type-2 AGN do not differ significantly. 

\onecolumn
\begin{longtable}{ccccccccccccc}
\caption{Summary of detection of X-ray variability in {\it Lockman Hole} sources}\\
\hline
\hline
$XMM$ & {\sl ROSAT} & R.A. & Dec & Class & redshift & Model & Flux var & Spec var & $\sigma_{{\rm intr}}$ \\
(1) & (2) & (3) & (4) & (5) & (6) & (7) & (8) & (9) & (10)\\
\hline
\endfirsthead
\caption{Continued}\\
\hline
\hline
$XMM$ & {\sl ROSAT} & R.A. & Dec & Class & redshift & Model & Flux var & Spec var & $\sigma_{{\rm intr}}$ \\
(1) & (2) & (3) & (4) & (5) & (6) & (7) & (8) & (9) & (10)\\
\hline
\endhead
\hline
\endfoot
 607&        -& 10 53 01.86& +57 15 00.69&    --&     -&       SPL&$>$99.99&$>$99.99&$\le$  0.65\\[0.5ex]
 599&      54A& 10 53 07.46& +57 15 05.84&  type-1 & 2.416&       SPL&$>$99.99&    45.9&$  0.25_{  0.11}^{  0.35}$\\[0.5ex]
 400&      13A& 10 52 13.29& +57 32 25.58&  type-1 & 1.873&       SPL&$>$99.99&    81.5&$  0.20_{  0.14}^{  0.25}$\\[0.5ex]
  63&        -& 10 52 36.49& +57 16 04.07&    --&     -&       APL&     5.5&    93.4&$\le$  0.11\\[0.5ex]
   5&      52A& 10 52 43.30& +57 15 45.95&  type-1 & 2.144&       SPL&$>$99.99&    25.1&$  0.17_{  0.10}^{  0.22}$\\[0.5ex]
   6& 504(51D)& 10 51 14.30& +57 16 16.88& type-2 & 0.528&       SPL&    60.9&    66.7&$\le$  0.26\\[0.5ex]
  16&        -& 10 51 46.64& +57 17 16.02&    --&     -&       APL&    44.0&    45.9&$\le$  0.31\\[0.5ex]
  21&      48B& 10 50 45.67& +57 17 32.60& type-2 & 0.498&       SPL&    -1.0&    -1.0&$     -_{     -}^{     -}$\\[0.5ex]
  26&        -& 10 52 32.99& +57 17 50.96&    --&     -&       SPL&    12.1&     1.3&$\le$  0.20\\[0.5ex]
  31&        -& 10 52 00.34& +57 18 08.24&    --&     -&      CAPL&$>$99.99&    99.8&$\le$  0.77\\[0.5ex]
  41&      46A& 10 51 19.14& +57 18 34.09&  type-1 & 1.640&       APL&       -&       -&$     -_{     -}^{     -}$\\[0.5ex]
  39&      45Z& 10 53 19.09& +57 18 53.58& type-2 & 0.711&       SPL&$>$99.99&    51.9&$  0.22_{  0.11}^{  0.31}$\\[0.5ex]
  53&      43A& 10 51 04.39& +57 19 23.90&  type-1 & 1.750&       APL&    19.3&    61.6&$\le$  0.10\\[0.5ex]
  65&        -& 10 52 55.46& +57 19 52.80&    --&     -&       APL&    87.6&    85.3&$  0.18_{  0.03}^{  0.32}$\\[0.5ex]
  74&     905A& 10 52 51.13& +57 20 15.70&    --&     -&       SPL&    83.9&    69.5&$\le$  0.28\\[0.5ex]
  72&      84Z& 10 52 16.94& +57 20 19.71& type-2 & 2.710&       APL&$>$99.99&    55.0&$  0.29_{  0.20}^{  0.36}$\\[0.5ex]
  85&      38A& 10 53 29.50& +57 21 06.22&  type-1 & 1.145&       SPL&$>$99.99&    70.7&$  0.39_{  0.27}^{  0.48}$\\[0.5ex]
  86&        -& 10 53 09.68& +57 20 59.58&  type-1 & 3.420&       SPL&     8.3&    54.1&$\le$  0.10\\[0.5ex]
  88&      39B& 10 52 09.37& +57 21 05.43&  type-1 & 3.279&       SPL&$>$99.99&    74.4&$  0.46_{  0.23}^{  0.59}$\\[0.5ex]
  90&      37A& 10 52 48.09& +57 21 17.43&  type-1 & 0.467&    SPL+SE&$>$99.99&$>$99.99&$  0.64_{  0.40}^{  0.81}$\\[0.5ex]
  96& 814(37G)& 10 52 44.87& +57 21 24.84&  type-1 & 2.832&       SPL&$>$99.99&    43.3&$  0.19_{  0.12}^{  0.25}$\\[0.5ex]
 107&        -& 10 52 19.49& +57 22 15.26& type-2 & 0.075&       APL&    89.6&    84.9&$\le$  0.20\\[0.5ex]
 120&        -& 10 52 25.17& +57 23 07.02&    --&     -&       SPL&$>$99.99&    69.9&$\le$  0.41\\[0.5ex]
 108&        -& 10 50 50.91& +57 22 15.65&    --&     -&      2SPL&    88.7&    49.0&$\le$  0.16\\[0.5ex]
 900&        -& 10 54 59.43& +57 22 18.84&    --&     -&       APL&    35.3&    34.6&$\le$  0.17\\[0.5ex]
 121&     434B& 10 52 58.08& +57 22 51.95& type-2 & 0.772&       APL&    98.2&    29.4&$  0.16_{  0.07}^{  0.24}$\\[0.5ex]
 135& 513(34O)& 10 52 54.39& +57 23 43.89&  type-1 & 0.761&       SPL&$>$99.99&    63.2&$  0.26_{  0.18}^{  0.32}$\\[0.5ex]
 124&     634A& 10 53 11.72& +57 23 09.07&  type-1 & 1.544&       SPL&    90.0&    64.0&$  0.12_{  0.03}^{  0.19}$\\[0.5ex]
 125& 607(36Z)& 10 52 19.90& +57 23 07.92&    --&     -&       SPL&$>$99.99&    28.1&$\le$  0.47\\[0.5ex]
 142&        -& 10 52 03.74& +57 23 39.62&    --&     -&       SPL&    88.5&    97.7&$  0.16_{  0.05}^{  0.27}$\\[0.5ex]
 133&      35A& 10 50 38.77& +57 23 39.67&  type-1 & 1.439&       SPL&    -1.0&    -1.0&$     -_{     -}^{     -}$\\[0.5ex]
 148&      32A& 10 52 39.66& +57 24 32.83&  type-1 & 1.113&    SPL+SE&$>$99.99&    65.7&$  0.30_{  0.22}^{  0.37}$\\[0.5ex]
 166&        -& 10 52 31.98& +57 24 30.82&    --&     -&       APL&    78.2&$>$99.99&$  0.10_{  0.01}^{  0.20}$\\[0.5ex]
 156&        -& 10 51 54.59& +57 24 09.28& type-2 & 2.365&       APL&    99.2&    95.6&$\le$  0.30\\[0.5ex]
 163&      33A& 10 51 59.88& +57 24 26.31&  type-1 & 0.974&       APL&$>$99.99&    89.7&$  0.33_{  0.21}^{  0.44}$\\[0.5ex]
 168&      31A& 10 53 31.72& +57 24 56.19&  type-1 & 1.956&       SPL&$>$99.99&    68.6&$  0.53_{  0.38}^{  0.66}$\\[0.5ex]
 172&        -& 10 53 15.71& +57 24 50.84& type-2 &  1.17&       APL&    60.2&$>$99.99&$\le$  0.22\\[0.5ex]
 183&      82A& 10 53 12.27& +57 25 08.28&  type-1 &  0.96&       SPL&$>$99.99&    74.4&$  0.28_{  0.18}^{  0.36}$\\[0.5ex]
 176&      30A& 10 52 57.25& +57 25 08.77&  type-1 & 1.527&       SPL&$>$99.99&    59.1&$  0.32_{  0.22}^{  0.41}$\\[0.5ex]
 179&        -& 10 52 31.64& +57 25 03.93&    --&     -&      CAPL&    85.5&$>$99.99&$  0.15_{  0.01}^{  0.31}$\\[0.5ex]
 174&        -& 10 51 20.63& +57 24 58.24&    --&     -&       SPL&$>$99.99&    53.3&$\le$  0.64\\[0.5ex]
 186&        -& 10 51 49.93& +57 25 25.13& type-2 & 0.676&       APL&$>$99.99&    41.0&$  0.34_{  0.18}^{  0.46}$\\[0.5ex]
 171&      28B& 10 54 21.22& +57 25 45.40& type-2 & 0.205&       APL&$>$99.99&    62.0&$\le$  0.17\\[0.5ex]
 200&        -& 10 53 46.81& +57 26 07.77&    --&     -&       APL&    90.6&    45.8&$  0.14_{  0.03}^{  0.25}$\\[0.5ex]
 187&        -& 10 50 47.96& +57 25 22.71&    --&     -&       SPL&$>$99.99&    81.6&$\le$  0.49\\[0.5ex]
 191&      29A& 10 53 35.03& +57 25 44.13&  type-1 & 0.784&       SPL&$>$99.99&    90.6&$  0.50_{  0.31}^{  0.64}$\\[0.5ex]
 199&        -& 10 52 25.28& +57 25 51.27&    --&     -&       APL&    99.9&$>$99.99&$  0.16_{  0.06}^{  0.25}$\\[0.5ex]
 217&        -& 10 51 11.60& +57 26 36.67&    --&     -&       APL&    99.3&    60.0&$  0.14_{  0.06}^{  0.23}$\\[0.5ex]
 214&        -& 10 53 15.09& +57 26 30.65&    --&     -&       APL&    98.8&    97.9&$  0.18_{  0.06}^{  0.30}$\\[0.5ex]
 222&        -& 10 53 51.67& +57 27 03.64& type-2 & 0.917&       APL&$>$99.99&$>$99.99&$  0.25_{  0.16}^{  0.33}$\\[0.5ex]
2020&      27A& 10 53 50.19& +57 27 11.61&  type-1 & 1.720&       SPL&$>$99.99&    53.6&$  0.43_{  0.30}^{  0.53}$\\[0.5ex]
 226&        -& 10 51 20.49& +57 27 03.47&    --&     -&       SPL&    99.9&    98.8&$  0.21_{  0.09}^{  0.30}$\\[0.5ex]
 243&        -& 10 51 28.14& +57 27 41.55&    --&     -&      CAPL&$>$99.99&$>$99.99&$  0.25_{  0.14}^{  0.33}$\\[0.5ex]
 254&     486A& 10 52 43.37& +57 28 01.49& type-2 & 1.210&       APL&$>$99.99&    13.6&$  0.30_{  0.16}^{  0.42}$\\[0.5ex]
 261&      80A& 10 51 44.63& +57 28 08.89&  type-1 & 3.409&       SPL&$>$99.99&    95.9&$  0.22_{  0.14}^{  0.30}$\\[0.5ex]
 259&        -& 10 53 05.60& +57 28 12.50& type-2 & 0.792&    APL+SE&    71.2&$>$99.99&$\le$  0.10\\[0.5ex]
 270&     120A& 10 53 09.28& +57 28 22.65&  type-1 & 1.568&    SPL+SE&$>$99.99&    97.7&$  0.15_{  0.10}^{  0.20}$\\[0.5ex]
 267&     428E& 10 53 24.54& +57 28 20.65&  type-1 & 1.518&       APL&$>$99.99&    32.1&$  0.28_{  0.20}^{  0.35}$\\[0.5ex]
 287&     821A& 10 53 22.04& +57 28 52.76&  type-1 & 2.300&       SPL&$>$99.99&    99.8&$  0.25_{  0.14}^{  0.33}$\\[0.5ex]
 268&        -& 10 53 48.09& +57 28 17.75&    --&     -&       APL&    96.0&    99.1&$  0.17_{  0.06}^{  0.27}$\\[0.5ex]
 277&      25A& 10 53 44.85& +57 28 42.24&  type-1 & 1.816&       SPL&$>$99.99&    77.1&$  0.51_{  0.38}^{  0.62}$\\[0.5ex]
 272&      26A& 10 50 19.40& +57 28 13.99& type-2 & 0.616&       APL&    80.9&    33.0&$\le$  0.21\\[0.5ex]
 290&     901A& 10 52 52.74& +57 29 00.81& type-2 & 0.204&    APL+SE&    98.3&$>$99.99&$  0.13_{  0.03}^{  0.22}$\\[0.5ex]
 369&        -& 10 51 06.50& +57 15 31.92&    --&     -&       SPL&    96.9&    15.7&$\le$  0.31\\[0.5ex]
 300&     426A& 10 53 03.64& +57 29 25.56&  type-1 & 0.788&      CAPL&$>$99.99&    99.0&$  0.35_{  0.21}^{  0.45}$\\[0.5ex]
 306&        -& 10 52 06.84& +57 29 25.43& type-2 & 0.708&       APL&    98.3&    77.0&$  0.19_{  0.06}^{  0.30}$\\[0.5ex]
 321&      23A& 10 52 24.74& +57 30 11.40&  type-1 & 1.009&       SPL&$>$99.99&$>$99.99&$  0.20_{  0.07}^{  0.29}$\\[0.5ex]
 326&     117Q& 10 53 48.80& +57 30 36.09& type-2 &  0.78&       APL&$>$99.99&    96.7&$  0.37_{  0.26}^{  0.46}$\\[0.5ex]
 350&        -& 10 52 41.65& +57 30 39.97&    --&     -&       SPL&    43.1&    94.1&$\le$  0.12\\[0.5ex]
 332&      77A& 10 52 59.16& +57 30 31.81&  type-1 & 1.676&       SPL&$>$99.99&    98.4&$  0.31_{  0.23}^{  0.38}$\\[0.5ex]
 411&      53A& 10 52 06.02& +57 15 26.41& type-2 & 0.245&      CAPL&    95.4&     4.8&$\le$  0.19\\[0.5ex]
2024&        -& 10 54 10.68& +57 30 56.73&    --&     -&       SPL&    98.9&    99.6&$  0.15_{  0.07}^{  0.23}$\\[0.5ex]
 343&        -& 10 50 41.22& +57 30 23.31&    --&     -&       SPL&    70.5&    96.3&$\le$  0.32\\[0.5ex]
 342&      16A& 10 53 39.62& +57 31 04.89&  type-1 & 0.586&    SPL+SE&$>$99.99&    34.7&$  0.44_{  0.31}^{  0.55}$\\[0.5ex]
 351&        -& 10 51 46.39& +57 30 38.14&    --&     -&       SPL&    99.1&    37.4&$  0.19_{  0.08}^{  0.29}$\\[0.5ex]
 353&      19B& 10 51 37.27& +57 30 44.43&  type-1 & 0.894&       SPL&    94.7&    77.0&$  0.12_{  0.03}^{  0.20}$\\[0.5ex]
 354&      75A& 10 51 25.25& +57 30 52.33&  type-1 & 3.409&       SPL&$>$99.99&    80.9&$  0.35_{  0.24}^{  0.44}$\\[0.5ex]
 358&      17A& 10 51 03.86& +57 30 56.65&  type-1 & 2.742&       SPL&    49.8&    99.3&$\le$  0.08\\[0.5ex]
 355&        -& 10 52 37.33& +57 31 06.67& type-2 & 0.708&       APL&$>$99.99&    18.4&$\le$  0.86\\[0.5ex]
 385&      14Z& 10 52 42.37& +57 32 00.64& type-2 & 1.380&       APL&    99.6&$>$99.99&$  0.14_{  0.07}^{  0.19}$\\[0.5ex]
 364&      18Z& 10 52 28.36& +57 31 06.57&  type-1 & 0.931&       SPL&    93.2&    99.7&$  0.11_{  0.02}^{  0.19}$\\[0.5ex]
 901&        -& 10 50 05.55& +57 31 09.01&    --&     -&       SPL&    86.3&    35.1&$\le$  0.21\\[0.5ex]
 902&      73C& 10 50 09.12& +57 31 46.29&  type-1 & 1.561&       SPL&    92.4&    99.5&$\le$  0.35\\[0.5ex]
 377&        -& 10 52 52.11& +57 31 38.02&    --&     -&       APL&$>$99.99&    62.0&$  0.43_{  0.16}^{  0.62}$\\[0.5ex]
 384&        -& 10 53 21.63& +57 31 49.44&    --&     -&       APL&    64.0&    71.6&$\le$  0.15\\[0.5ex]
 387&      15A& 10 52 59.78& +57 31 56.69&  type-1 & 1.447&       SPL&$>$99.99&    86.2&$  0.34_{  0.22}^{  0.44}$\\[0.5ex]
 394&        -& 10 52 51.40& +57 32 02.03& type-2 & 0.664&       APL&    98.8&    21.0&$  0.24_{  0.10}^{  0.37}$\\[0.5ex]
 406&     828A& 10 53 57.16& +57 32 44.00&  type-1 & 1.282&       SPL&$>$99.99&    28.9&$  0.38_{  0.24}^{  0.49}$\\[0.5ex]
 419&        -& 10 54 00.46& +57 33 22.19&    --&     -&       APL&    97.3&    54.2&$  0.16_{  0.06}^{  0.27}$\\[0.5ex]
 407&      12A& 10 51 48.69& +57 32 50.07& type-2 & 0.990&      CAPL&$>$99.99&$>$99.99&$  0.30_{  0.22}^{  0.37}$\\[0.5ex]
 424&        -& 10 52 37.93& +57 33 22.65& type-2 & 0.707&    APL+SE&$>$99.99&$>$99.99&$  0.28_{  0.15}^{  0.39}$\\[0.5ex]
 427&        -& 10 52 27.88& +57 33 30.65& type-2 & 0.696&       SPL&$>$99.99&$>$99.99&$\le$  0.89\\[0.5ex]
 430&      11A& 10 51 08.19& +57 33 47.06&  type-1 & 1.540&       APL&    91.0&    95.3&$\le$  0.11\\[0.5ex]
 458&        -& 10 51 06.22& +57 34 36.67&    --&     -&      CAPL&$>$99.99&    94.7&$\le$  0.34\\[0.5ex]
 442&     805A& 10 53 47.28& +57 33 50.41&  type-1 & 2.586&       SPL&    95.7&    52.1&$  0.21_{  0.05}^{  0.36}$\\[0.5ex]
 443&        -& 10 52 36.89& +57 33 59.80& type-2 & 1.877&       APL&    69.7&    99.1&$\le$  0.20\\[0.5ex]
 474&        -& 10 51 28.13& +57 35 04.20&    --&    --&       SPL&    90.8&    99.9&$  0.11_{  0.02}^{  0.20}$\\[0.5ex]
 451&        -& 10 52 07.87& +57 34 17.48&    --&     -&       APL&    67.4&    97.0&$\le$  0.19\\[0.5ex]
 450&     477A& 10 53 05.98& +57 34 26.70&  type-1 & 2.949&       SPL&$>$99.99&$>$99.99&$  0.19_{  0.11}^{  0.25}$\\[0.5ex]
 453&     804A& 10 53 12.24& +57 34 27.39&  type-1 & 1.213&       SPL&    99.6&    45.2&$  0.17_{  0.08}^{  0.25}$\\[0.5ex]
 456&       9A& 10 51 54.30& +57 34 38.66&  type-1 & 0.877&       SPL&$>$99.99&    90.0&$  0.53_{  0.38}^{  0.66}$\\[0.5ex]
 491&        -& 10 51 41.91& +57 35 56.00&    --&     -&       APL&    91.3&    87.9&$  0.14_{  0.02}^{  0.26}$\\[0.5ex]
 469&        -& 10 54 07.21& +57 35 24.89&    --&     -&       SPL&$>$99.99&    38.8&$  0.43_{  0.29}^{  0.53}$\\[0.5ex]
 475&       6A& 10 53 16.51& +57 35 52.23&  type-1 & 1.204&       SPL&$>$99.99&     5.4&$  0.45_{  0.35}^{  0.55}$\\[0.5ex]
 476&     827B& 10 53 03.43& +57 35 30.80& type-2 & 0.607&       SPL&$>$99.99&    67.9&$  0.27_{  0.15}^{  0.37}$\\[0.5ex]
 505&     104A& 10 52 41.54& +57 36 52.85& type-2 & 0.137&      CAPL&$>$99.99&    99.6&$  0.47_{  0.32}^{  0.59}$\\[0.5ex]
 504&        -& 10 54 26.22& +57 36 49.05&    --&     -&    APL+SE&$>$99.99&    66.0&$\le$  0.54\\[0.5ex]
 511&        -& 10 53 38.50& +57 36 55.47& type-2 & 0.704&    APL+SE&     7.0&$>$99.99&$\le$  0.14\\[0.5ex]
 518&        -& 10 53 36.33& +57 37 32.14&    --&    --&       APL&    67.4&    60.8&$\le$  0.20\\[0.5ex]
 523&        -& 10 51 29.98& +57 37 40.71&    --&     -&       SPL&$>$99.99&    99.8&$  0.49_{  0.30}^{  0.62}$\\[0.5ex]
 529&        -& 10 51 37.30& +57 37 59.11&    --&     -&      CAPL&    99.0&    71.2&$  0.13_{  0.04}^{  0.21}$\\[0.5ex]
 532&     801A& 10 52 45.36& +57 37 48.69&  type-1 & 1.677&       SPL&    76.9&    38.7&$\le$  0.23\\[0.5ex]
 527&       5A& 10 53 02.34& +57 37 58.62&  type-1 & 1.881&       SPL&    92.6&     4.3&$  0.07_{  0.02}^{  0.12}$\\[0.5ex]
 537&        -& 10 50 50.04& +57 38 21.79&    --&     -&       SPL&$>$99.99&    84.2&$\le$  1.26\\[0.5ex]
 548&     832A& 10 52 07.53& +57 38 41.40&  type-1 & 2.730&       SPL&    95.1&$>$99.99&$  0.20_{  0.04}^{  0.33}$\\[0.5ex]
 557&        -& 10 52 07.75& +57 39 07.49&    --&     -&       APL&     1.1&     4.5&$\le$  0.12\\[0.5ex]
 553&       2A& 10 52 30.06& +57 39 16.81&  type-1 & 1.437&       APL&$>$99.99&    99.9&$  0.32_{  0.24}^{  0.40}$\\[0.5ex]
 555&        -& 10 51 52.07& +57 39 09.41&    --&     -&       SPL&    98.8&    17.9&$  0.18_{  0.02}^{  0.29}$\\[0.5ex]
 594&        -& 10 52 48.40& +57 41 29.14&    --&     -&       SPL&    99.9&    80.0&$\le$  0.58\\[0.5ex]
2045&        -& 10 52 04.47& +57 41 15.65&    --&     -&       SPL&$>$99.99&    99.9&$\le$  0.59\\[0.5ex]
 584&        -& 10 52 06.28& +57 41 25.53&    --&     -&       SPL&$>$99.99&    97.7&$  0.28_{  0.13}^{  0.39}$\\[0.5ex]
 591&        -& 10 52 23.17& +57 41 24.62&    --&    --&       SPL&    88.6&    89.5&$  0.11_{  0.02}^{  0.20}$\\[0.5ex]
 601&        -& 10 51 15.91& +57 42 08.59&    --&     -&       SPL&    76.9&    81.3&$\le$  0.21\\[0.5ex]
\hline
\hline
\label{summary_var_det}
\end{longtable}
 Columns are as follows: (1) Source X-ray identification number; 
(2) {\sl ROSAT} identification number;
(3) Right ascension (J2000); (4) Declination (J2000);
(5) object class based on optical spectroscopy;
(6) redshift;
(7) Best fit model of the co-added spectrum of each individual source (from Mateos et al.(2005b)~\cite{Mateos05b});
(8) Significance of 0.2-12 keV flux variability (in \%);
(9) Significance of spectral variability (in \%);
(10) Variability amplitude. 
Errors correspond to the 1$\sigma$ confidence interval.
For the objects where the lower error bar reached zero 68\% upper limits are given.
\twocolumn

We have calculated the mean probability distribution of the excess variance, 
$P(\sigma_{{\rm intr}})$, for our sample of sources as the unweighted average of the probability distributions of 
$\sigma_{{\rm intr}}$ obtained from each light curve. As we mentioned before, because we want to describe the 
X-ray flux variability properties for the sample as a whole, we have included in the calculation the probability 
distributions of $\sigma_{intr}$ for both variable and non variable sources. These mean probability distributions are shown in 
Fig.~\ref{dist_qdp_all} for all sources (top) and for the two AGN samples (bottom). 
The most probable values (modes) and the corresponding 1$\sigma$ uncertainties obtained from these 
distributions are listed in column 6 of Table~\ref{tab_sigmaq}.
The mode for the whole sample of sources 
is $\sigma_{{\rm intr}}$$\sim$0.15 (68\% upper limit =0.36), consistent with the value obtained from the arithmetic mean 
of the values for each source, as expected.

We see that the average probability distributions of $\sigma_{{\rm intr}}$ for type-1 and type-2 AGN do not differ significantly, although the most probable value of $\sigma_{{\rm intr}}$ 
is marginally lower for type-2 AGN than for type-1 AGN. 

The flux variability properties for all our sources (significance of detection and amplitude 
of variability) are listed 
in Table~\ref{summary_var_det}. The first and second columns list the {\sl XMM-Newton} and {\sl ROSAT}
identification number of the sources. Columns 3 and 4 show the coordinates of the sources while 
columns 5 and 6 list the optical class and spectroscopic redshift. Column 7 lists the best fit model 
from the spectral analysis of the co-added spectra of our objects (from Mateos et al.~\cite{Mateos05b}). The next three columns list the results of the detection of flux and spectral variability.   

\subsection{Probability distribution of the excess variance}
\label{sims_sigma} 
\begin{figure}[!tb]
    \hbox{
    \hspace{-1cm}
    \includegraphics[angle=90,width=0.55\textwidth]{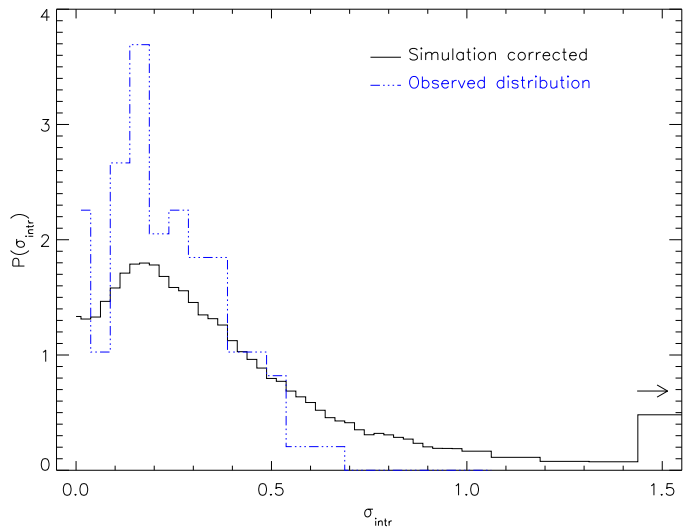}}
    \caption{Distribution of intrinsic amplitude of flux variability, $P(\sigma_{{\rm intr}})$, obtained from our simulations (solid line). Note that in our simulations (see Appendix~\ref{appendix_B}) we referred to the measured amplitude of flux variability as $\sigma_{obs}$ or $S_o$ instead of $\sigma_{intr}$.
    The distribution of values obtained for our sample 
    of sources (excluding objects with less than 5 points in their light curves, see Sec.~\ref{sims_sigma}) 
    is shown for comparison (dashed line). 
      }
    \label{psi_dist}
\end{figure}

There are a number of factors that could be affecting the values and 
distribution of measured excess variances that we utilised to model the flux variability. The most clear 
example is the low quality in our light curves. 
We have made simulations in order to obtain the 
true probability distribution of excess variance for our sources, after accounting for all the selection 
effects in our sample of objects. The details on how the simulations were carried out are given in Appendix~\ref{appendix_B}. These simulations assume that the flux variability properties are not dependent on the source's flux (or count 
rate), an assumption that is consistent with our analysis. 

Our simulations have shown that the observed amplitude of variability depends strongly on the number of 
points in the light curves, in the sense that systematically lower amplitudes of variability are measured 
in objects with smaller number of points in the light curves. Therefore in this section we use only the 103 
sources with at least 5 points in their light curves. The ``corrected'' probability distribution, 
$P(\sigma_{{\rm intr}})$, obtained using our simulations is shown in 
Fig.~\ref{psi_dist} (solid line). To witness the effects that go into this, we also have included the 
observed probability distribution of $\sigma_{{\rm intr}}$ values (dashed-line) for our 
sources, excluding objects with less than 5 points in their light curves (see Fig.~\ref{hist_excess_var}).

We see that both distributions peak at similar values of $\sigma_{{\rm intr}}$$\sim$0.2 and therefore 
effects as the ones listed before, are not affecting significantly the mode of the distribution
of amplitudes of flux variability.
However the corrected distribution of intrinsic excess variances shows a long tail towards high 
values of $\sigma_{{\rm intr}}$ ($\ge$0.6), which we failed to detect in the real data.
During our simulations work we found that for sources with light curves with the lowest number of 
bins, we always measured values of the variability amplitude significantly lower than the input value, no matter 
how large this value was (see Fig.~\ref{bad_srcs}). Hence, we can explain the scarcity of sources 
in which we have detected large ($\ge$0.6) variability amplitudes 
as being due to a small number of points in the light curve.

\subsection{Dependence of flux variability with luminosity and redshift}
\label{var_amplitude_vs_lumin}
\begin{figure*}[!tb]
    \hbox{
    \includegraphics[angle=90,width=0.5\textwidth]{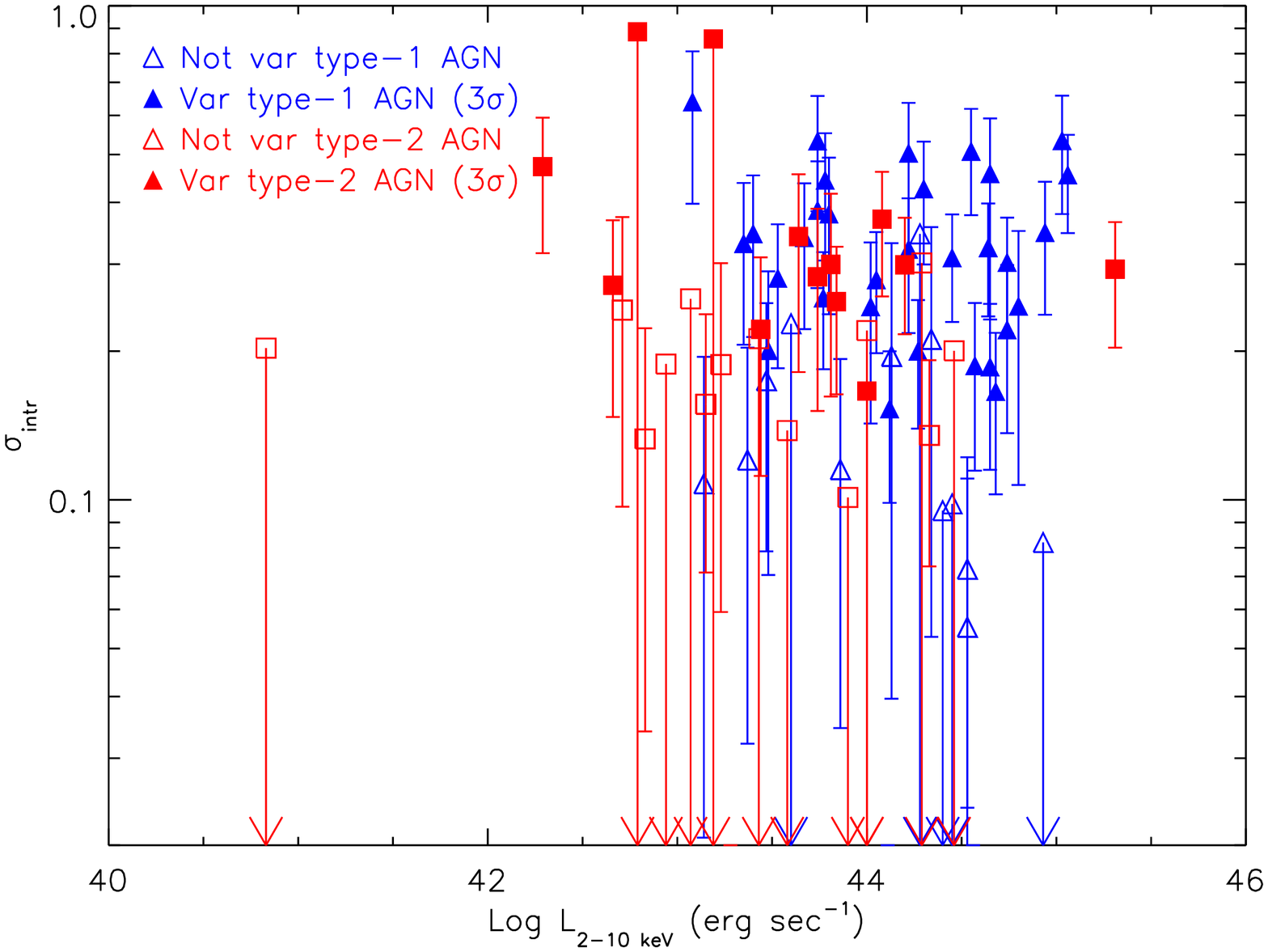}
    \includegraphics[angle=90,width=0.5\textwidth]{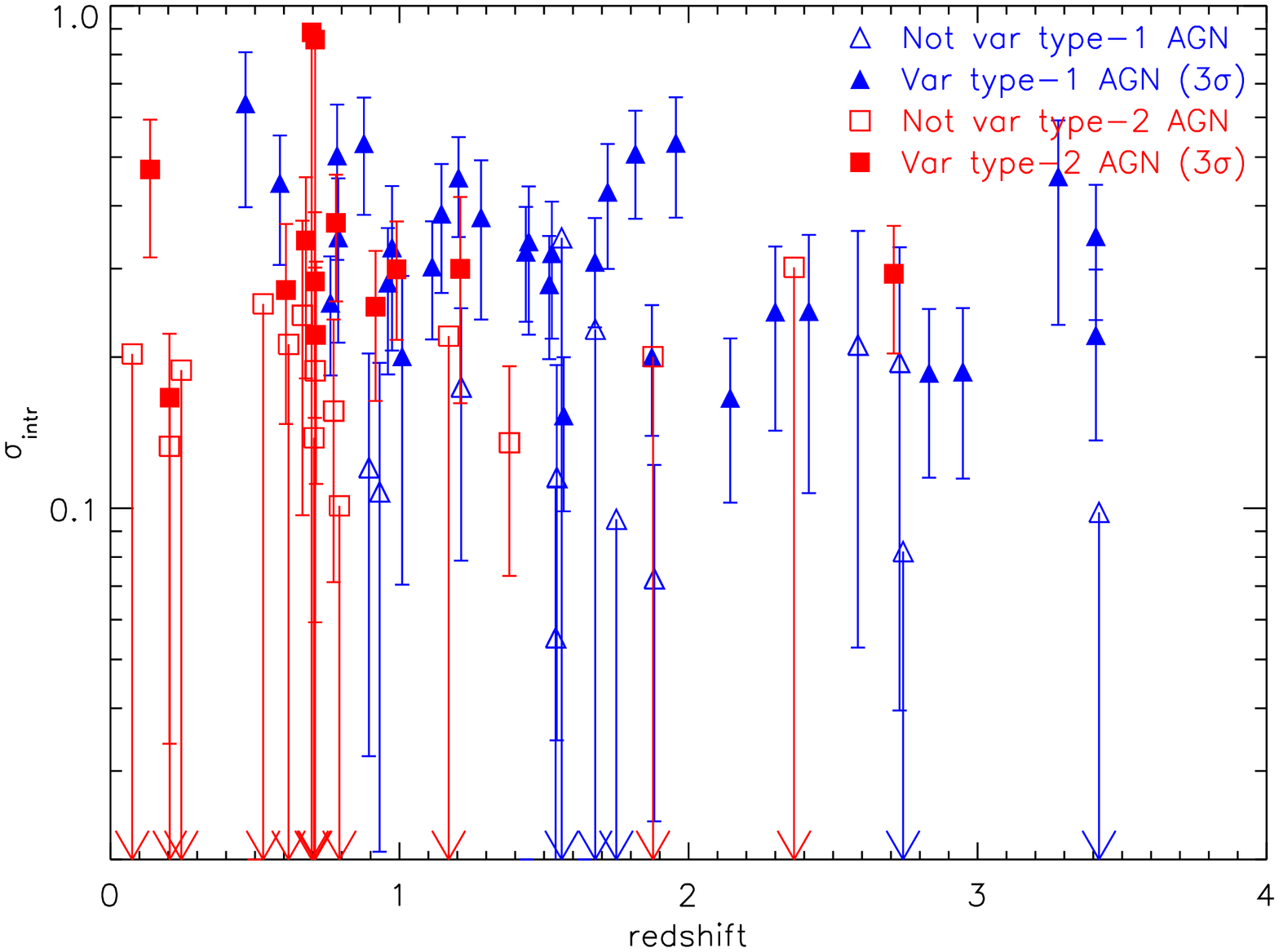}}
    \caption{
      Dependence of the excess variance, $\sigma_{{\rm intr}}$, 
      on the absorption corrected 2-10 keV luminosity (obtained from the best fit model of each object, left) and 
      redshift (right) for our sample of AGN. Errors correspond to the 1$\sigma$ confidence interval. 
      }
    \label{var_evolution}
\end{figure*}

It has been suggested that flux variability amplitude, when measured on a fixed temporal frequency, 
correlates inversely with X-ray luminosity (this has been confirmed on short time scales, $\sim$1 day) 
for nearby Seyfert 1 galaxies 
(Barr et al.~\cite{Barr86}; Lawrence et al.~\cite{Lawrence93}; Nandra et al.~\cite{Nandra97}), in the sense that 
more luminous sources show lower variability amplitudes than less luminous sources. 
However there is significant scatter in this correlation on both short and long time scales, 
which has been attributed for example, to a dependence of the amplitude of variability on the spectral 
properties of the sources (Fiore et al.~\cite{Fiore98}; Green et al.~\cite{Green93}). 
Furthermore, there are also 
indications that the strength of the correlation 
decreases towards longer time scales (Markowitz et al.~\cite{Markowitz04}) and might be weaker 
for sources at higher redshifts (Almaini et al.~\cite{Almaini00}; Manners et al.~\cite{Manners02}).

\begin{table}[!tb]
\caption{Amplitudes of variability obtained from the mean probability distributions of $\sigma_{{\rm intr}}$ for 
type-1 AGN at different luminosities and redshifts.}
\begin{center}
\begin{tabular}{lcccccccc}
\hline
Group & $\sigma_{{\rm intr}}$ \\
(1) & (2) \\
\hline
\hline
 type-1 AGN z$<$1.5 & $0.32_{-0.23}^{+0.17}$ \\[0.5ex]
 type-1 AGN z$>$1.5 & $0.18_{-0.17}^{+0.17}$ \\[0.5ex]
 type-1 AGN $L_{\rm (2-12\,keV)}>10^{44}$ & $\le$0.38 \\[0.5ex]
 type-1 AGN $L_{\rm (2-12\,keV)}<10^{44}$ & $0.29_{-0.20}^{+0.20}$  \\

\hline 
\end{tabular}
\label{tab_amp_sigma_agn1}
\begin{list}{}{}
Columns are as follows: (1) type-1 AGN group; (2) variability amplitude, $\sigma_{{\rm intr}}$, obtained from 
the mode of the mean probability distributions, $P(\sigma_{{\rm intr}})$, of each group of sources.
Errors correspond to the 1$\sigma$ confidence interval.
In the cases where the lower error bound of the integrals reached zero we calculated 90\% upper limits 
for $\sigma_{intr}$.    
\end{list}
\end{center}
\end{table}

\begin{figure*}[!tb]
    \hbox{
    \hspace{0.0\textwidth}\includegraphics[angle=90,width=0.5\textwidth]{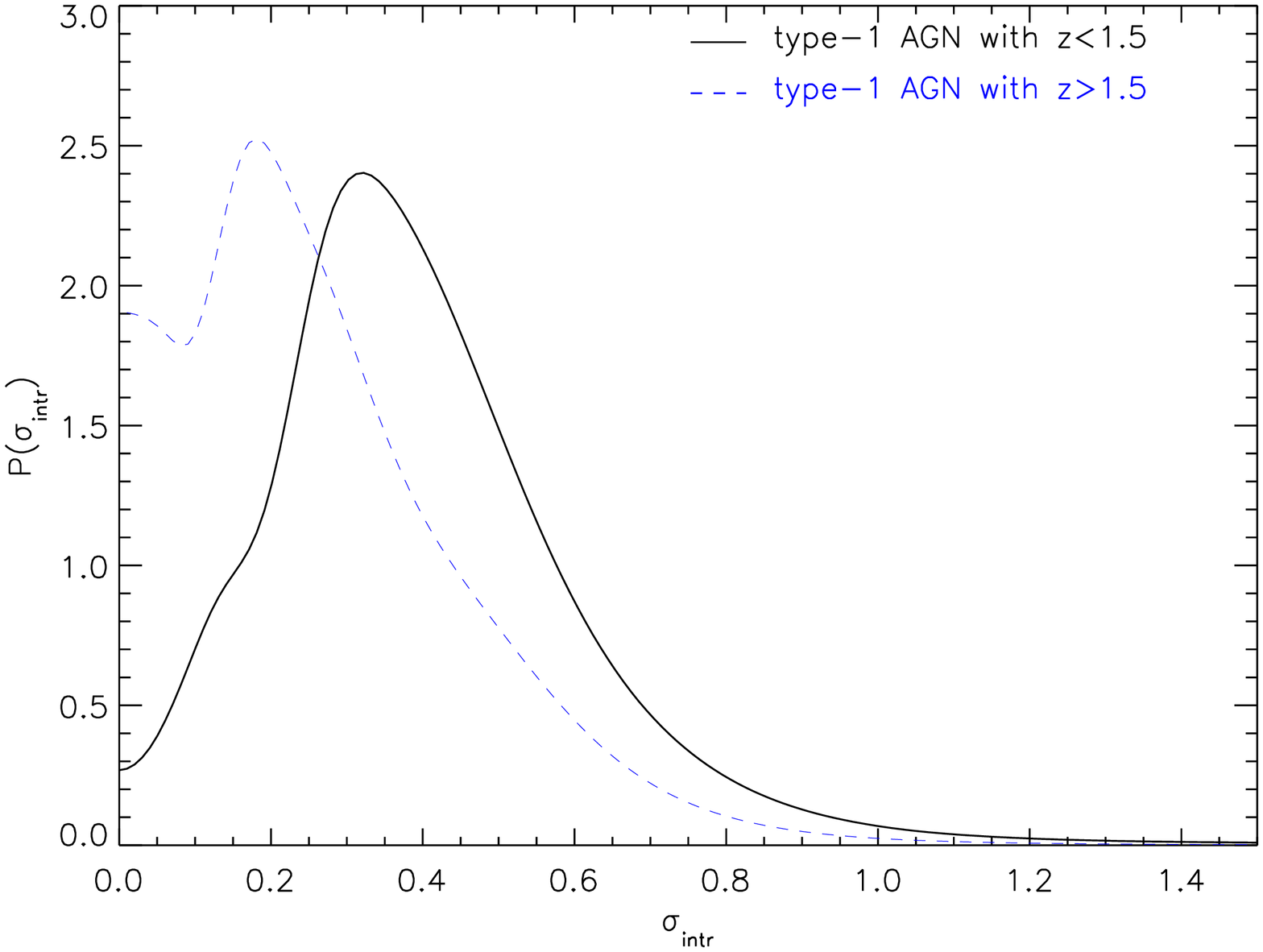}
    \hspace{0.0\textwidth}\includegraphics[angle=90,width=0.5\textwidth]{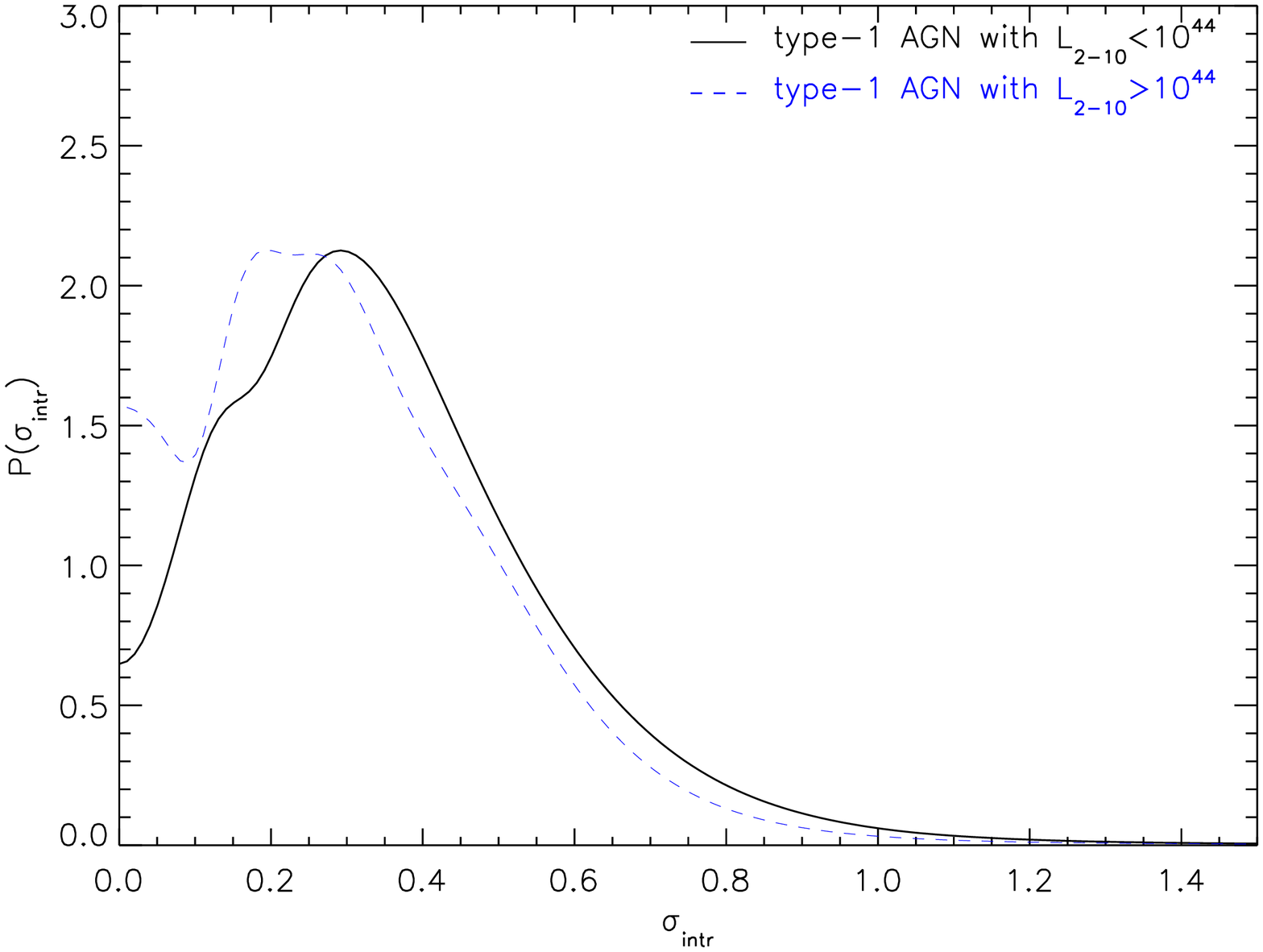}}
    \caption{
      Left: Average probability distributions of $\sigma_{{\rm intr}}$ 
      for sources with redshifts below 1.5 and above 1.5. 
      Right: Average probability distributions of $\sigma_{{\rm intr}}$ 
      for sources at 2-10 keV luminosities above and below 
      $10^{44}\,{\rm erg\,s^{-1}}$.
      The probability distributions were obtained using all AGN, variable and non variable in terms of 
      the $\chi^2$ test. 
    }
    \label{evolv_dist_qdp}
\end{figure*}

We have analysed whether there exists any dependence of the detected flux variability in our sources 
with their X-ray luminosity and redshift. 
Fig.~\ref{var_evolution} (left) shows the measured variability amplitudes for our AGN
as a function of the 2-10 keV luminosity. 
We have used the absorption-corrected 2-10 keV luminosities obtained from the best fit of the co-added spectrum 
of each object. The values we used are reported in Table 8 of Mateos et al. (\cite{Mateos05b}).
In order to determine the significance of any correlation between the two quantities we used the
version of the generalised Kendall's tau test provided in the Astronomy Survival Analysis package 
(ASURV; Lavalley et al. \cite{Lavalley92}) which implements the methods described in Isobe et al. \cite{Isobe86} 
for the case of censored data. The results of the Kendall's $\tau$ test including the probability that a correlation is present are 
$\tau=0.05\,(4\%)$ for $\sigma_{intr}$-${\rm L_{2-10 keV}}$ 
and $\tau=2.25\,(97.7\%)$ for $\sigma_{intr}$-redshift. 
We have derived linear regression parameters using the ''estimate and maximise" (EM) and 
the Buckley-James regression methods also included in the ASURV package, however as 
both methods agreed within the errors, we only give the results from the EM test.
We found the relations between $\sigma_{intr}$-${\rm L_{2-10 keV}}$ and $\sigma_{intr}$-redshift to be 

\begin{equation}\sigma_{intr}=(-0.012\pm0.044)\times\,\log({\rm L_{2-10\,keV}})+ (1.054\pm1.937) \end{equation}
and
\begin{equation}\sigma_{intr}=(-0.062\pm 0.028)\times\,{\rm z}+ (0.378\pm 0.053) \label{eq_corr} \end{equation}

We see that we do not find any anticorrelation between excess variance and luminosity. 
However it is important to note that the known anticorrelation was found when using the same rest-frame  
frequency interval for all sources, while the light curves of our sources are not uniformly 
distributed in time and cover much longer time scales than the ones used in those works. 

\begin{figure*}[!htb]
    \hbox{
    \includegraphics[angle=90,width=0.33\textwidth]{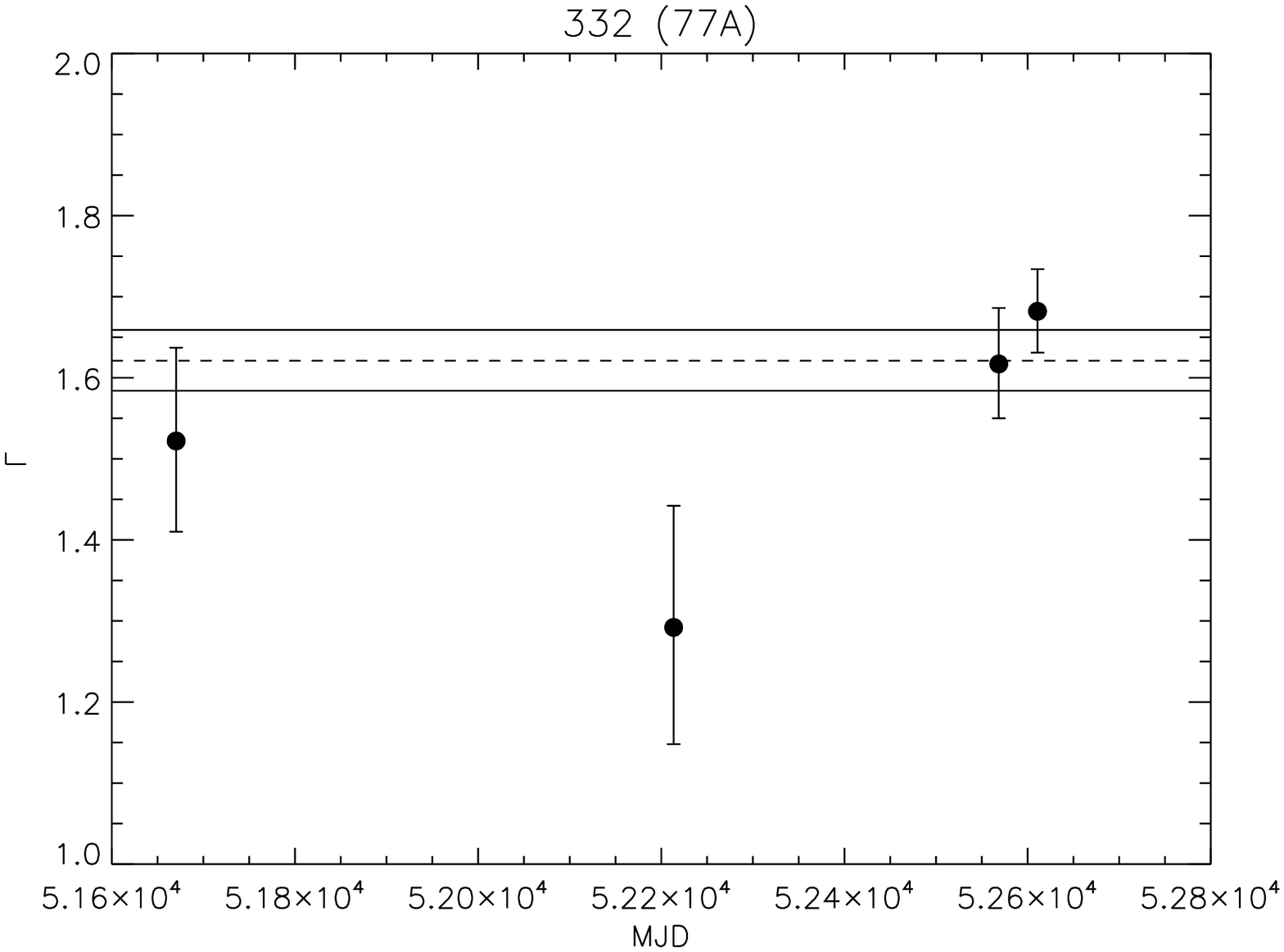}
    \includegraphics[angle=90,width=0.33\textwidth]{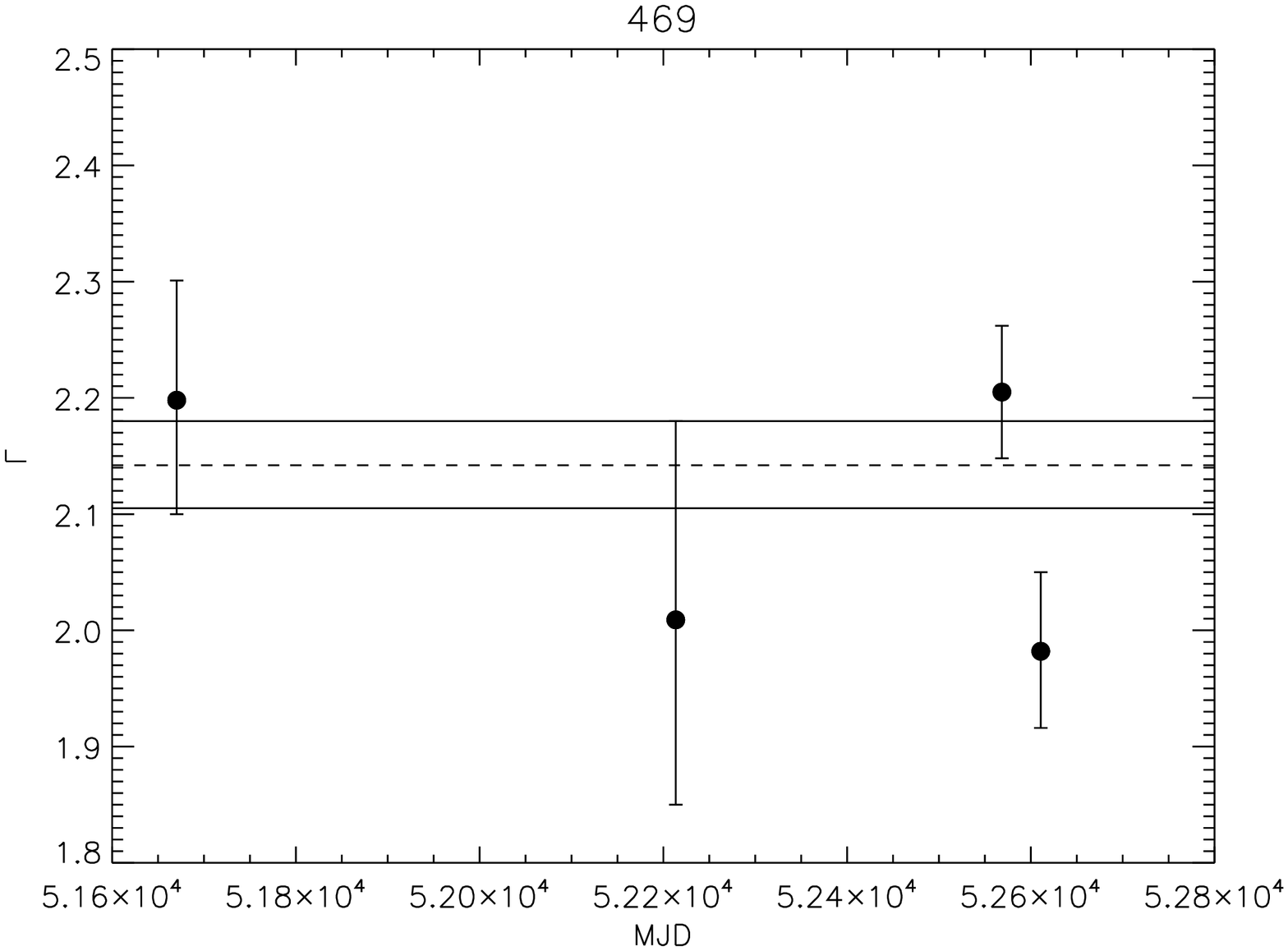}
    \includegraphics[angle=90,width=0.33\textwidth]{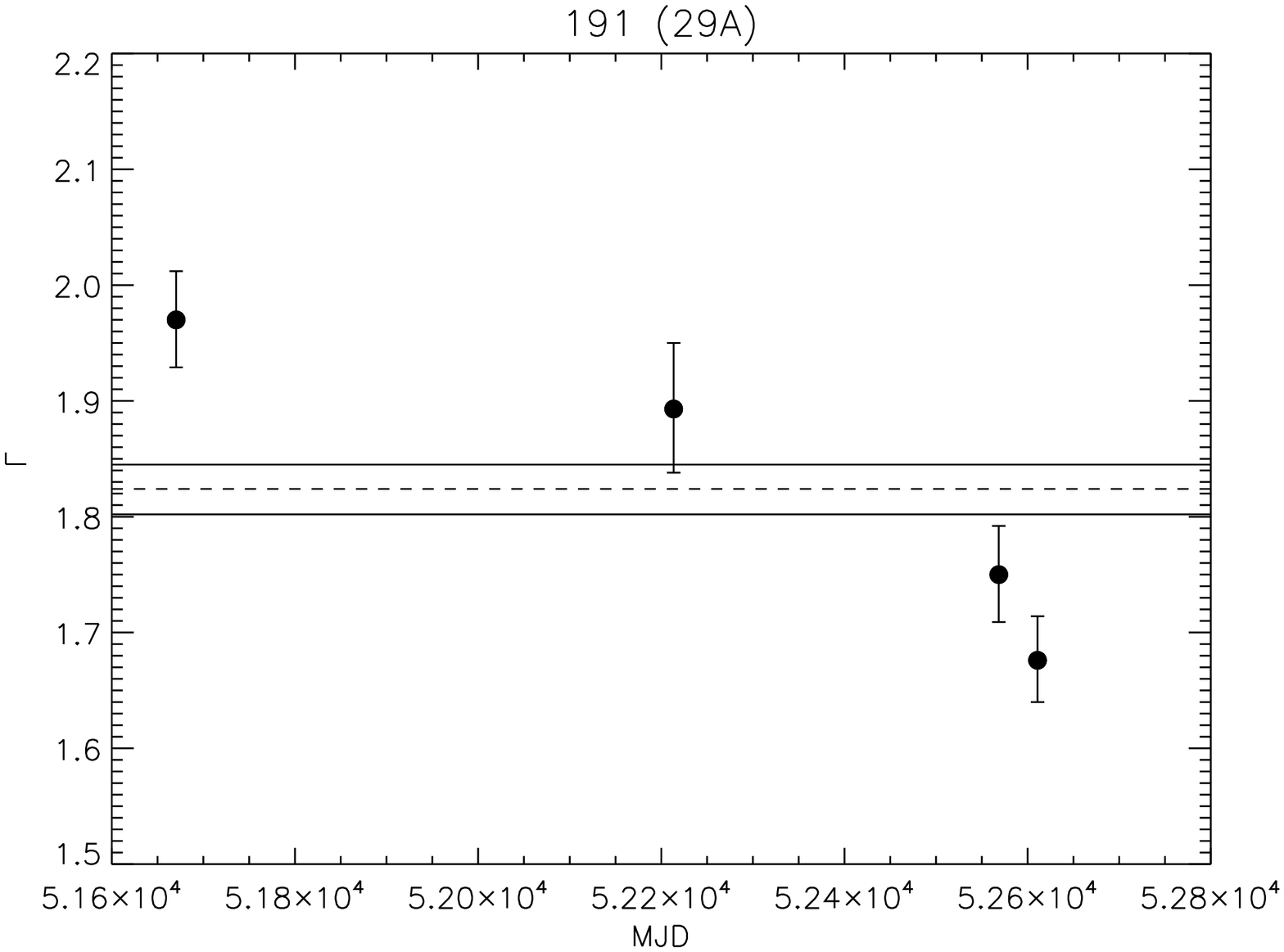}}
    \caption{Observed spectral variability of three objects in our sample for which their 
    co-added spectra were best fitted with a simple power law. Variation of the continuum shape 
    ($\Gamma$) with time (errors correspond to the 90\% confidence interval). 
    Horizontal lines indicate the best fit continuum shape (dashed lines) and 
    the corresponding 90\% confidence intervals (solid lines) measured in the co-added spectra.
    } 
    \label{var1}
\end{figure*}

We have used the same observed energy band to study the variability 
properties of our sources, but because our sources span a broad range in redshifts we are sampling 
different rest-frame energies (harder energies for higher redshift sources). 
This could be a problem when comparing observed variability properties between sources, if there exists 
a dependence of variability properties with energy (stronger variability in the soft band compared to 
the hard band, has been observed in a number of Seyfert 1 galaxies).
By plotting the excess variance versus 
redshift we should be able to see whether this effect is present in our data.
This dependence of $\sigma_{{\rm intr}}$ with redshift is shown in 
Fig.~\ref{var_evolution} (right). The results from the generalised Kendall's tau test (probability 
of detection of 97.7\% ) and from the linear regression analysis (Eq.\ref{eq_corr}) suggest that there is 
a weak correlation between $\sigma_{{\rm intr}}$ and the redshift.

To enhance any underlying correlation between the amplitude of variability and the X-ray 
luminosity or redshift, we have obtained average probability distributions for our sample of type-1 AGN (we cannot repeat the 
experiment for type-2 AGN as the sample size is too small) in bins of redshift and luminosity. In order 
to have enough objects per bin and enough data points we have used only two bins in 
redshift and luminosity (all bins having the same number of objects). 
The results are shown in Fig.~\ref{evolv_dist_qdp} while the values of $\sigma_{{\rm intr}}$ obtained from 
these distributions (modes) are shown in Table~\ref{tab_amp_sigma_agn1}. 
The variability amplitude appears to be independent of the 2-10 keV luminosity, confirming the 
above results. The same result holds for the dependence of variability 
amplitude with redshift as we see that there is some indication that the amplitude of flux variability 
is lower for higher redshift sources, although the effect is not very significant.

In addition, we do not find that the detection of flux variability changes with redshift.
In summary, neither the fraction of sources with variability or its amplitude change with redshift in our sample. 

\section{Spectral variability}
\label{sp_var}
Comptonization models are most popular in explaining the X-ray emission in 
AGN (Haardt et al.~\cite{Haardt97}). In these models the X-ray emission is 
produced by Compton up-scattering of optical/UV photons in a hot electron corona above the accretion disc. 
One of the predictions of these models is that, if the nuclear emission increases, the Compton cooling in the corona 
should increase too, and therefore the corona will become colder, and the X-ray spectrum becomes softer.

This flux-spectral behaviour is similar to the one observed in different states 
of BHXRB (McClintock et al. \cite{McClintock03}): 
in the {\it low}/{\it hard state}\footnote{The {\it soft state} is usually seen at higher 
2-10 keV luminosity than the {\it hard state} motivating the names {\it high}/{\it soft} and {\it low}/{\it hard} 
states for BHXRBs.}, 
most commonly observed in these objects, their X-ray fluxes are low 
and the 2-10 keV X-ray spectra are dominated by non thermal emission, best reproduced by 
a hard ($\Gamma$$\sim$1.5-2.1) power law, along with a weak or undetected thermal component 
(multi-temperature accretion disk component). In the {\it high}/{\it soft state} their fluxes are high 
and their 2-10 keV spectra are soft, dominated by the thermal component with a temperature $\sim$1 keV. 

The accretion disk-static hot corona model of Haardt et al. (\cite{Haardt97}) predicts that, if the corona is 
dominated by electron-positron pairs, an intensity variability of a factor of 10 over the 2-10 keV rest-frame band 
will correspond to a change in the spectral slope of $\Delta\Gamma$$\sim$0.2. The relation between spectral shape 
and intensity is less clear for a corona not dominated by pairs, but qualitatively
significant spectral variations could be obtained for small changes in the observed count rate in this case.

In order to provide more insight into the nature of the X-ray variability in our objects,
we have studied which fraction of the sources in our sample show spectral variability and in these cases whether 
there exists some correlation between the observed flux and spectral variability.

\begin{figure*}[!tb]
    \hbox{
    \hspace{-0.03\textwidth}\includegraphics[angle=90,width=0.50\textwidth]{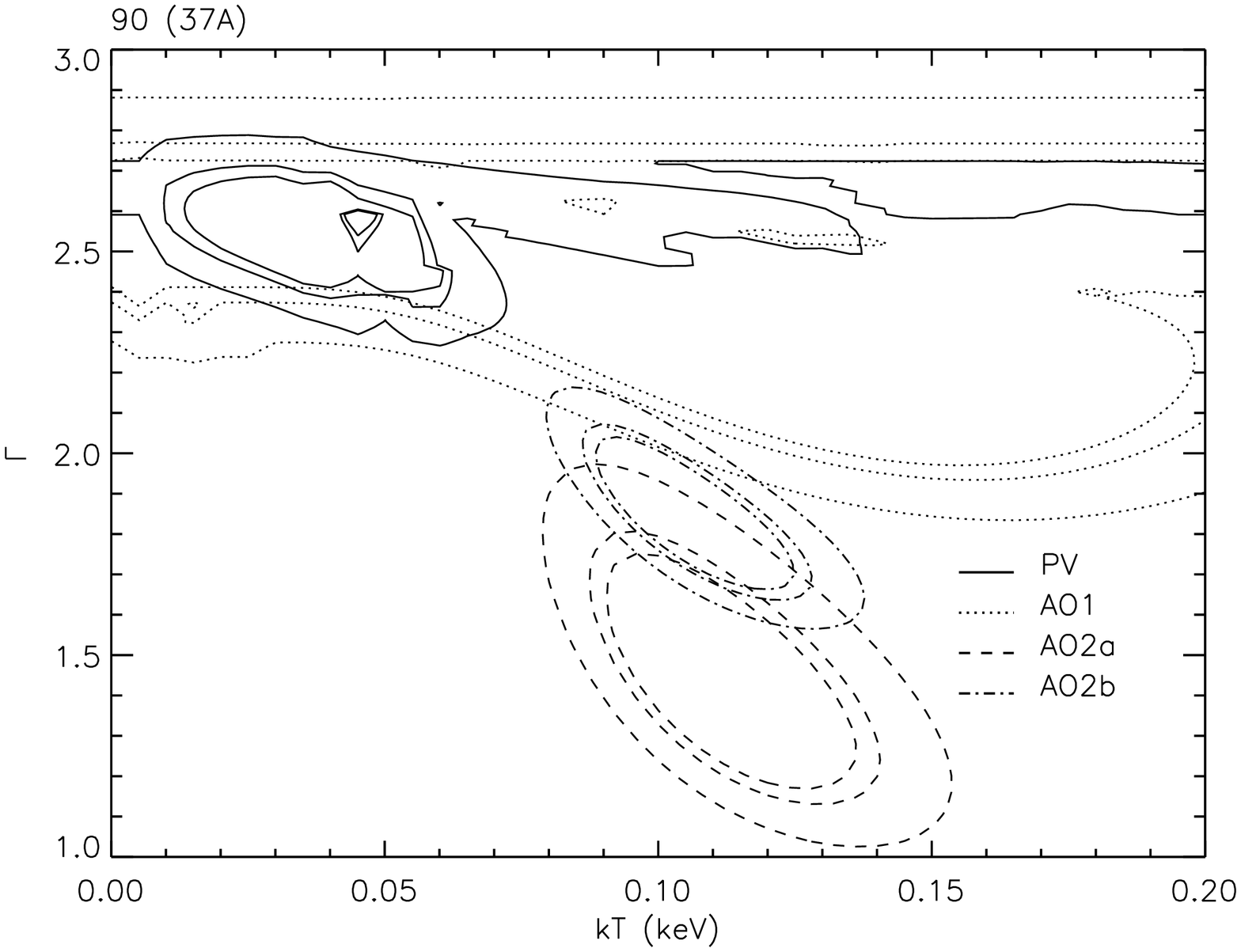}
    \hspace{-0.03\textwidth}\includegraphics[angle=90,width=0.50\textwidth]{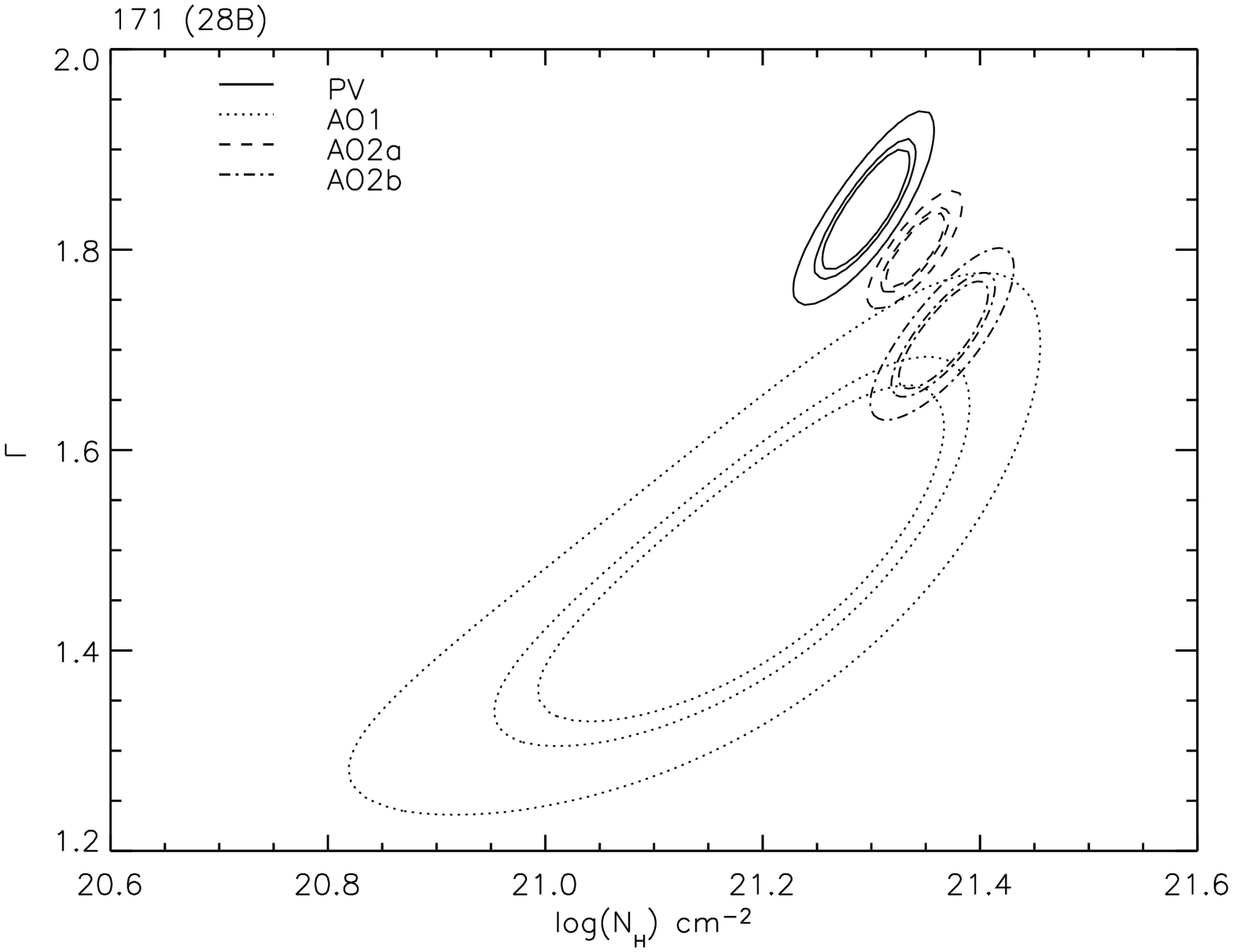}}
    \caption{
      Contour diagrams showing the observed spectral variability in type-1 and type-2 AGN.
      Contours correspond to 1$\sigma$, 2$\sigma$ and 3$\sigma$ confidence}
    \label{var2}
\end{figure*}

\subsection{Variability of the broad band X-ray colours}
\label{sp_var_colours}
Because our sources are typically faint, we cannot use their 
spectra from each revolution, as the uncertainties in the measured spectral parameters will be in most cases too 
large. Instead, we have used a broad band X-ray colour 
or hardness ratio ({\it HR}), that allows us to search for changes in the broad band spectral shape
of the sources. 
We calculated for each source a ``colour'' curve where each point 
is the X-ray hardness ratio from each observation which is obtained as 
\begin{equation} {\it HR^r=(CR_h^r-CR_s^r)/(CR_h^r+CR_s^r)} \end{equation}
where ${\it CR_s^r}$ and ${\it CR_h^r}$ are 
corrected count rates (see Appendix~\ref{appendix_A}) in the 0.5-2 keV (soft) and 2-12 keV (hard) energy bands for 
revolution {\it r} respectively. We used these two energy band definitions to separate better 
the most important spectral components found in the co-added spectra of our objects.
Changes in the intrinsic absorbing column density or in the properties of the soft excess emission component  
will affect mostly the soft count rate, while changes in the shape of the broadband continuum will affect 
both soft and hard band count rates. 

In order to search for objects with spectral variability, we 
used the $\chi^2$ test (see Sec.~\ref{flux_var}) and a threshold in the significance of detection of 3$\sigma$.
Results are summarised in columns 5 and 6 of 
Table~\ref{tab_fracc_det_var}. The fractions of sources with detected spectral variability 
have been corrected for the expected spurious detections at the selected 
confidence level as shown in Mateos et al. (\cite{Mateos05a}) and Stevens et al. (\cite{Stevens05}). 
We have detected spectral variability above a confidence level of 3$\sigma$
in 24 out of the 120 objects with light curves with at least 2 points. 
Spectral variability was detected in 6 out of 45 type-1 AGN and 9 out of 27 type-2 AGN. Among sources 
still not identified we detected spectral variability in 9 out of 48 (but note again the 
possible identification bias). 
These results indicate that spectral variability is much less common than flux variability. This result 
also holds for both samples of
 type-1 and type-2 AGN. This is not an unexpected result, as we know that with a limited 
number of counts, spectral variability is more difficult to detect than flux variability. 
As we will see in Sec.~\ref{origin_var} (see Fig.~\ref{pivoting}), for a spectral pivoting model, 
a change in spectral slope of $\Delta\Gamma$$\sim$0.2 
corresponds to a change in the observed X-ray colour of $\Delta${\it HR}$\sim$0.1 while $\Delta${\it HR}$\sim$0.2 
requires $\Delta\Gamma$$\sim$0.4-0.6. The typical dispersion in {\it HR} values observed in our sources with detected 
spectral variability is $\sim$0.1-0.2 while most of the measurement errors in {\it HR} in a single 
observation are of similar size.
Therefore changes in the continuum shape $\Delta\Gamma$$\le$0.3 will be detectable only in a small 
number of objects in our sample. We will come back to this point in more detail in Sec.~\ref{fracc_var} and Sec.~\ref{origin_var}. 

Finally, we have not found evolution with redshift of the fraction of objects with detected spectral variability, indicating that the effect of sampling harder rest-frame energy bands at higher redshifts does not reduce 
the detection of spectral variability. 

We have compared the fractions of type-1 and type-2 AGN with detected spectral variability using the 
method described in Mateos et al. (\cite{Mateos05a}). We found that 
the significance of these fractions being different is 99\%. 
This constitutes marginal evidence that spectral variability might be more common in type-2 AGN, although 
we should recall that we are dealing with a small number of sources.
If true, this might be due to the contribution to the X-ray emission from 
scattered radiation being higher in type-2 AGN. This scattered component 
will not be variable, as it will be related to the very distant reflector. 
The detected X-ray emission in type-2 AGN is dominated by the hard
X-ray component (i.e., in the nuclear emission transmitted through the absorbing material) while most of the non-variable soft X-ray emission will not be significantly absorbed. Therefore, changes in the intensity of 
the hard X-ray component alone would result in larger changes in the observed X-ray colour.

\subsection{Variability of the X-ray emission components}
\label{sp_var_sp}
In order to give more insight into the nature of the detected spectral variability in our objects,
we grouped the {\sl XMM-Newton} observations in four different periods of time 
corresponding to the {\sl XMM-Newton} phases PV, AO1 and AO2a and AO2b (see second column in 
Table~\ref{tab_observations}). As we have said before, most of the objects in our sample are too faint to 
search for spectral variability comparing the emission properties measured on the X-ray spectrum from each 
revolution. A significant fraction of the exposure time was 
achieved during the AO2 phase, so we separated the AO2 data into two contiguous groups and still obtained
data with enough quality for this analysis. 
We built for each object a co-added spectrum 
using the data from each observational phase. This approach allowed us to conduct standard spectral 
fitting and study spectral variability 
directly on time scales of months and years. We were able to 
carry out this study in 109 out of 123 objects for which we had more than one spectrum available. 
In order to search for spectral variability, 
we fitted the spectra of each observation phase with the best fit model (see column 7 in 
Table.~\ref{summary_var_det}) and best fit parameters obtained 
from the overall spectrum of each object (see Table 8 in Mateos et al.~\cite{Mateos05b}). 

\begin{figure}[!tb]
    \hbox{
    \includegraphics[angle=90,width=0.5\textwidth]{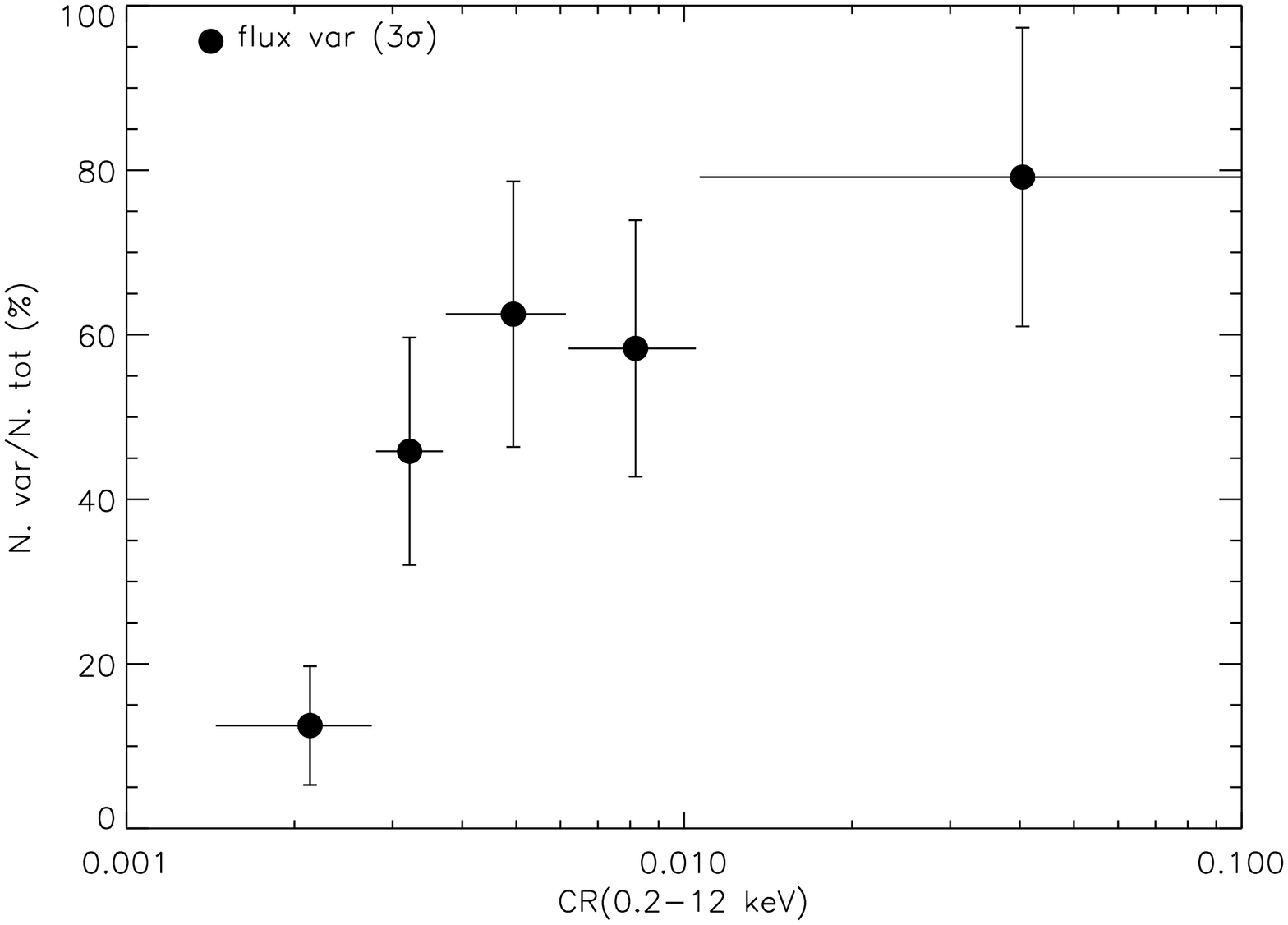}}
    \hbox{
    \includegraphics[angle=90,width=0.5\textwidth]{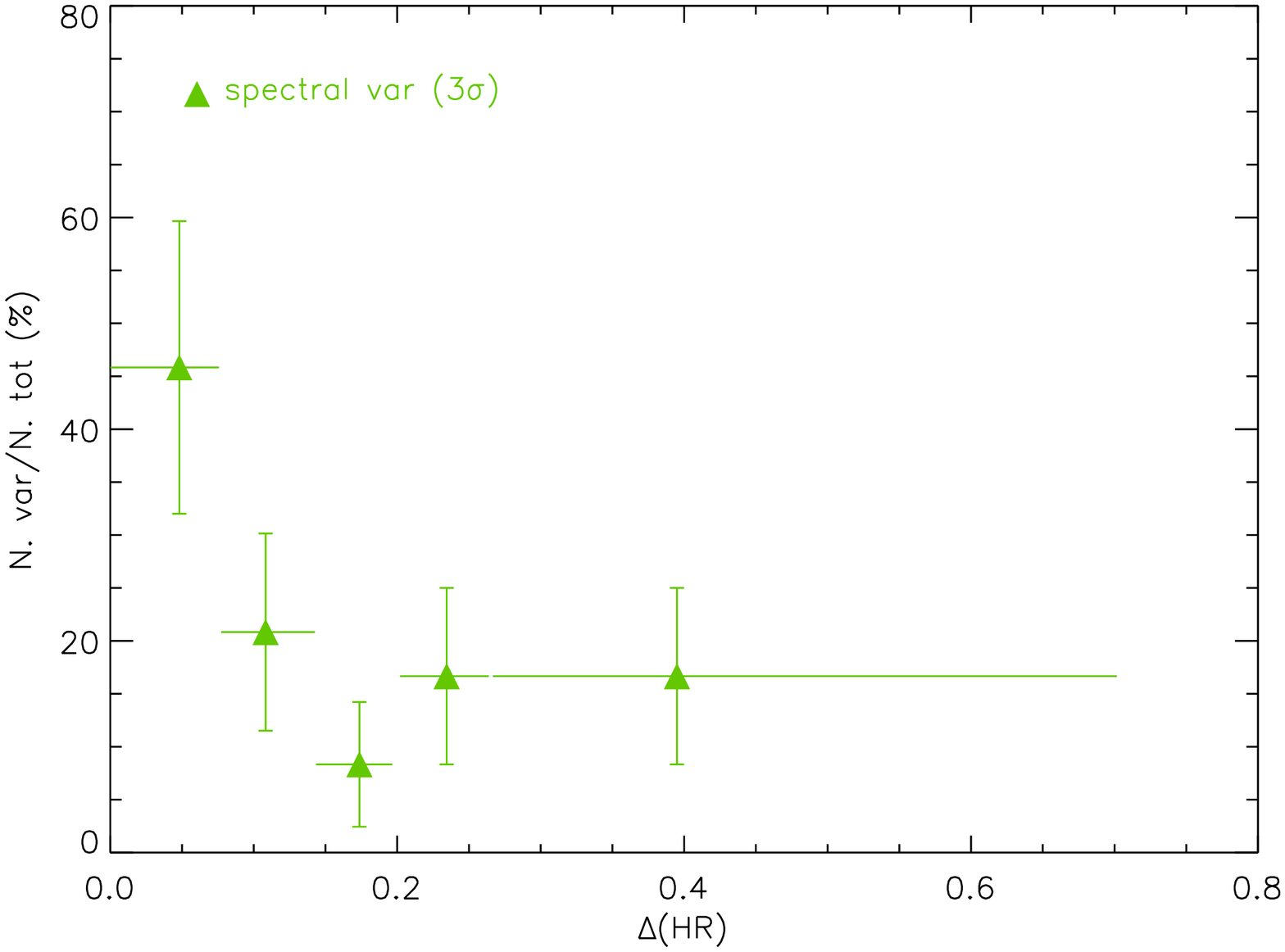}}
    \caption{Top: Fraction of sources with detected flux variability as a function of 
    the mean 0.2-12 keV count rate of the objects. Bottom: Fraction of sources with detected spectral 
    variability as a function of the error in their mean X-ray colour.}     
    \label{var_vs_qual2}
\end{figure}
We left free the normalisation of the 
continuum emission since we are only interested here in searching for spectral variability, and we have
seen that flux variability is common. For sources with detected 
soft excess emission, we kept fixed the ratio of the soft excess to the power law normalisations that we found 
in the best fit of the co-added spectra. 
We then repeated the fits using the same spectral models, but allowing all fitting parameters to vary.
We compared the $\chi^2$ of the two fits and searched for all cases 
where {\rm $\Delta\chi^2$}=9 for one parameter and {\rm $\Delta\chi^2$}=11.8 for two parameters, i.e. 
where we found a significant ($\ge$3$\sigma$) improvement in the quality of the fits varying the spectral shape.
From this analysis we detected spectral variability 
in 8 (7\%) sources (XMM-Newton identification numbers 400, 90, 148, 171, 191, 332, 342 and 469, see 
Table~\ref{summary_var_det}), including 6 type-1 and 1 type-2 AGN. 
The fraction of objects with detected spectral variability increased 
to $\sim$17\% if we used instead a confidence level of 2$\sigma$.
We illustrate the observed spectral variability properties of the sources 
with detected spectral variability from this analysis in Fig.~\ref{var1} for 
the 3 type-1 AGN best fitted with a power law (absorbed by the Galaxy). 
Fig.~\ref{var2} shows some examples of contour diagrams of $\Gamma$-kT and $\Gamma-{\rm N_H}$ for variable AGN where absorption 
or soft excess were detected. We found that in all objects with 
detected spectral variability 
(including all AGN), this was associated mostly with changes in the shape of the broad band continuum 
($\Delta \Gamma\sim$0.2-0.3 for all objects except source 90, where the observed maximum change 
in $\Gamma\sim$1 although with very large uncertainties). We did not see important changes in other 
spectral components such as the temperature of the soft excess component (when modelled with a black body) or the 
amount of intrinsic absorption. This is the case also for the one type-2 AGN in our 
sample with detected variability in its spectral fitting parameters.

Note that among the objects with detected spectral variability from this spectral fitting, we only 
detected spectral variability using the broad band X-ray colour in one 
source, 90 (see Table~\ref{summary_var_det}). These results support the argument that 
the low fraction of objects with detected spectral variability could be due to the fact that typical
changes in the continuum shape are of the order of $\Delta\Gamma$$\sim$0.2-0.3, which corresponds to 
$\Delta${\it HR}$\sim$0.1, undetectable in most of the sources in our sample.

If this is the case, then the results shown in this section support the idea that the main driver (but 
not the only one, see later) for spectral variability on month-years 
scales in AGN might be changes in the mass accretion rate, resulting in changes in the underlying 
power law $\Gamma$. We will discuss this point in more detail in Sec.~\ref{origin_var}.

\section{What is the true fraction of sources with X-ray variability?}
\label{fracc_var}
In Fig.~\ref{var_vs_qual2} (top) we show the fraction of objects in our sample with detected flux 
variability as a function of the mean 0.2-12 keV count rate.
We see that it is a strong function of the 
mean count rate, and hence the number we obtained for the whole sample of objects ($\sim$50\%) has to be taken as a lower limit. 
Indeed the plot suggests that the fraction of sources with flux variability 
might be 80\% (the maximum fraction) or higher. 
\begin{figure*}[!htb]
    \hbox{
    \includegraphics[angle=90,width=0.32\textwidth]{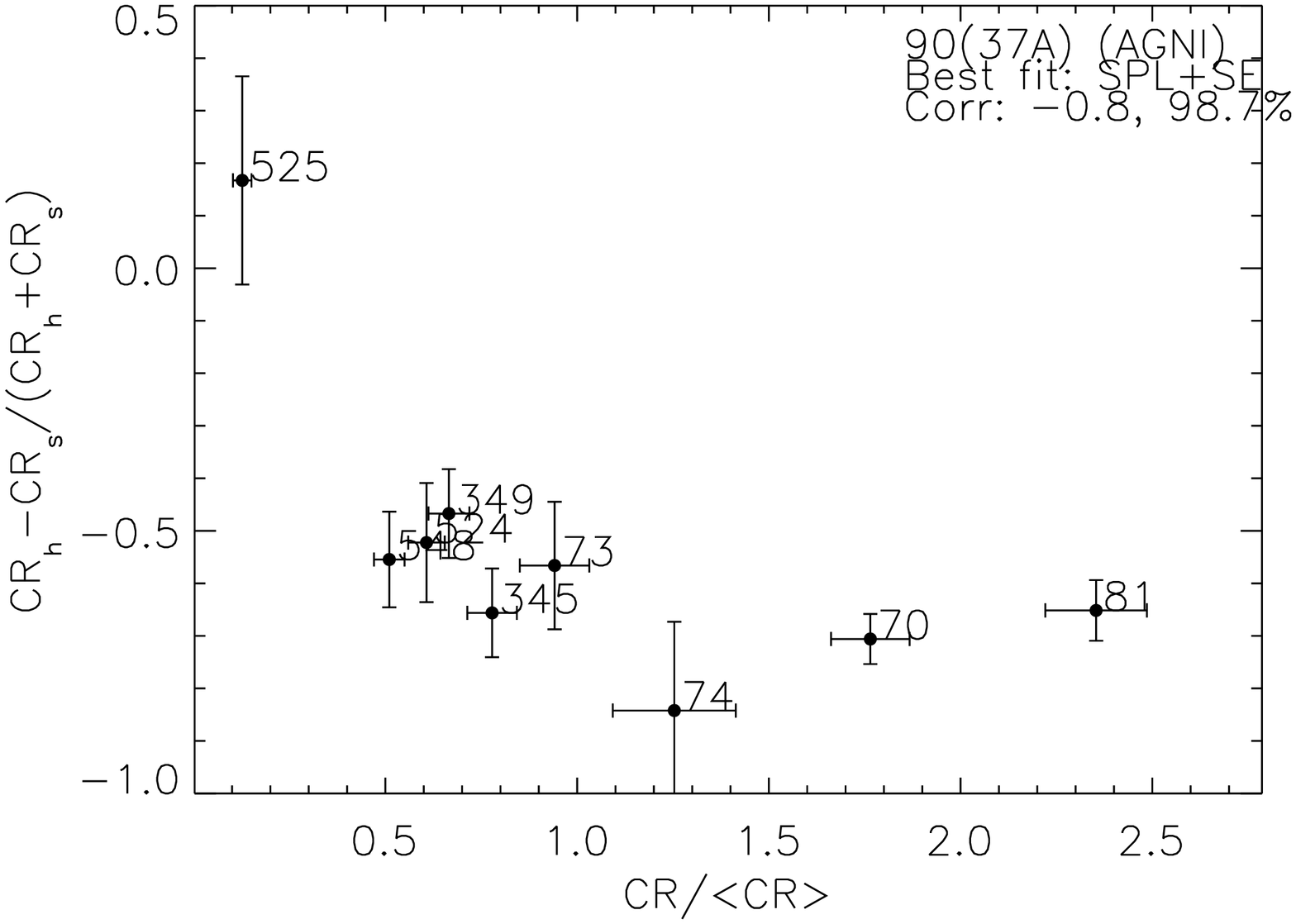}
    \includegraphics[angle=90,width=0.32\textwidth]{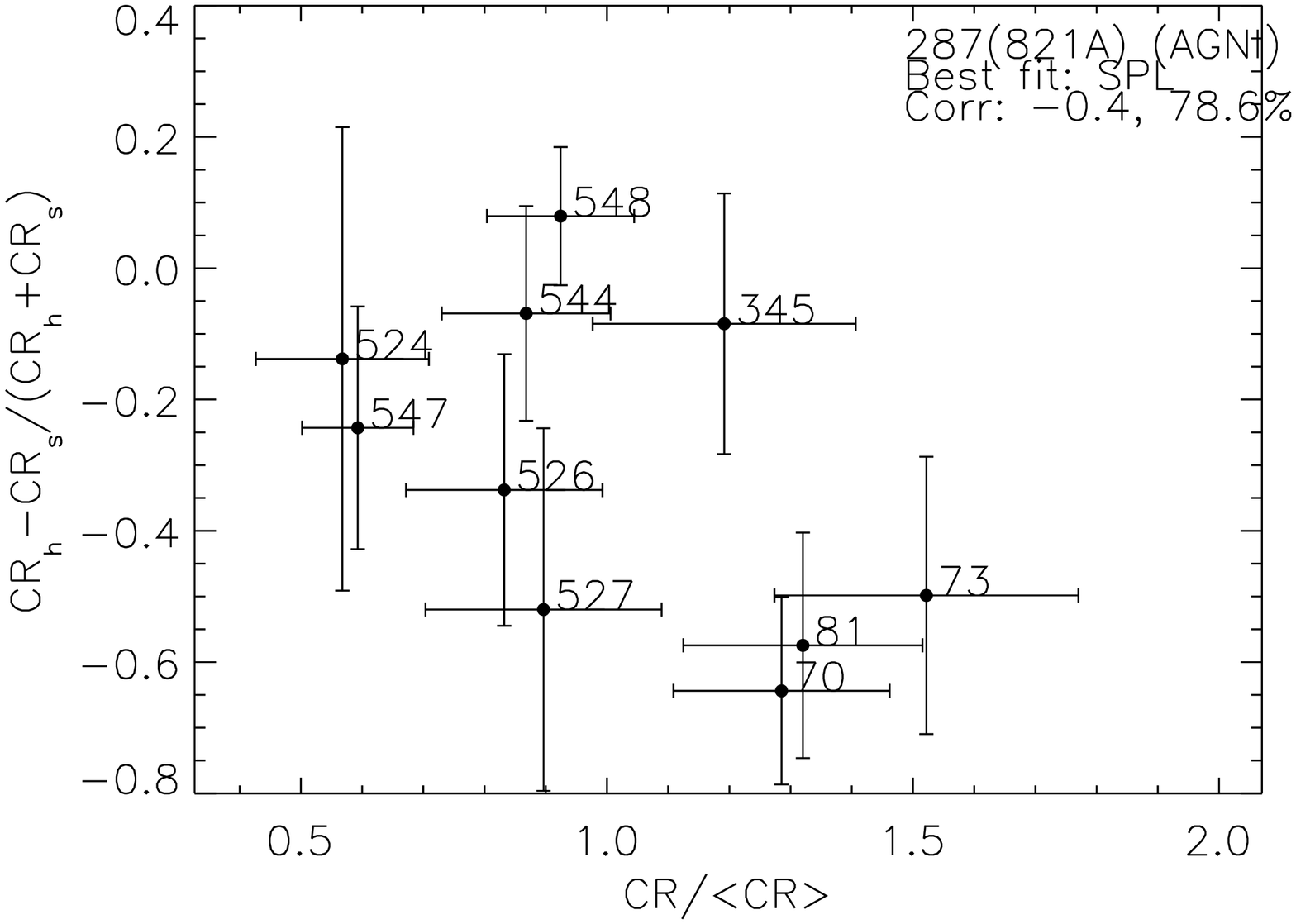}
    \includegraphics[angle=90,width=0.32\textwidth]{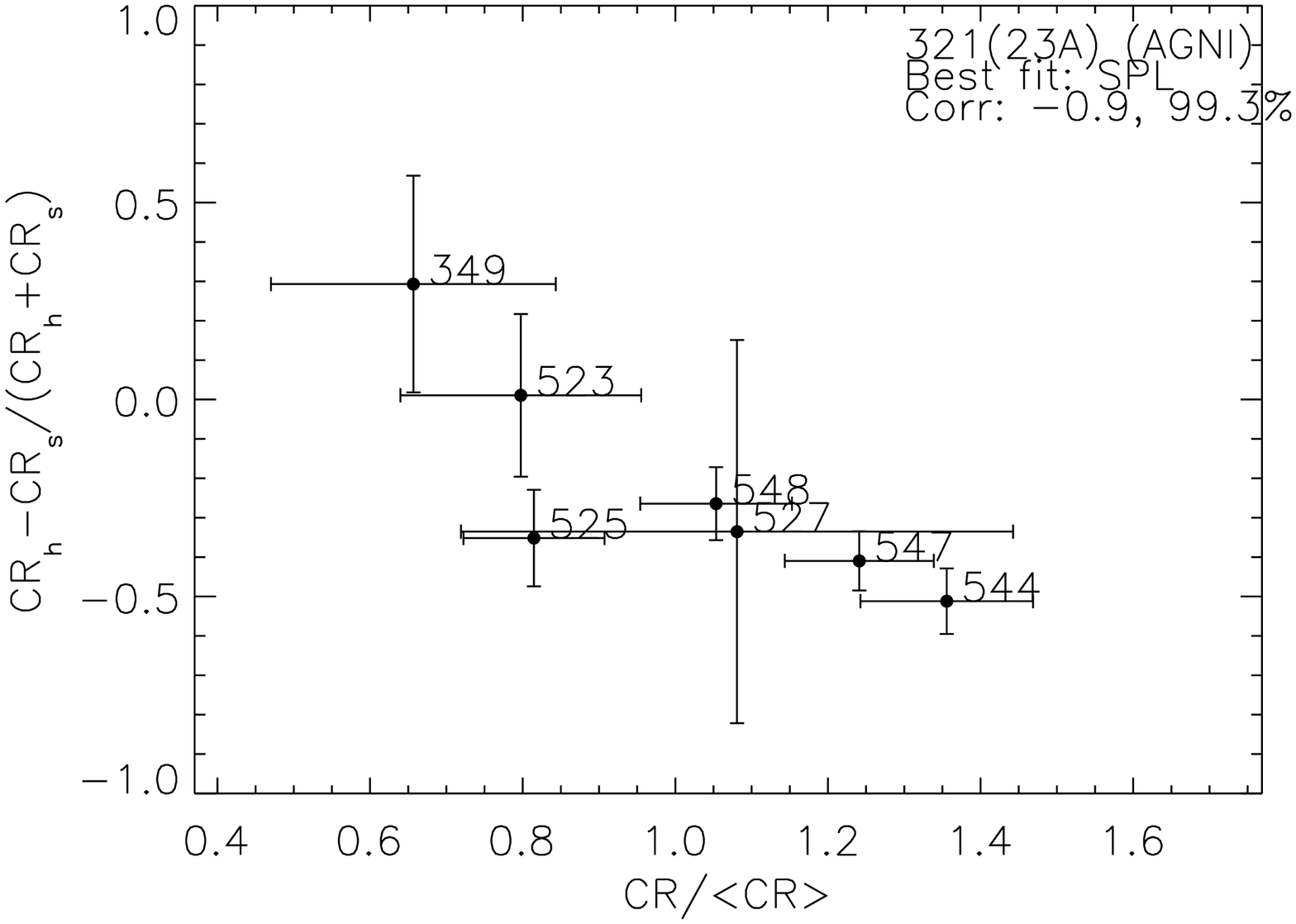}}
    \hbox{
    \includegraphics[angle=90,width=0.32\textwidth]{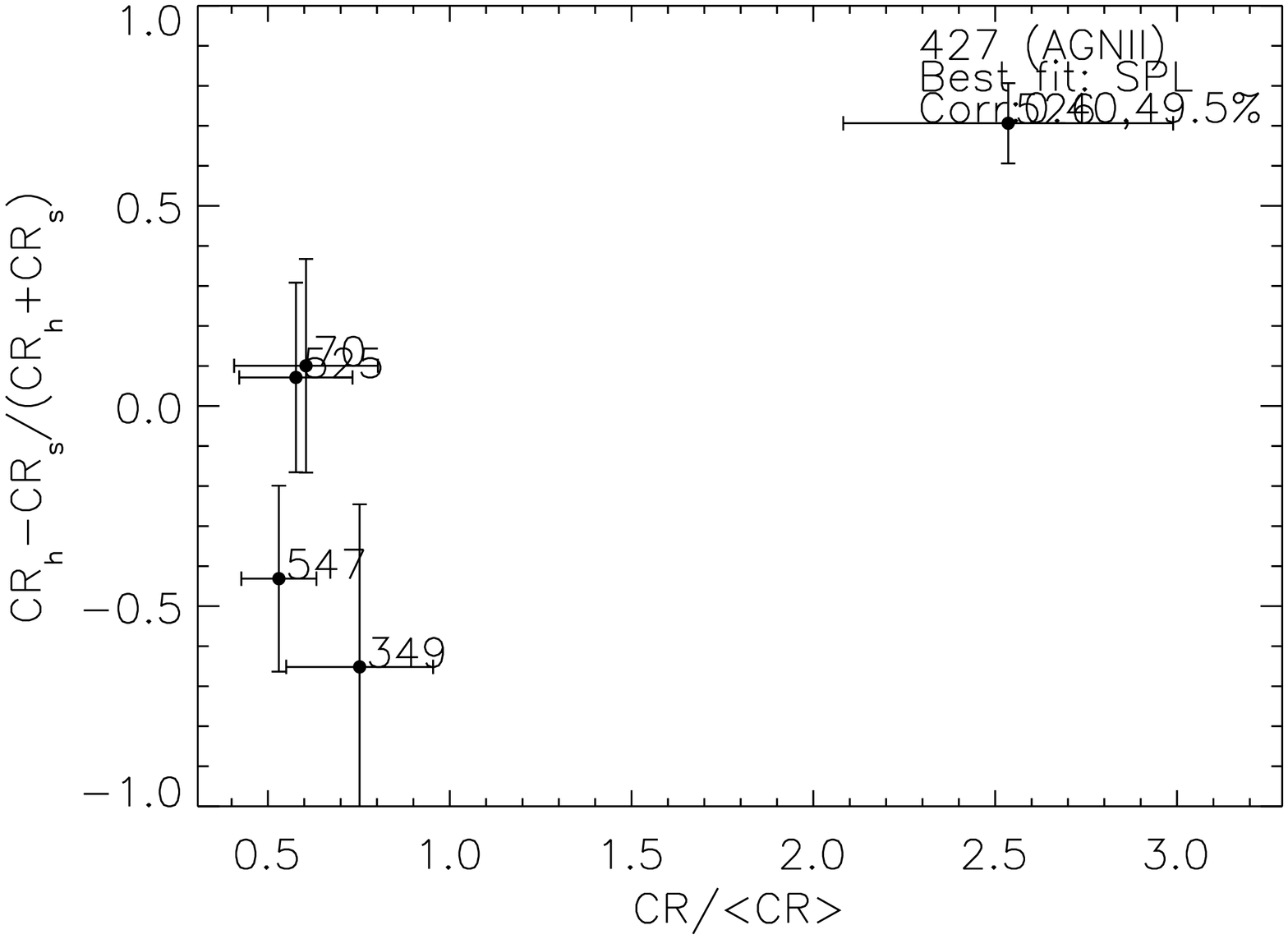}
    \includegraphics[angle=90,width=0.32\textwidth]{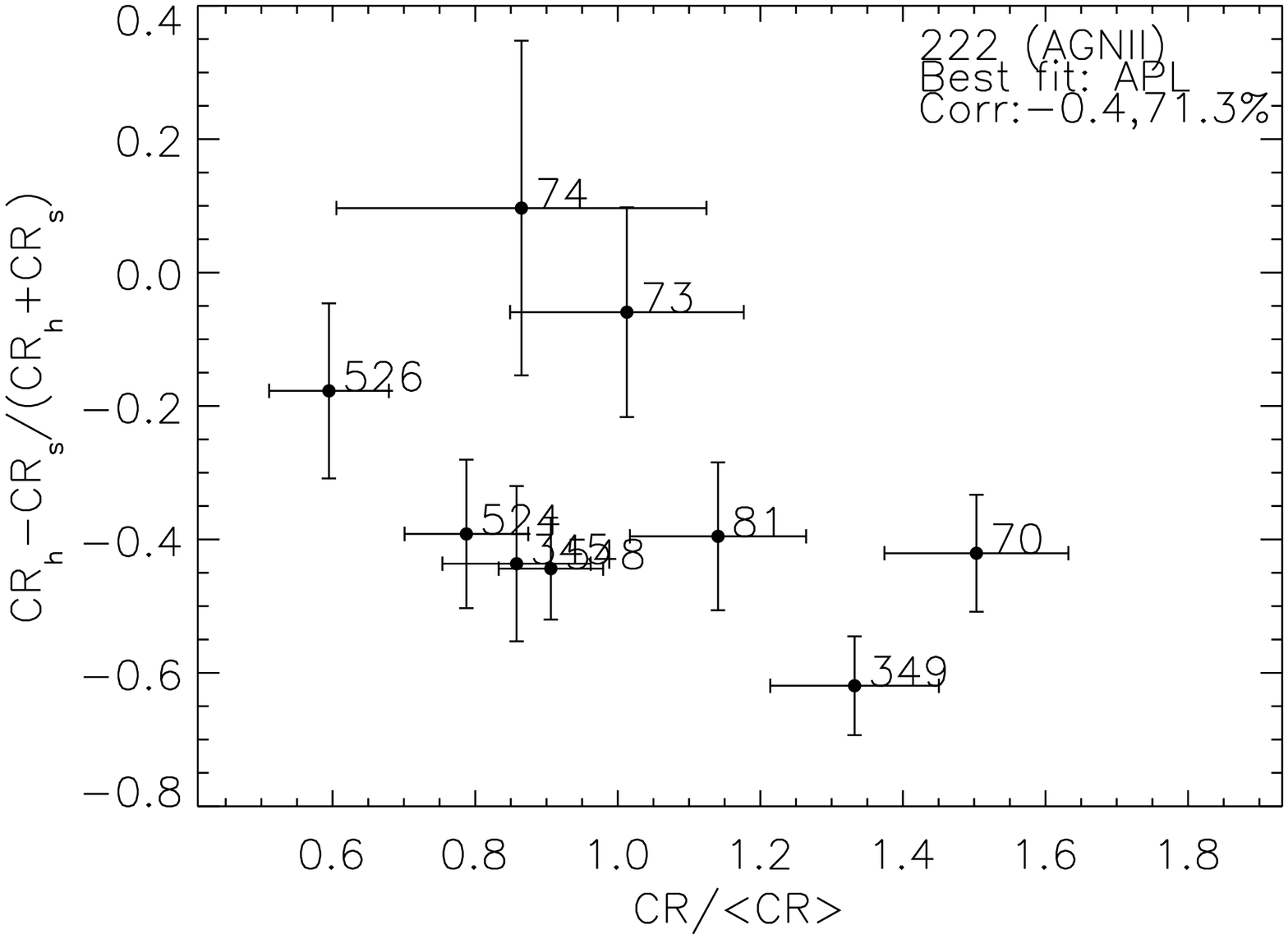}
    \includegraphics[angle=90,width=0.32\textwidth]{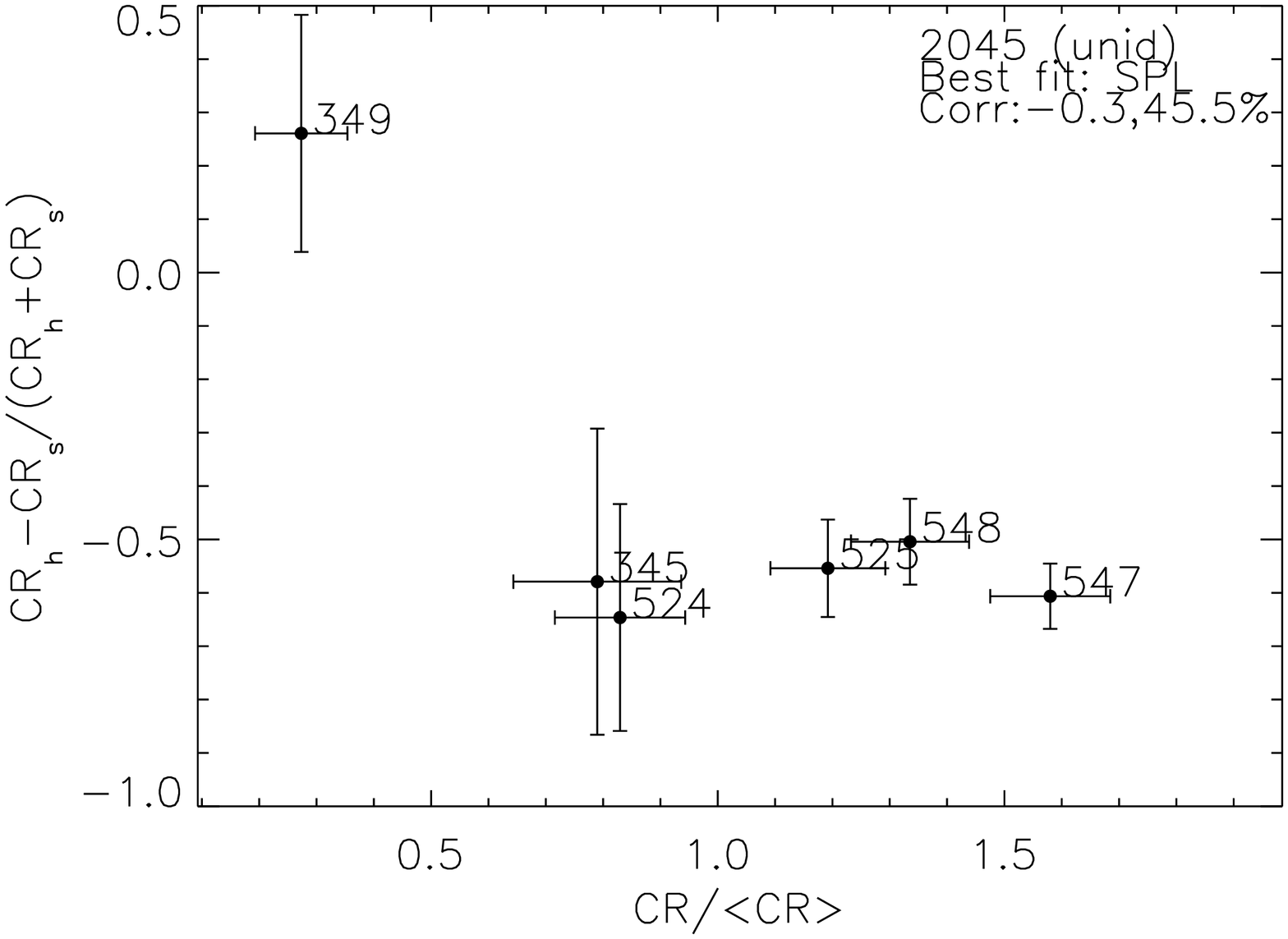}}
    \hbox{
    \includegraphics[angle=90,width=0.32\textwidth]{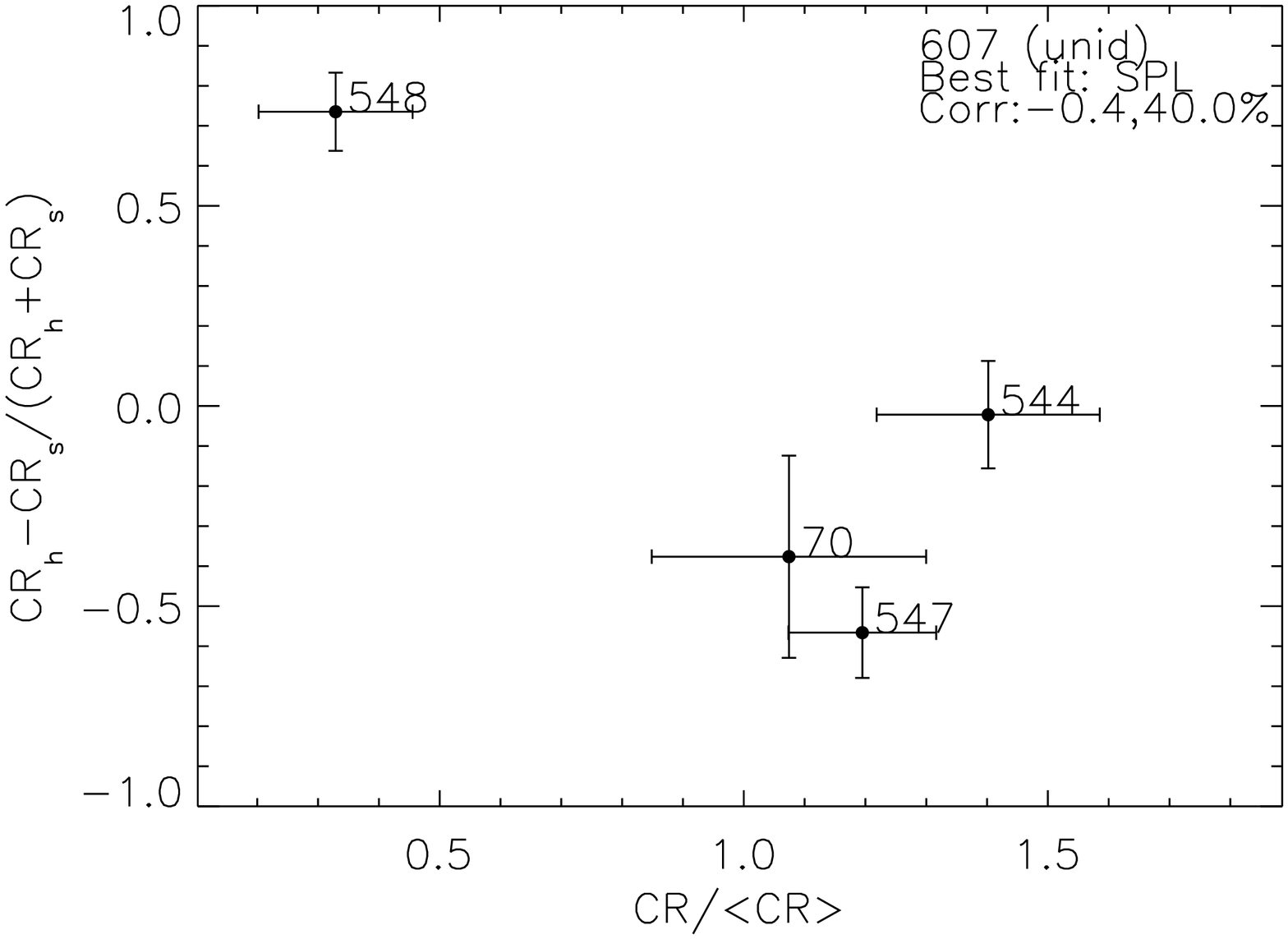}
    \includegraphics[angle=90,width=0.32\textwidth]{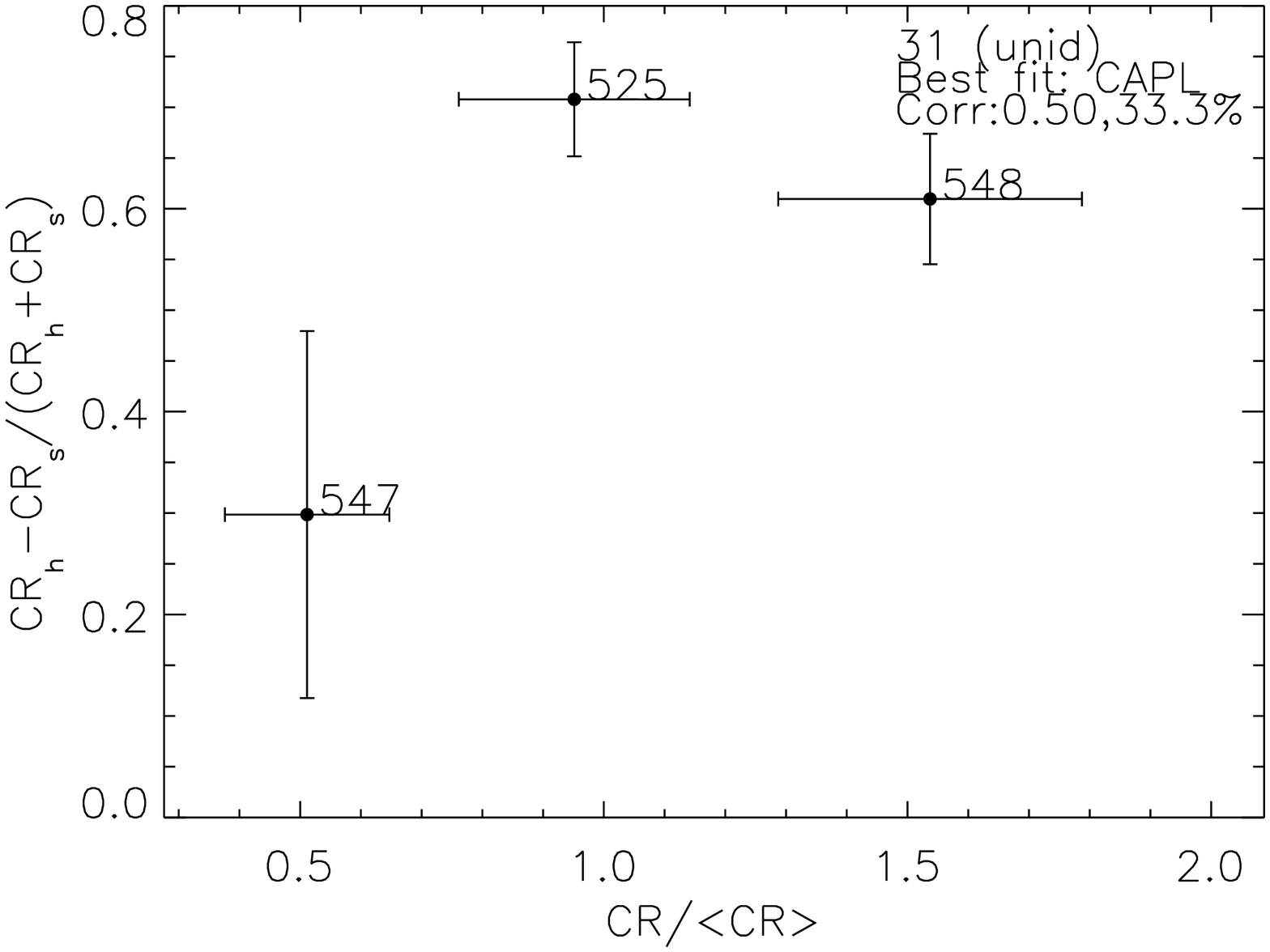}
    \includegraphics[angle=90,width=0.32\textwidth]{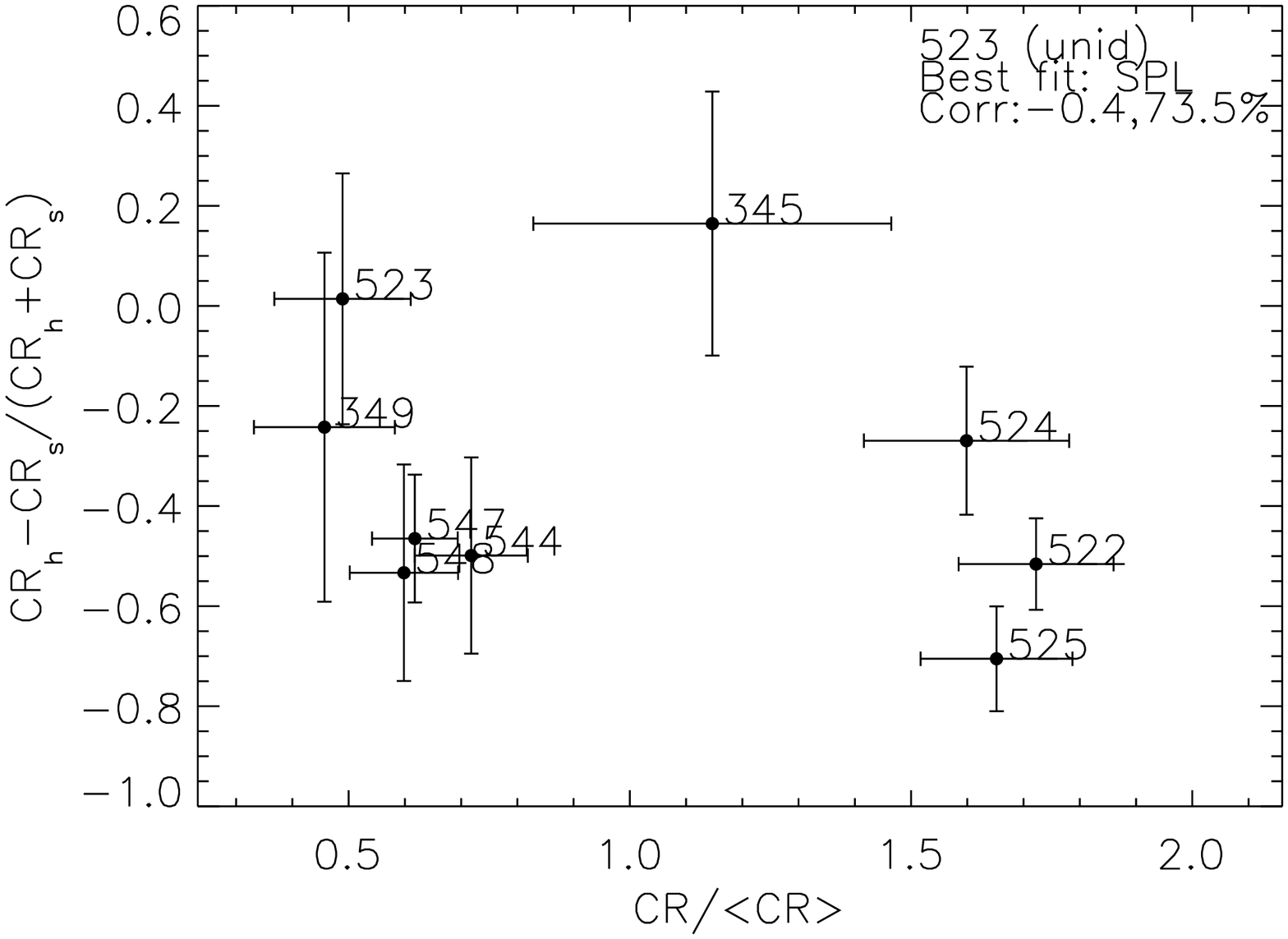}}
    \hbox{
    \includegraphics[angle=90,width=0.32\textwidth]{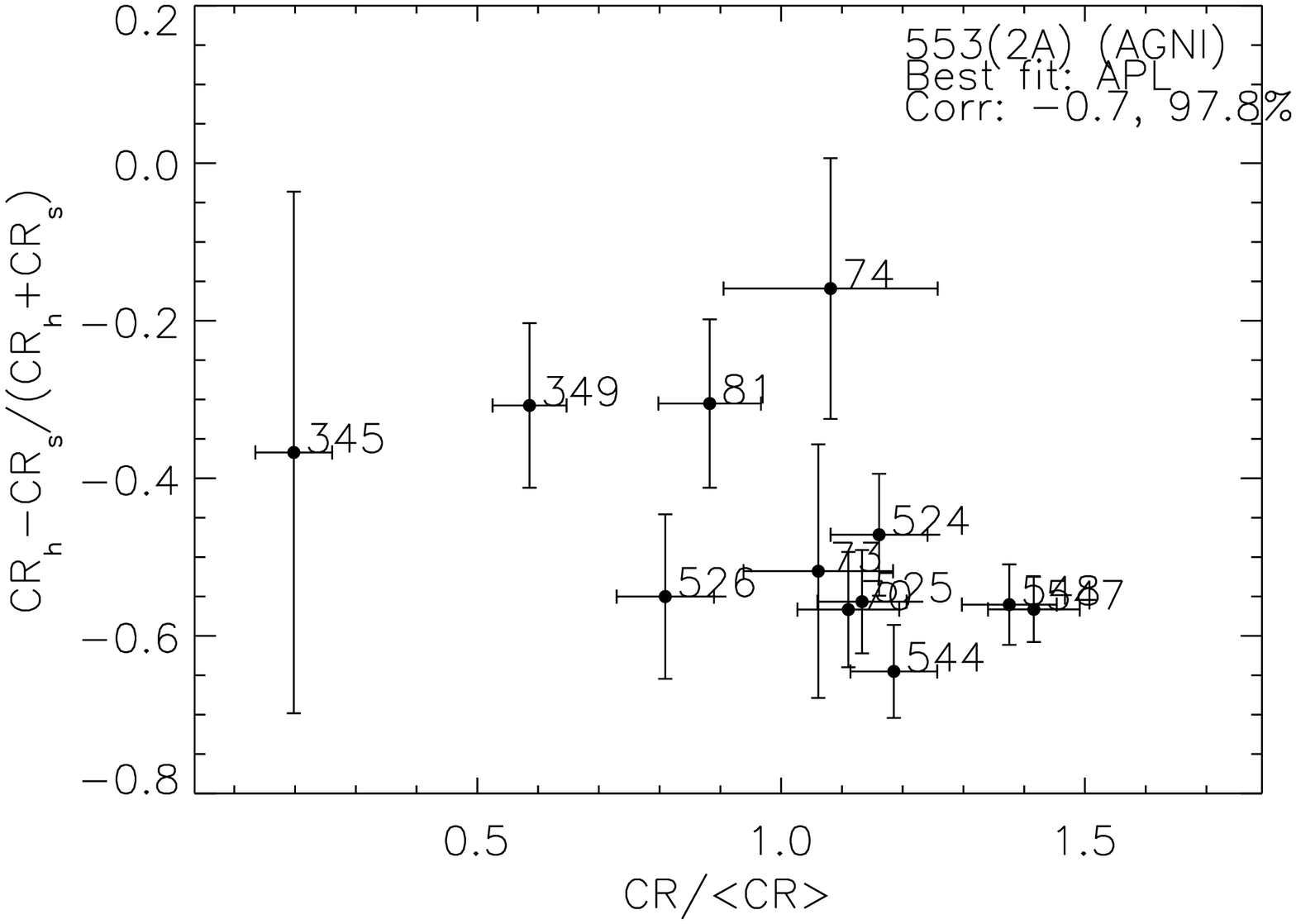}
    \includegraphics[angle=90,width=0.32\textwidth]{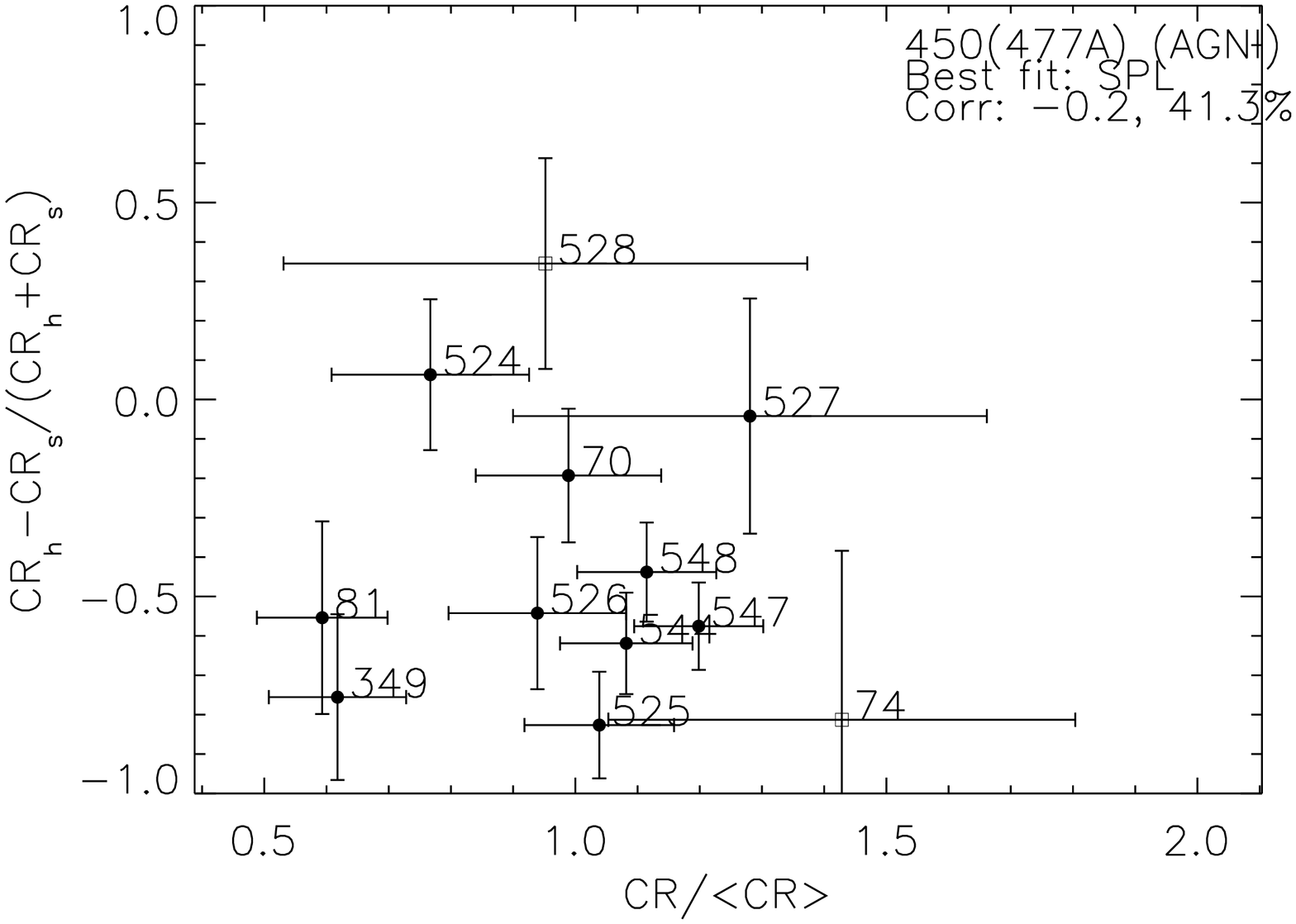}
    \includegraphics[angle=90,width=0.32\textwidth]{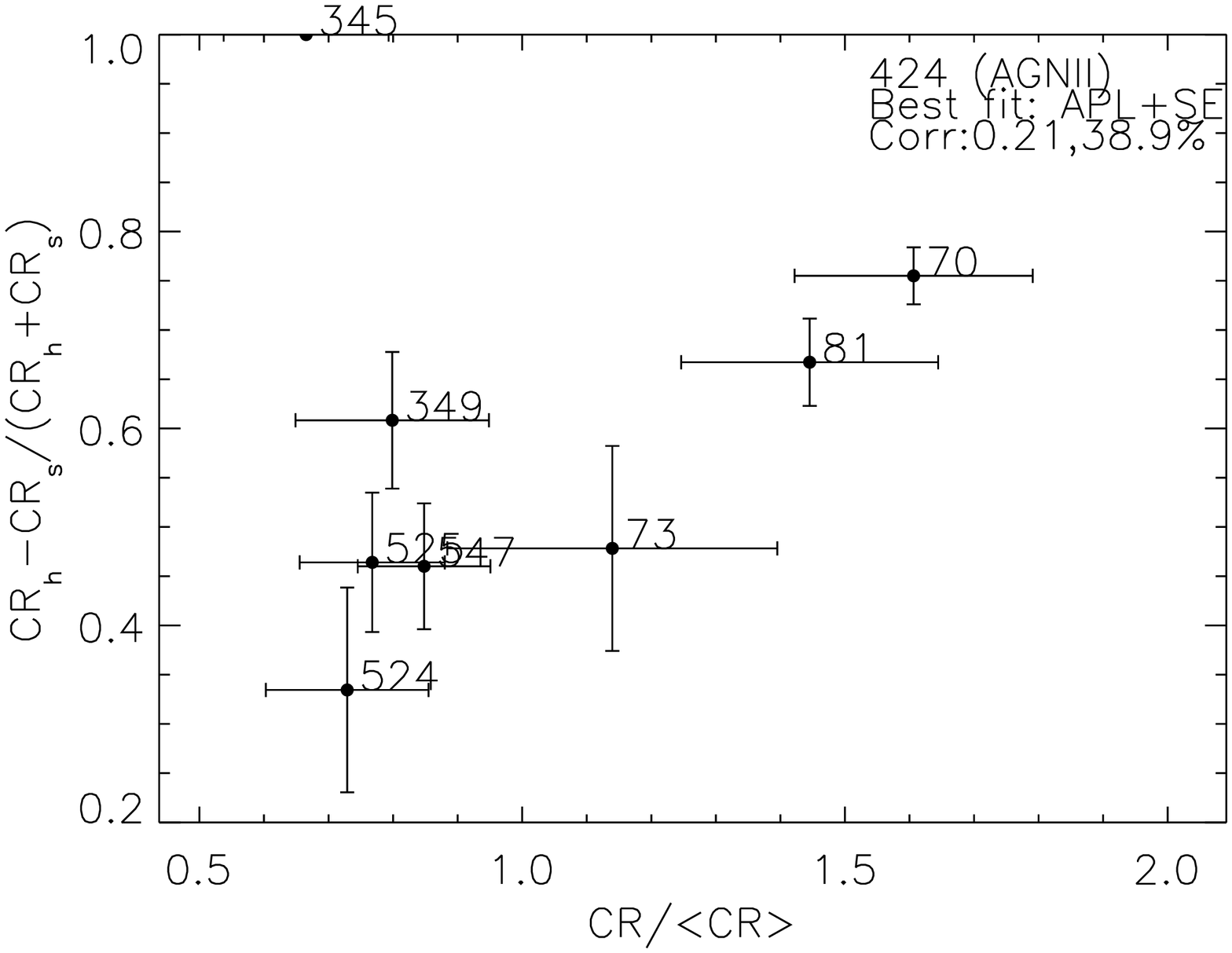}}
    \hbox{
    \includegraphics[angle=90,width=0.32\textwidth]{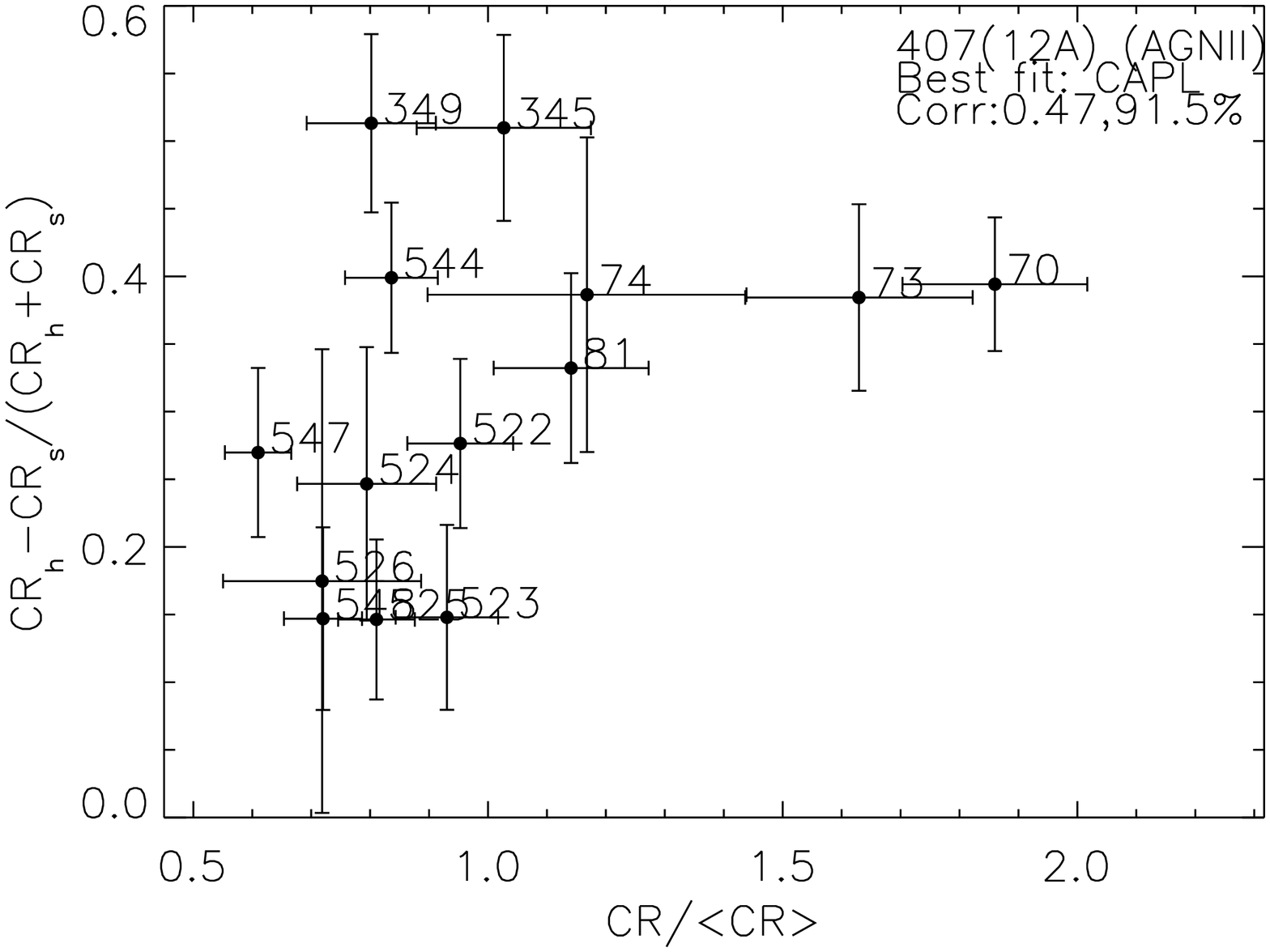}
    \includegraphics[angle=90,width=0.32\textwidth]{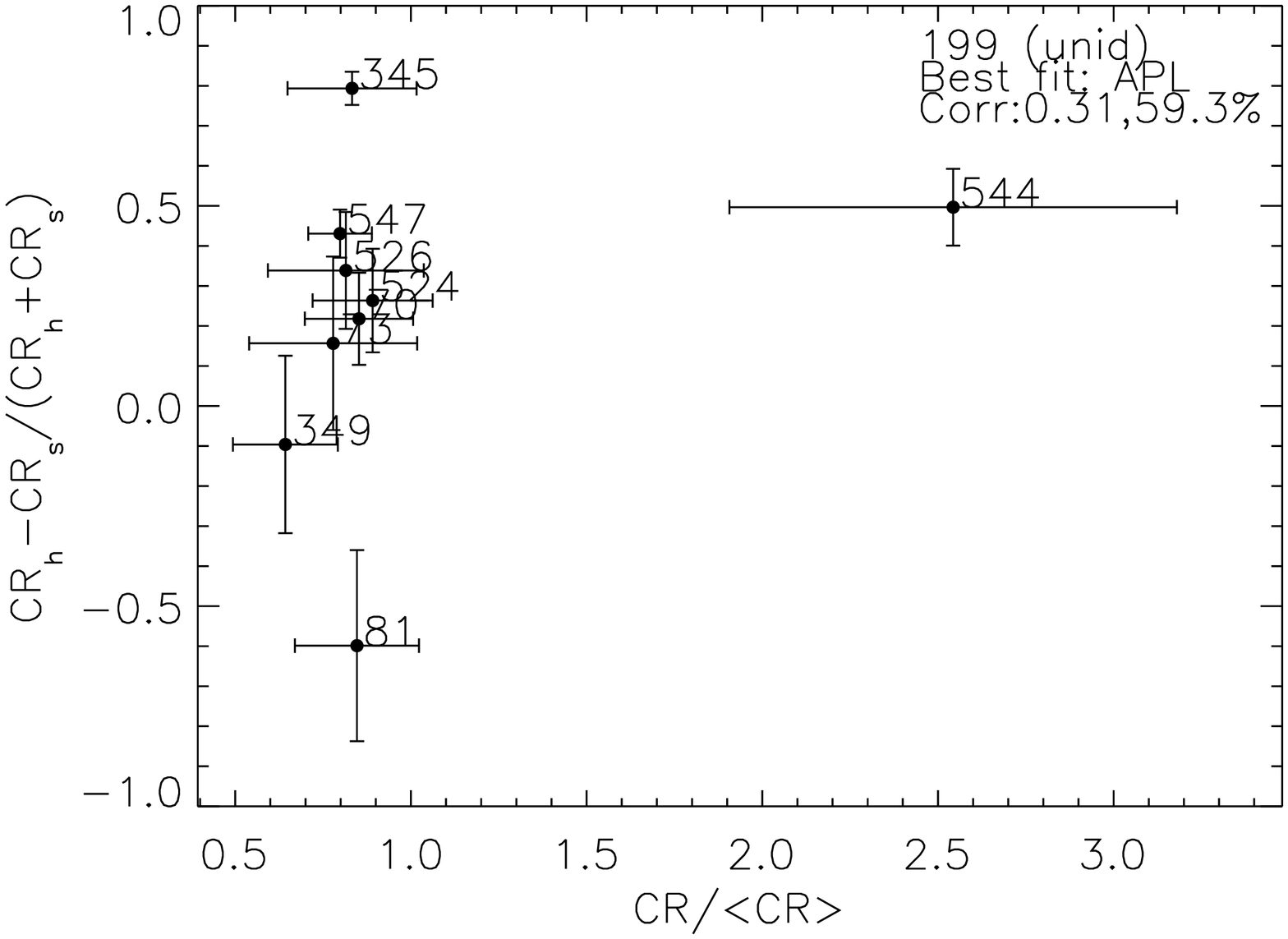}
    \includegraphics[angle=90,width=0.32\textwidth]{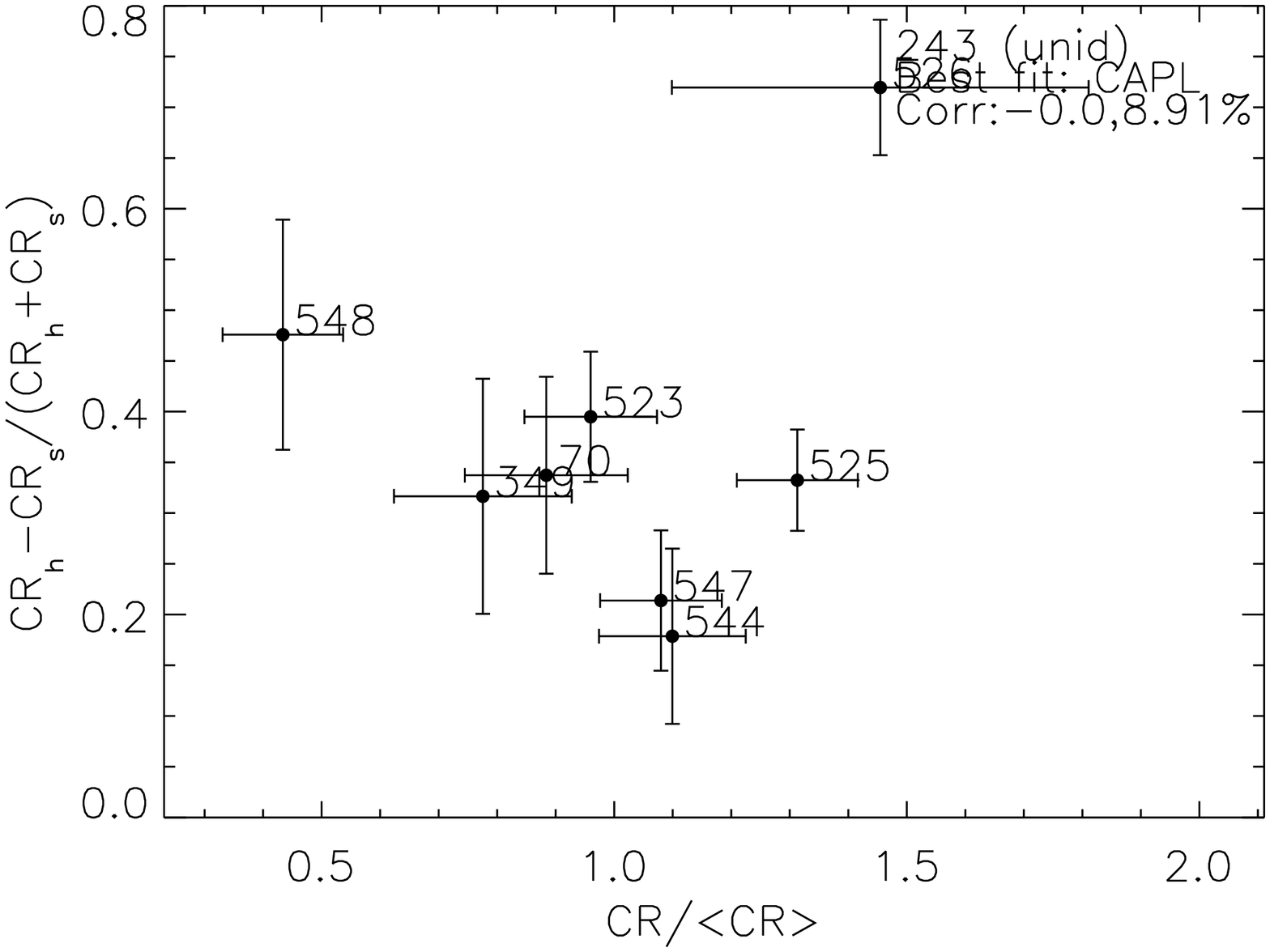}}
    \caption{
    Correlation of flux and spectral variability properties for the 15 objects in our sample with detected 
    flux and spectral variability with a confidence of more than 3$\sigma$.
    Errors correspond to the 1$\sigma$ confidence interval.}
    \label{sp_var_fl_var}
\end{figure*}

\begin{figure}[!htb]
    \hbox{
    \includegraphics[angle=90,width=0.5\textwidth]{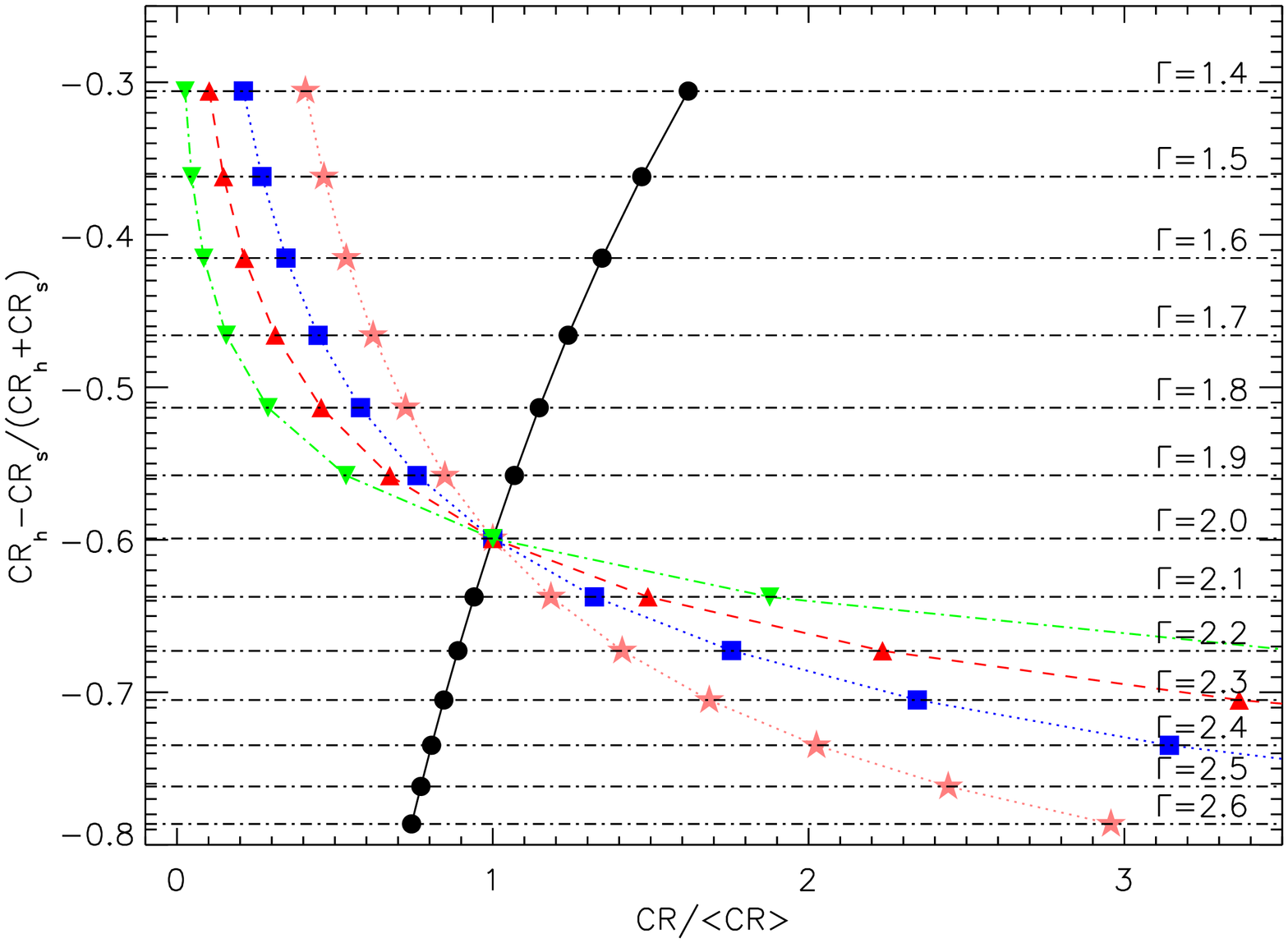}}
    \hbox{
    \includegraphics[angle=90,width=0.5\textwidth]{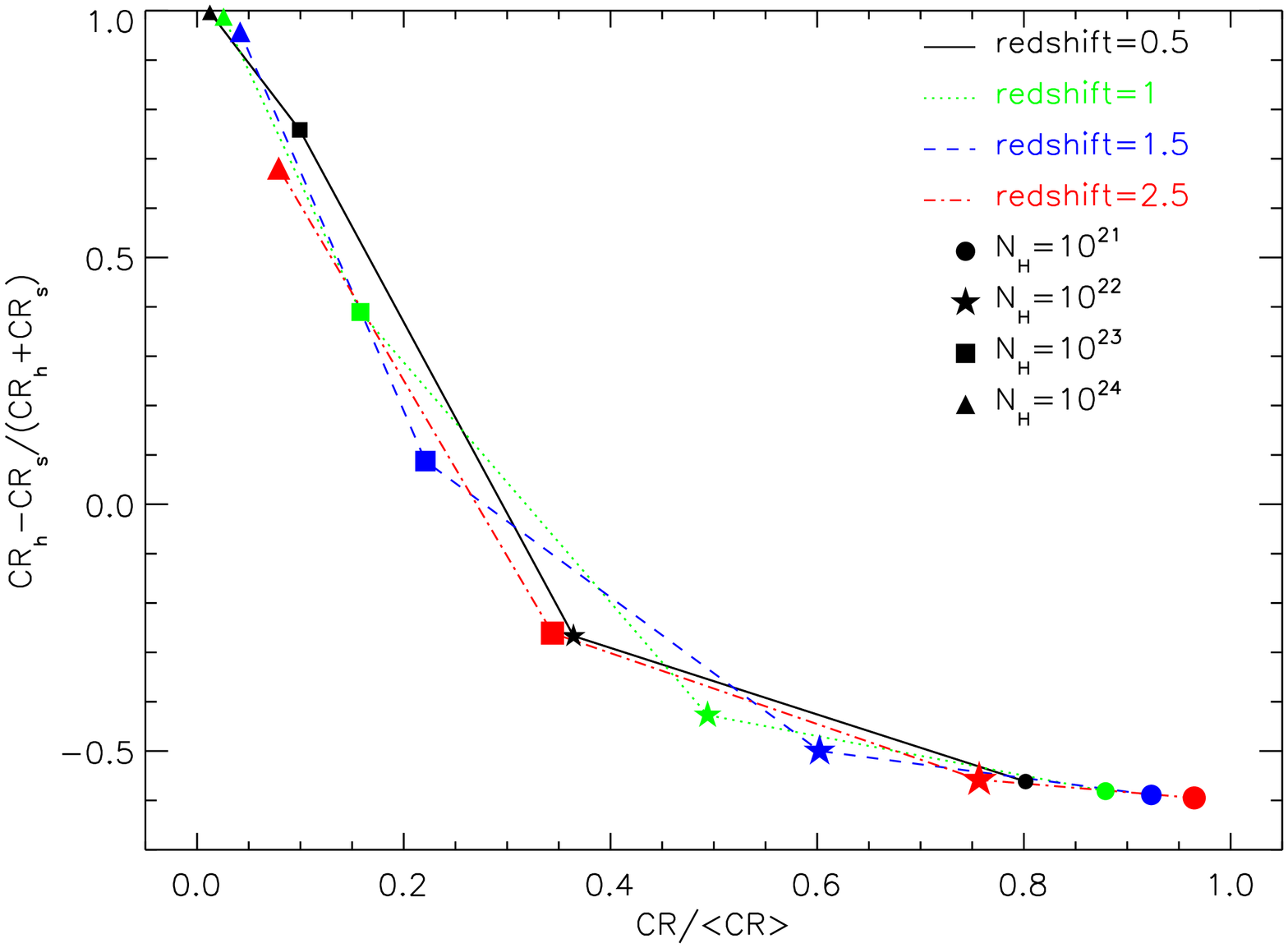}}
    \caption{Left: Correlation of flux and spectral variability for a source at redshift of 1 
      associated with changes in the continuum 
      shape alone for different rest-frame pivoting energies: 1 keV (circles), 10 keV (stars), 30 keV (squares), 
      100 keV (triangles) and 1000 keV (upside down triangles).
      Right: Correlation of flux and spectral variability associated with changes in the amount of intrinsic 
      absorption alone at different redshifts: ${\rm N_H=10^{21},10^{22},10^{23},10^{24} cm^{-2}}$. For each absorption the size of the symbols represents the redshift, from redshift 0.5 (smallest symbol size) to redshift 2.5 (largest symbol size).
      }
    \label{pivoting}
\end{figure}
The result that the majority of our sources are variable in flux is in agreement with previous variability 
analyses, that found long time scale variability in AGN to be more common than 
short time scale variability (see e.g. Ciliegi et al.~\cite{Ciliegi97}; Grandi et al.~\cite{Grandi92}). However not all sources in our sample are expected to be variable as some of the objects still not identified might either
not be AGN, or be heavily obscured AGN.

Fig.~\ref{var_vs_qual2} (bottom) shows the fraction of objects in our sample with detected 
spectral variability as a function of the mean error in the hardness ratio. 
We see that the detection of spectral variability varies with the quality of 
our data, although in this case a clear dependence is only evident for the first bin, which suggests that the fraction 
of sources in our sample with spectral variability could be $\sim$40\% or higher. 

Even if the fraction of sources with spectral variability is as high as $\sim$40\%, 
this fraction is still significantly lower than the fraction of sources with 
flux variability. If spectral variability is less common than flux variability in our 
sample, then for many sources flux variability will not be accompanied by 
spectral variability.

\begin{figure*}[!tb]
    \hbox{
    \hspace{0.40cm}\vspace{0.25cm}\includegraphics[angle=90,width=0.245\textwidth]{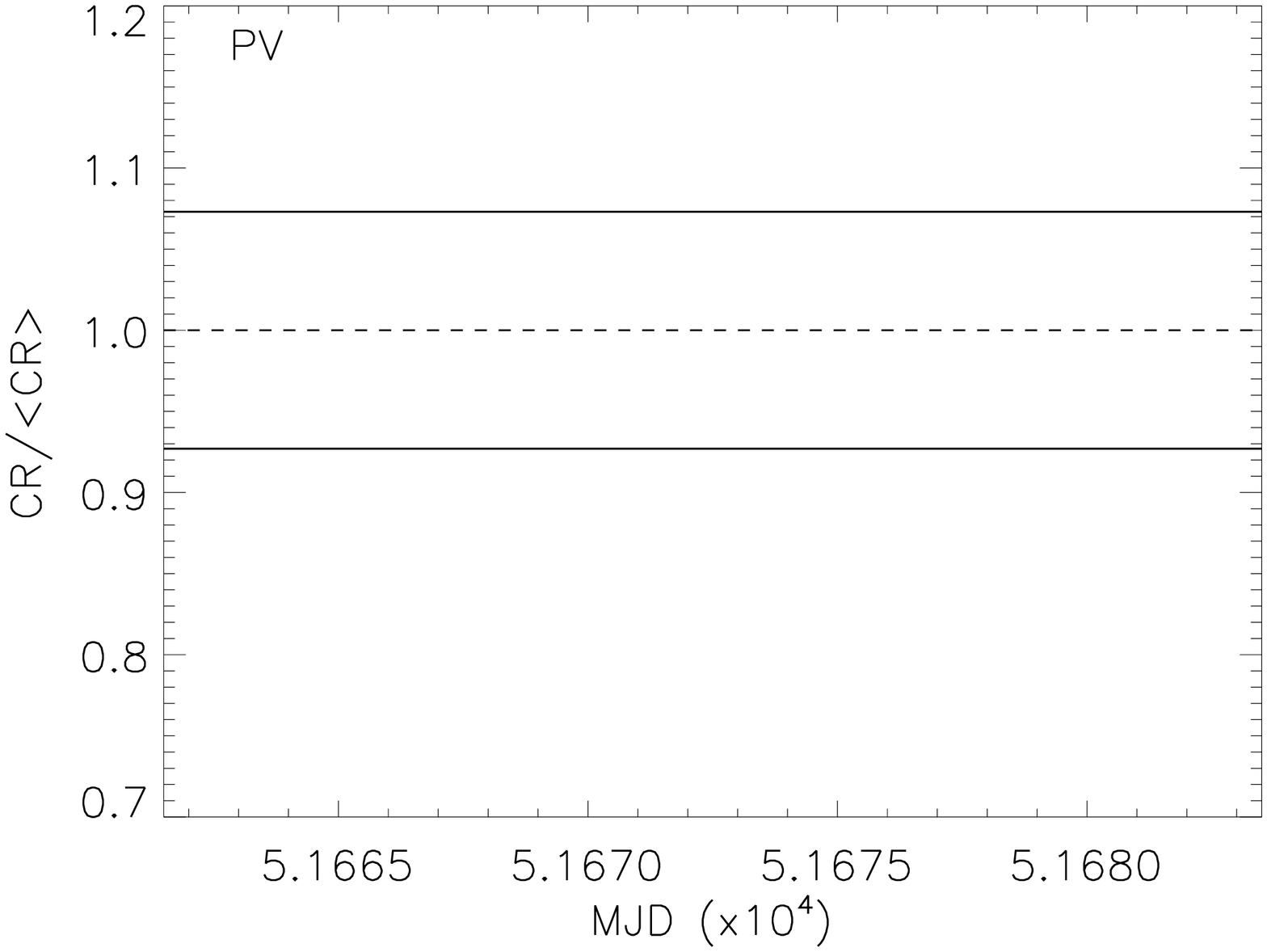}
    \includegraphics[angle=90,width=0.245\textwidth]{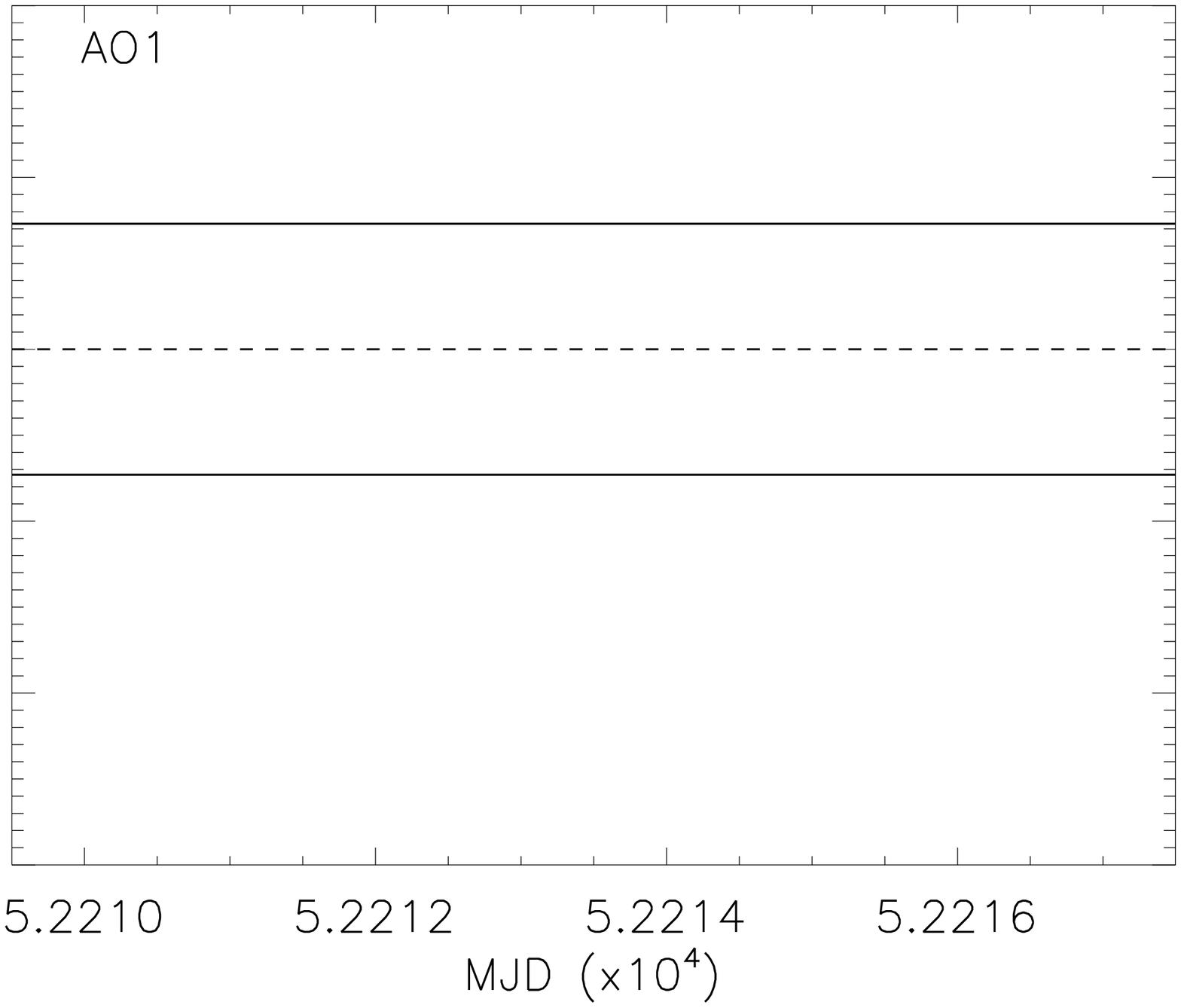}
    \includegraphics[angle=90,width=0.245\textwidth]{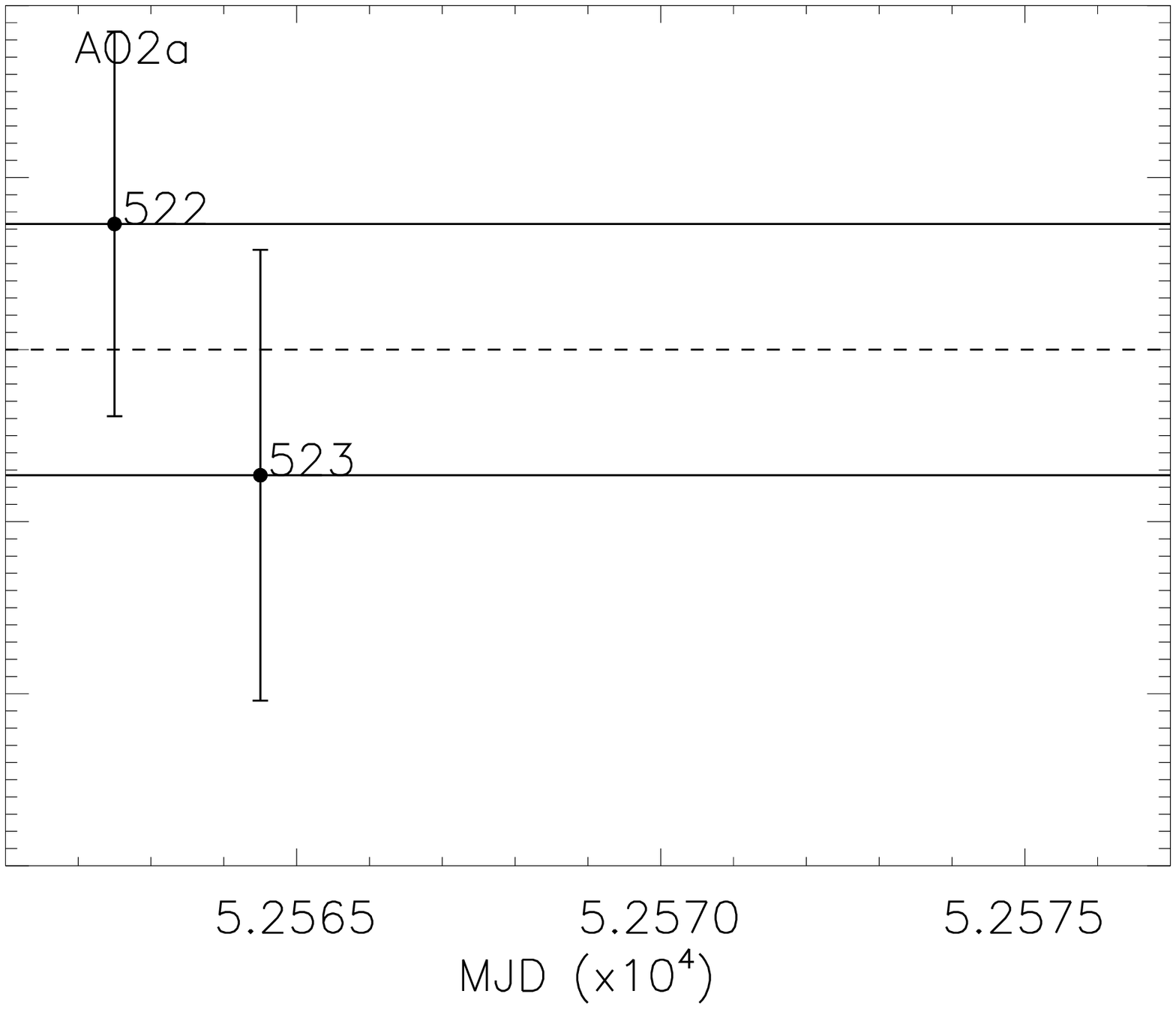}
    \includegraphics[angle=90,width=0.245\textwidth]{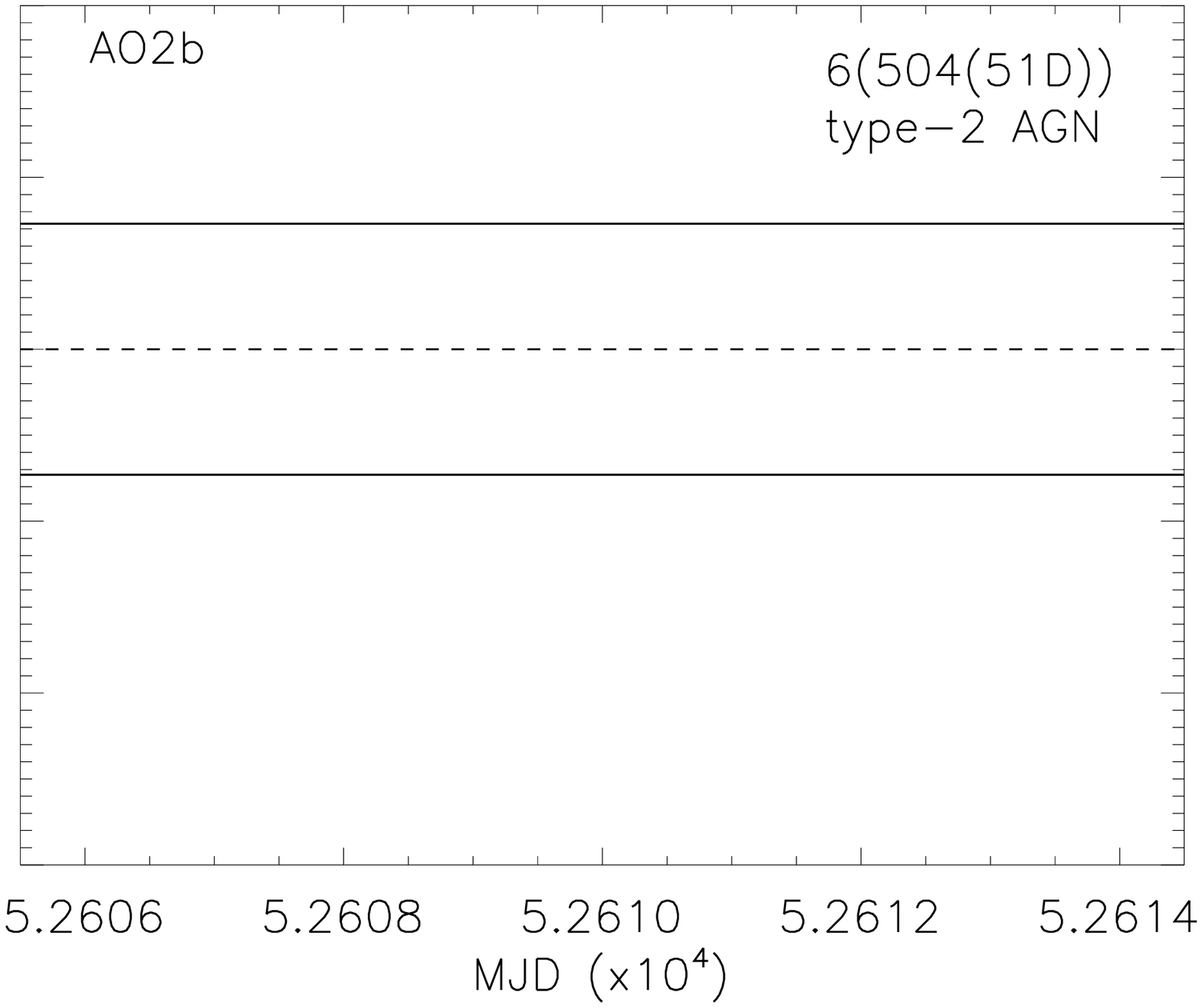}}
    \hbox{
    \hspace{0.40cm}\vspace{0.25cm}\includegraphics[angle=90,width=0.245\textwidth]{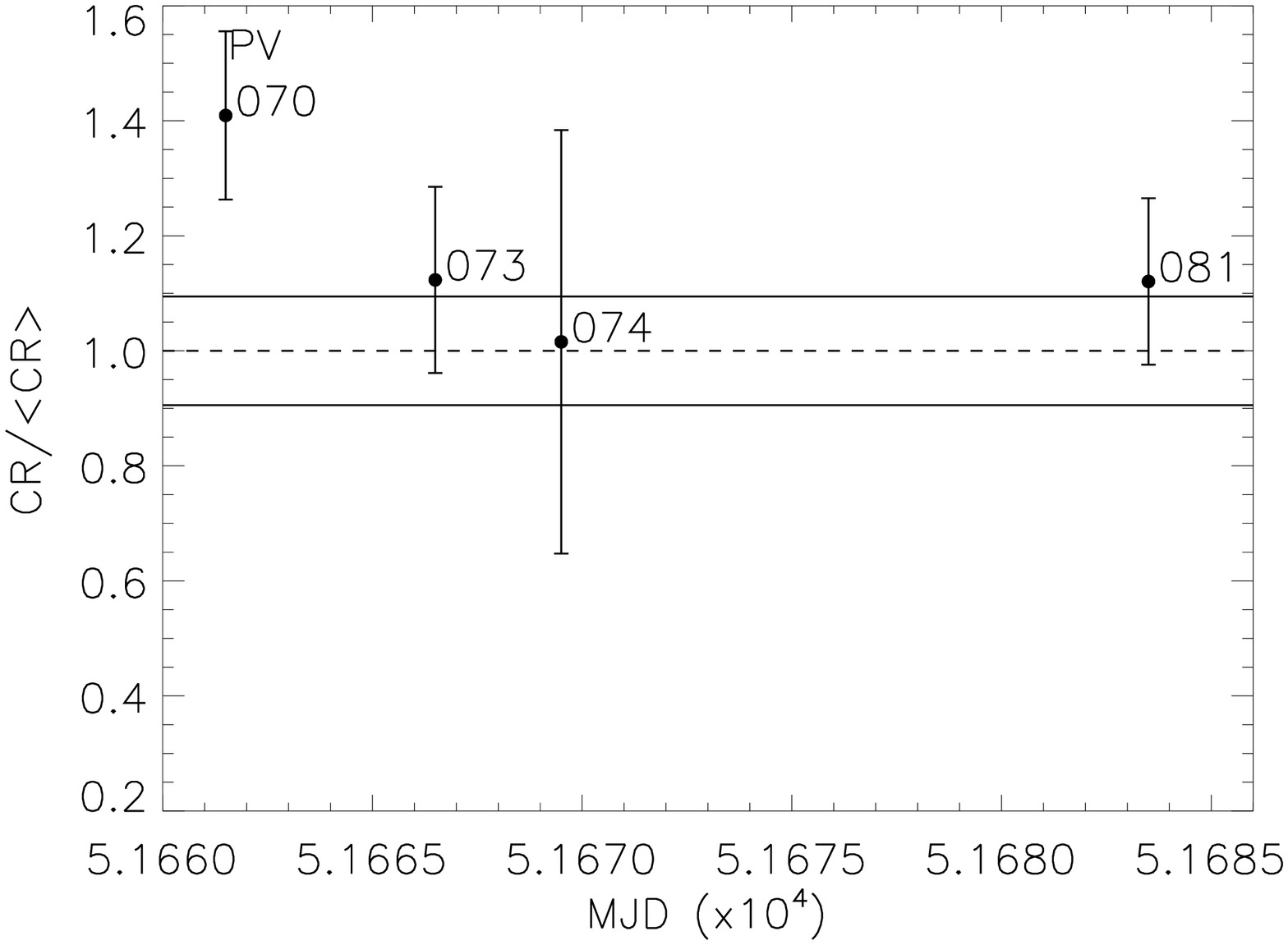}
    \includegraphics[angle=90,width=0.245\textwidth]{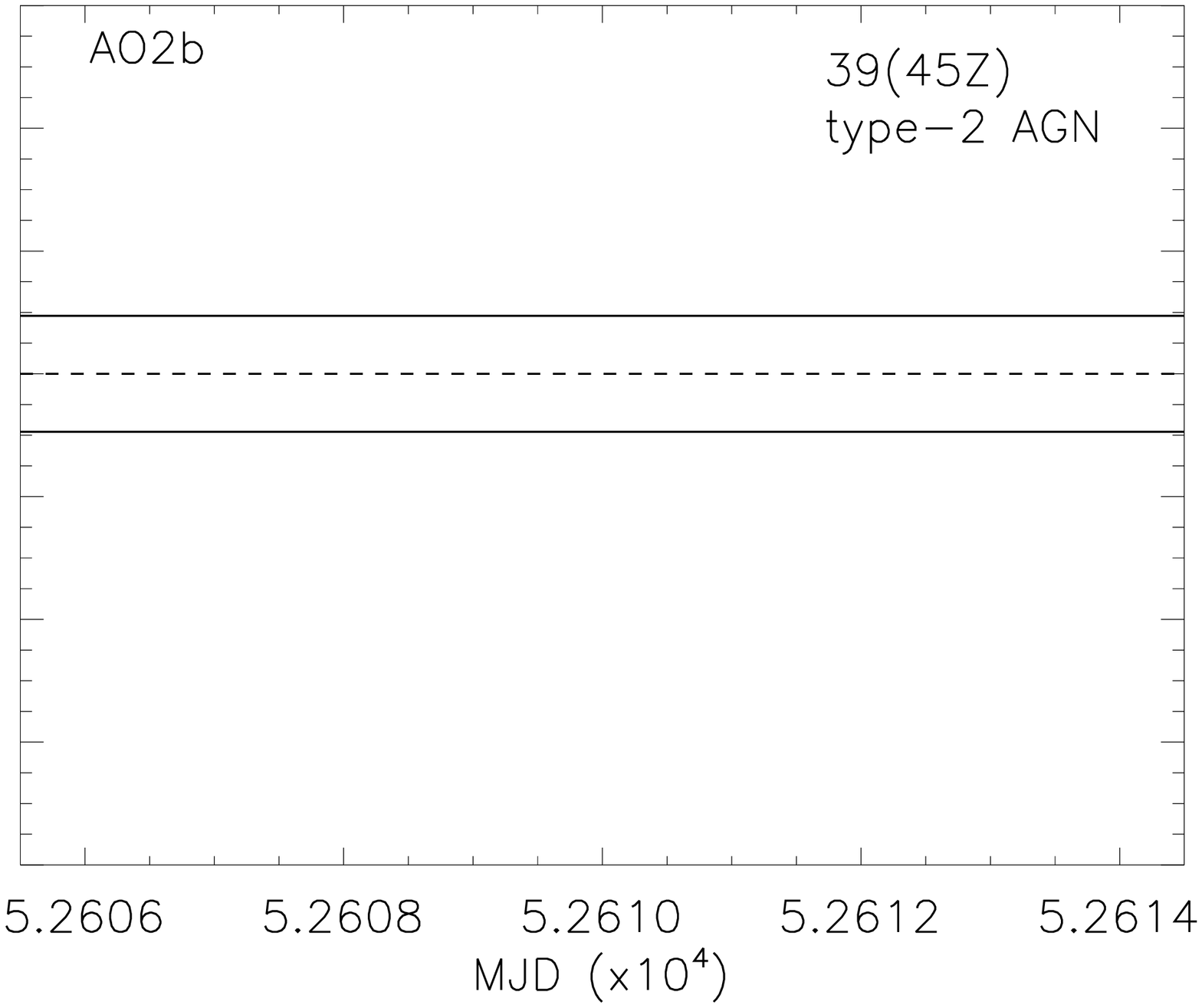}
    \includegraphics[angle=90,width=0.245\textwidth]{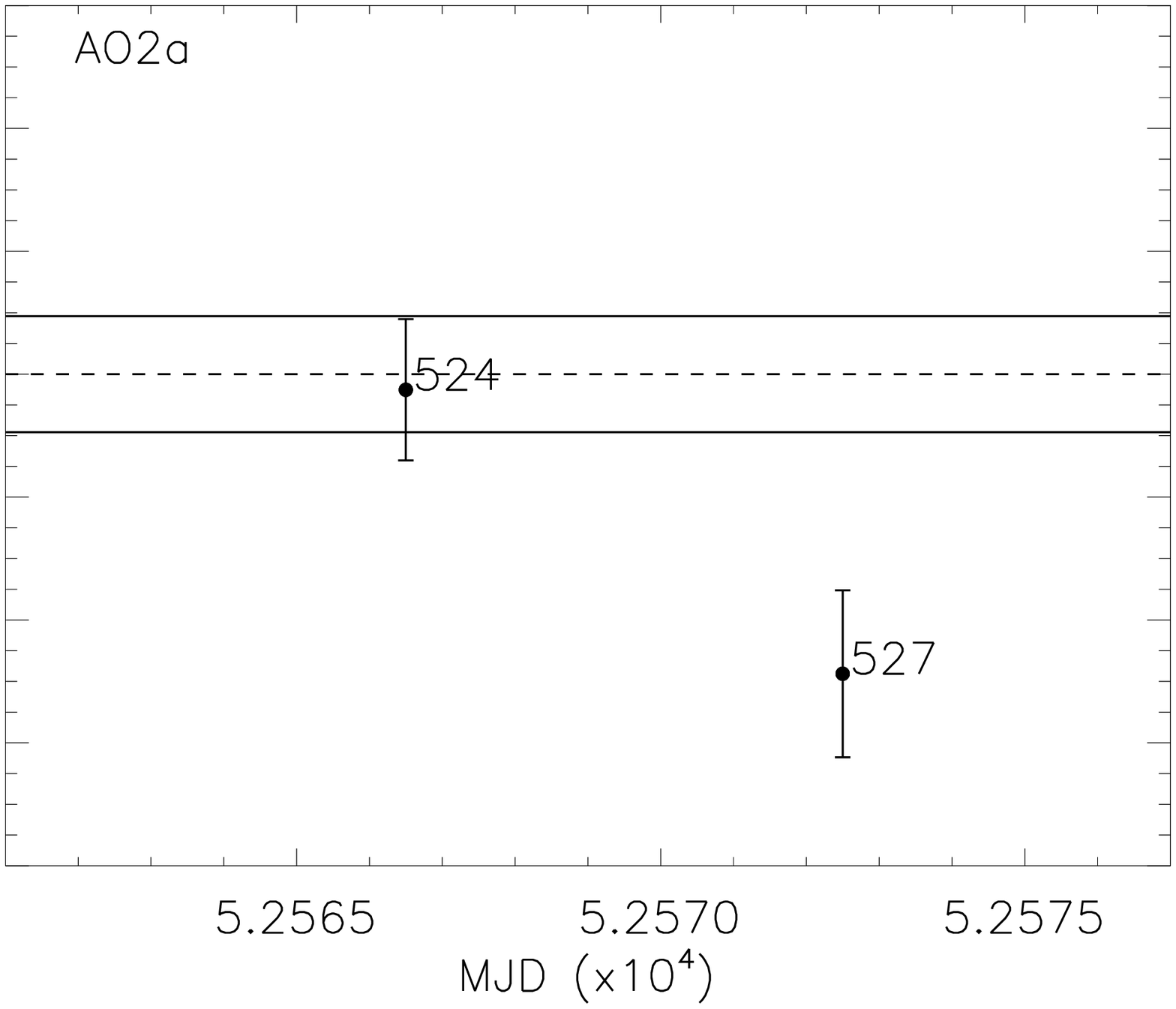}
    \includegraphics[angle=90,width=0.245\textwidth]{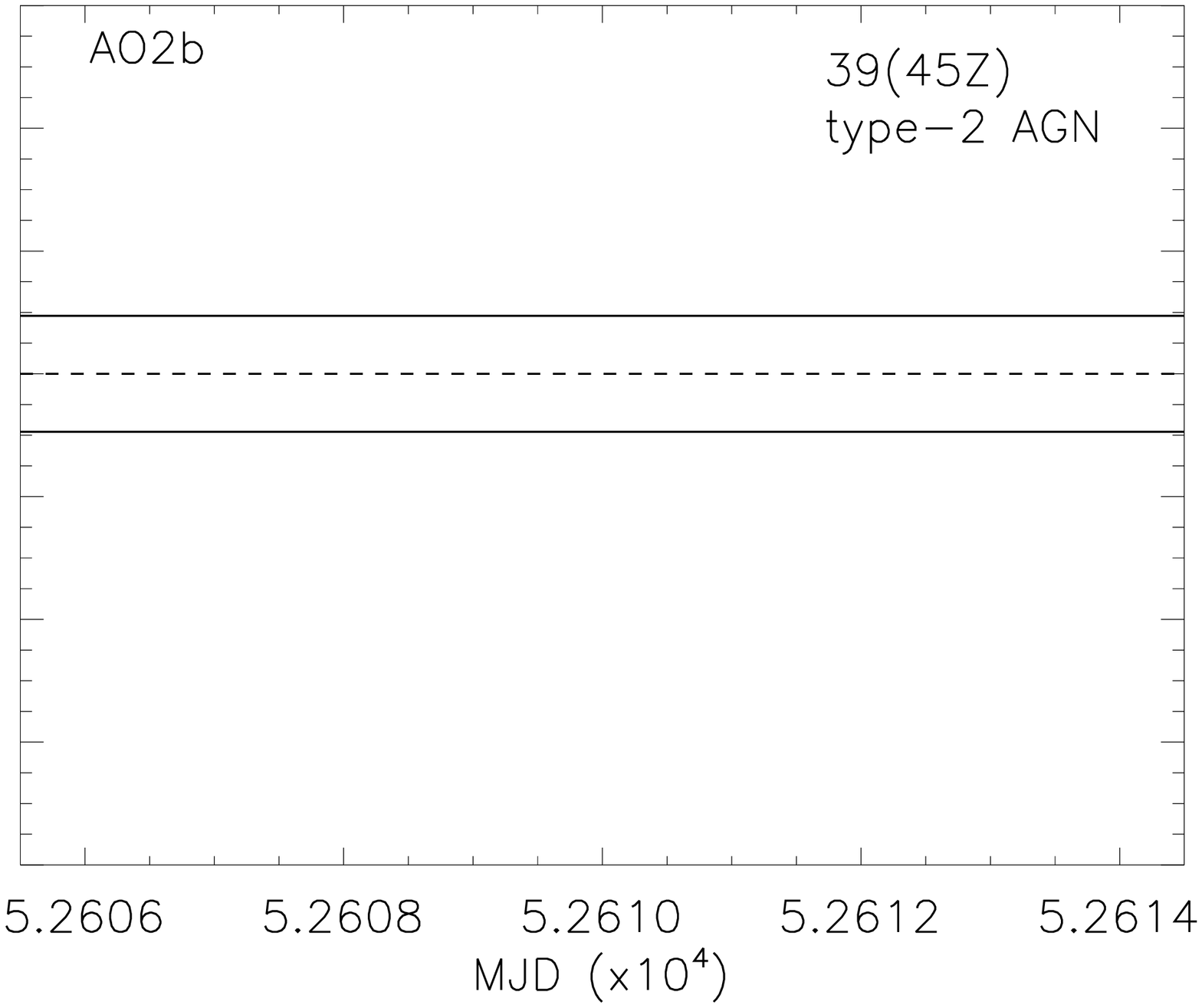}}
    \hbox{
    \hspace{0.40cm}\vspace{0.25cm}\includegraphics[angle=90,width=0.245\textwidth]{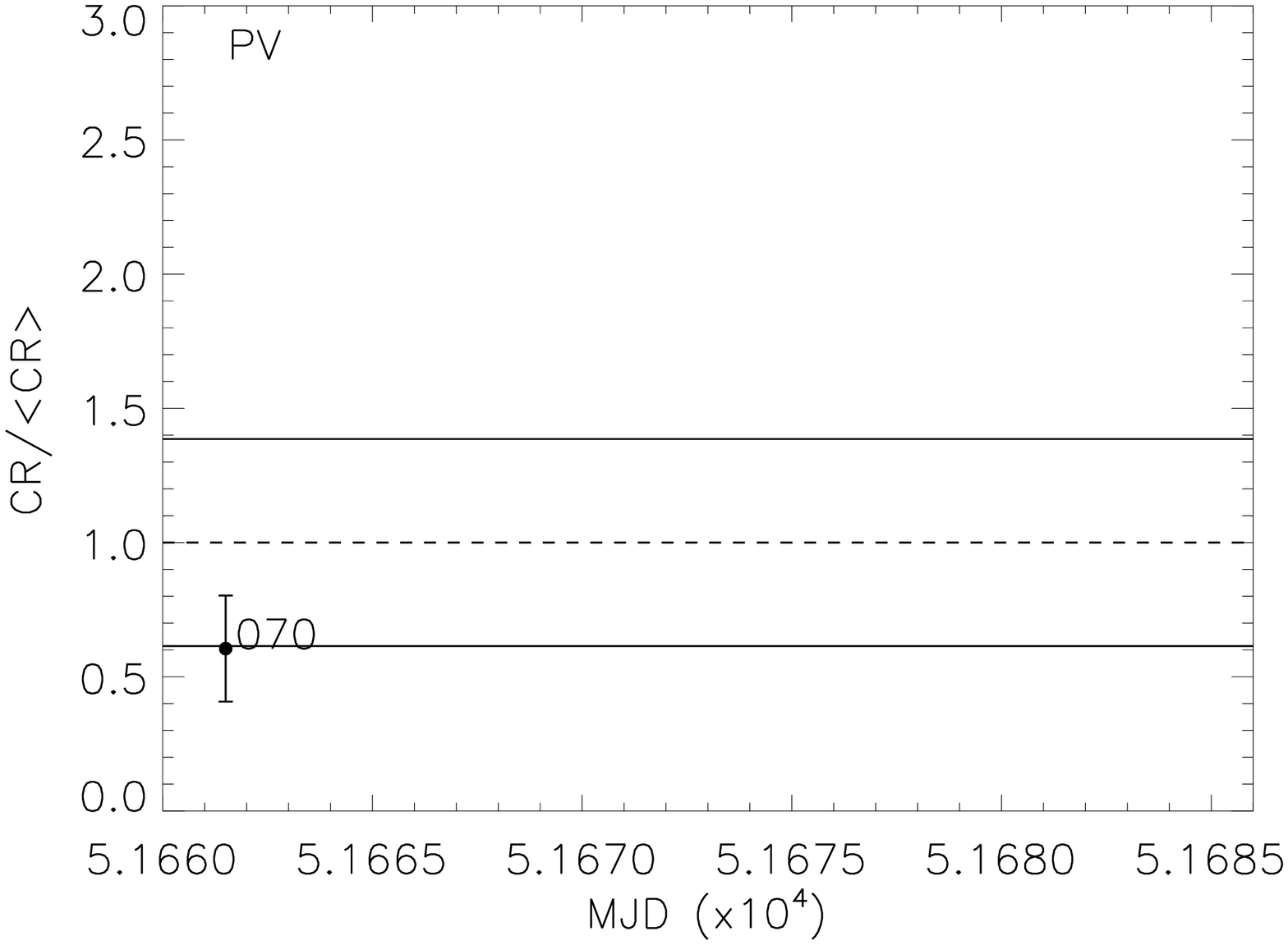}
    \includegraphics[angle=90,width=0.245\textwidth]{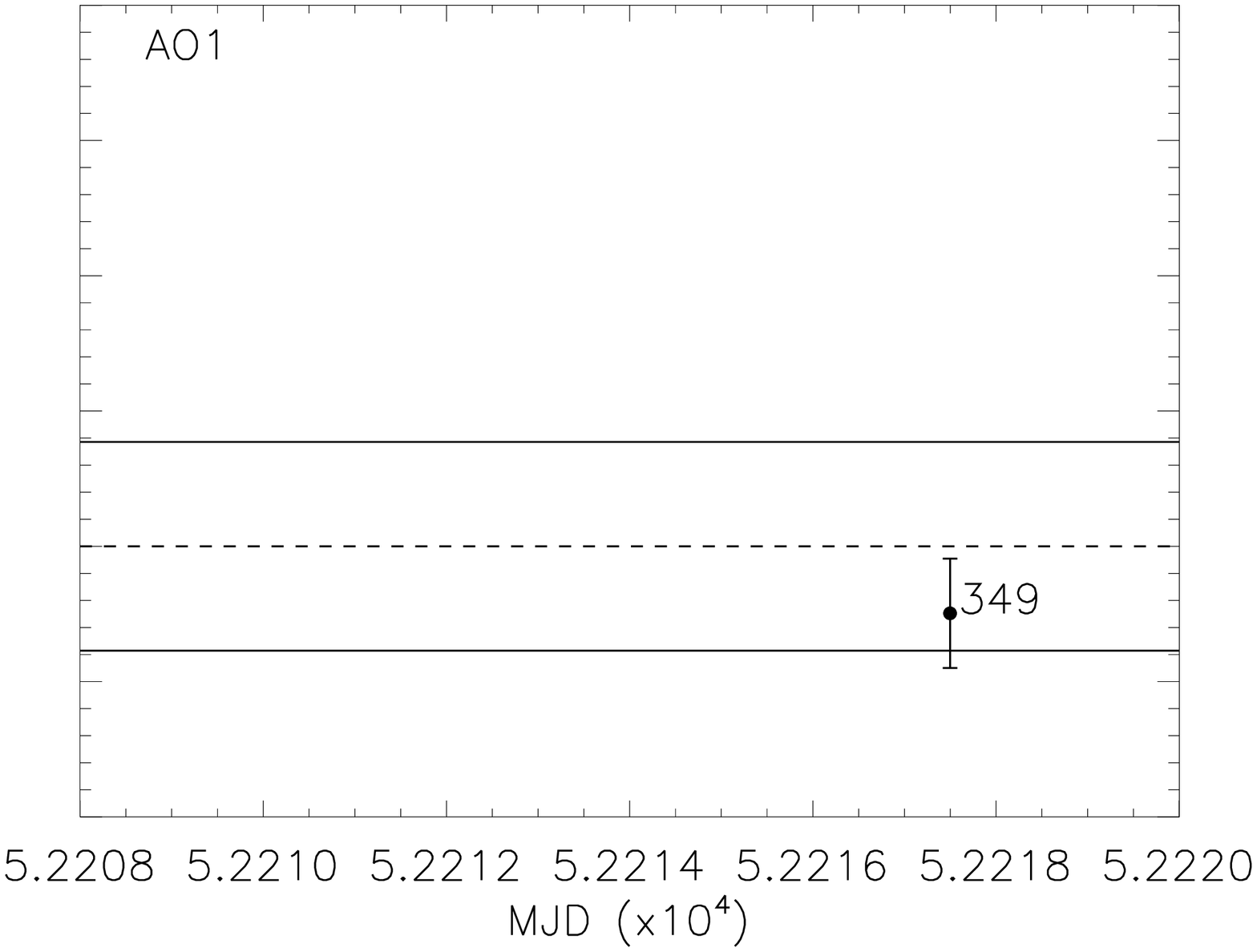}
    \includegraphics[angle=90,width=0.245\textwidth]{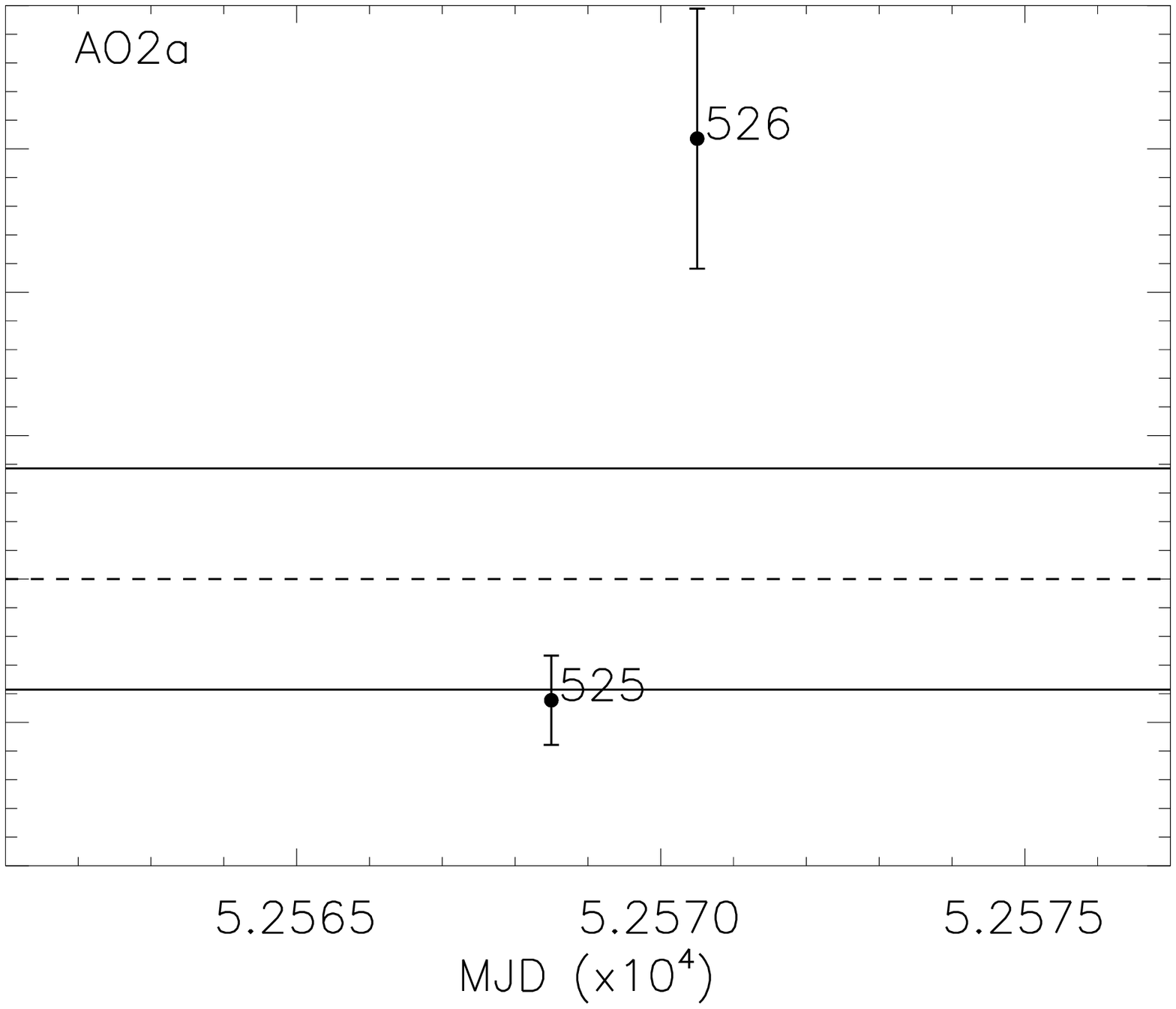}
    \includegraphics[angle=90,width=0.245\textwidth]{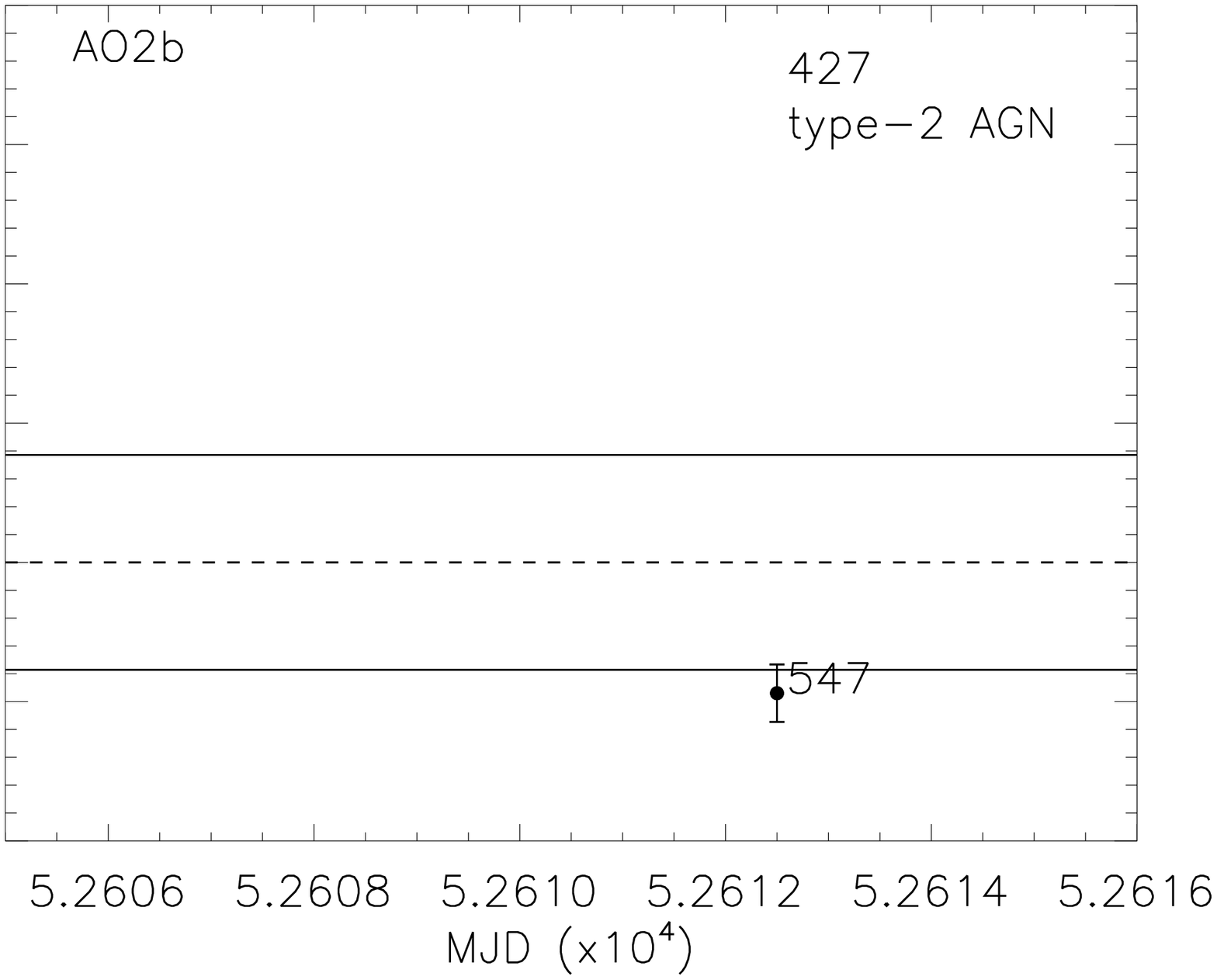}}
    \hbox{
    \hspace{0.40cm}\vspace{0.25cm}\includegraphics[angle=90,width=0.245\textwidth]{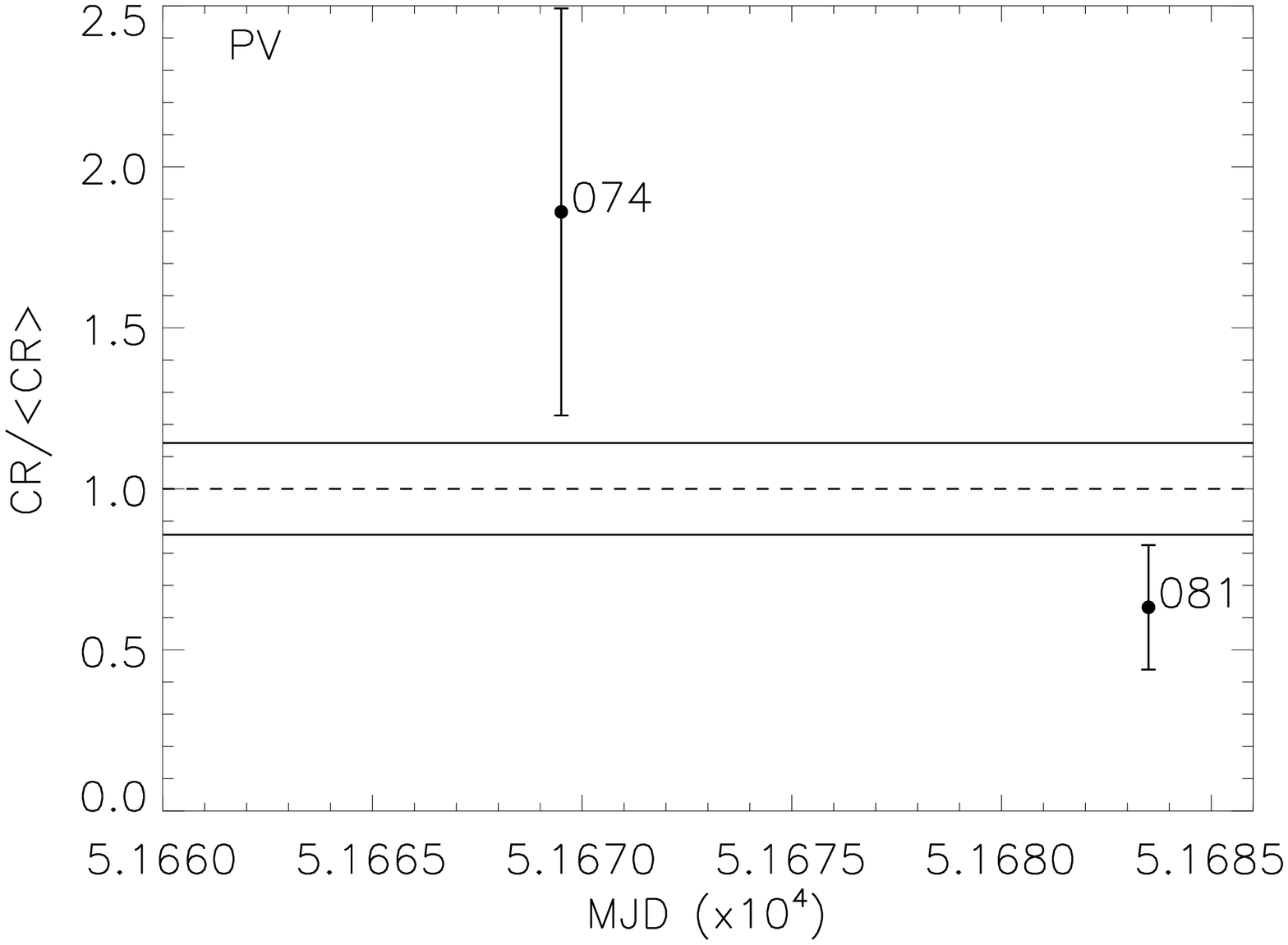}
    \includegraphics[angle=90,width=0.245\textwidth]{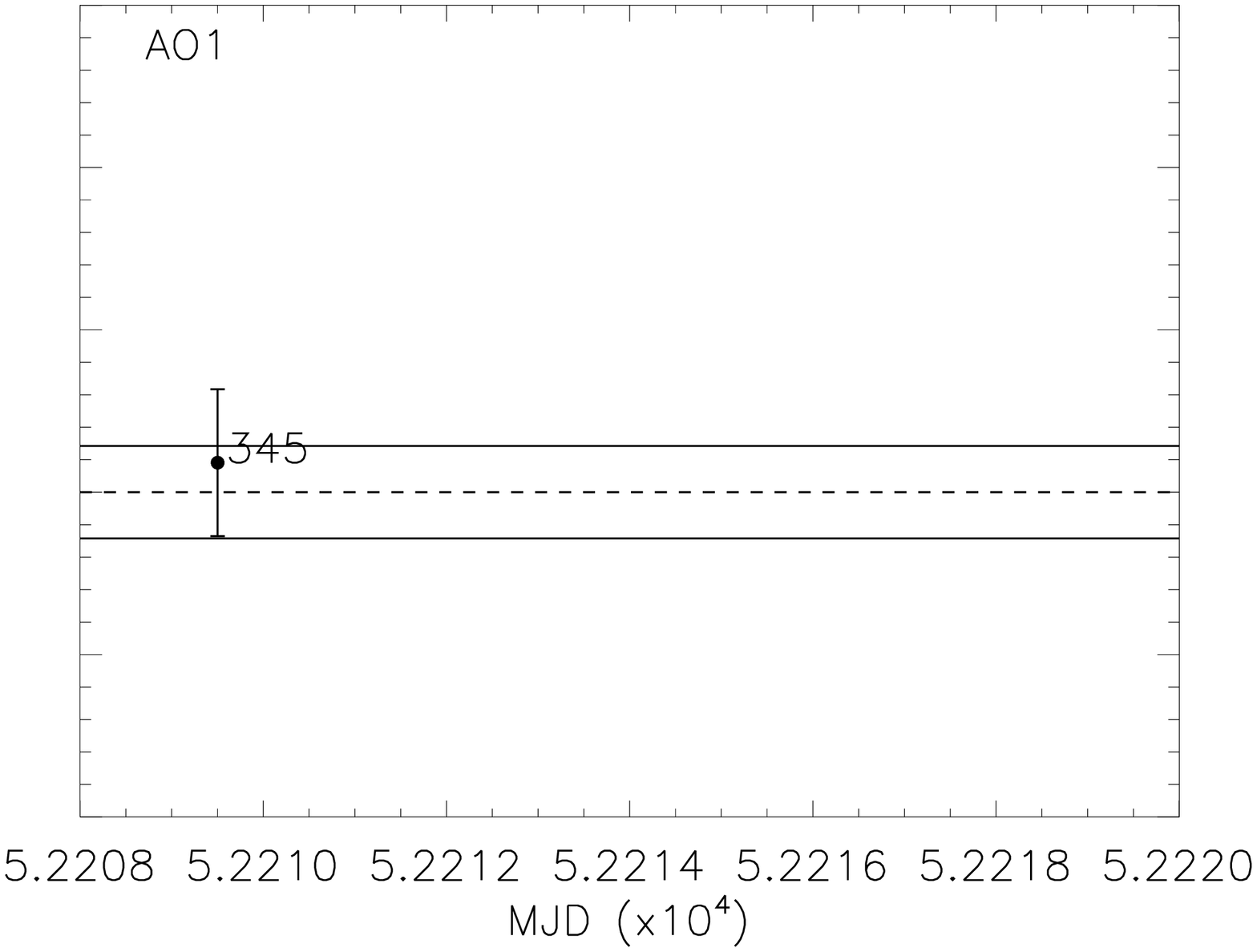}
    \includegraphics[angle=90,width=0.245\textwidth]{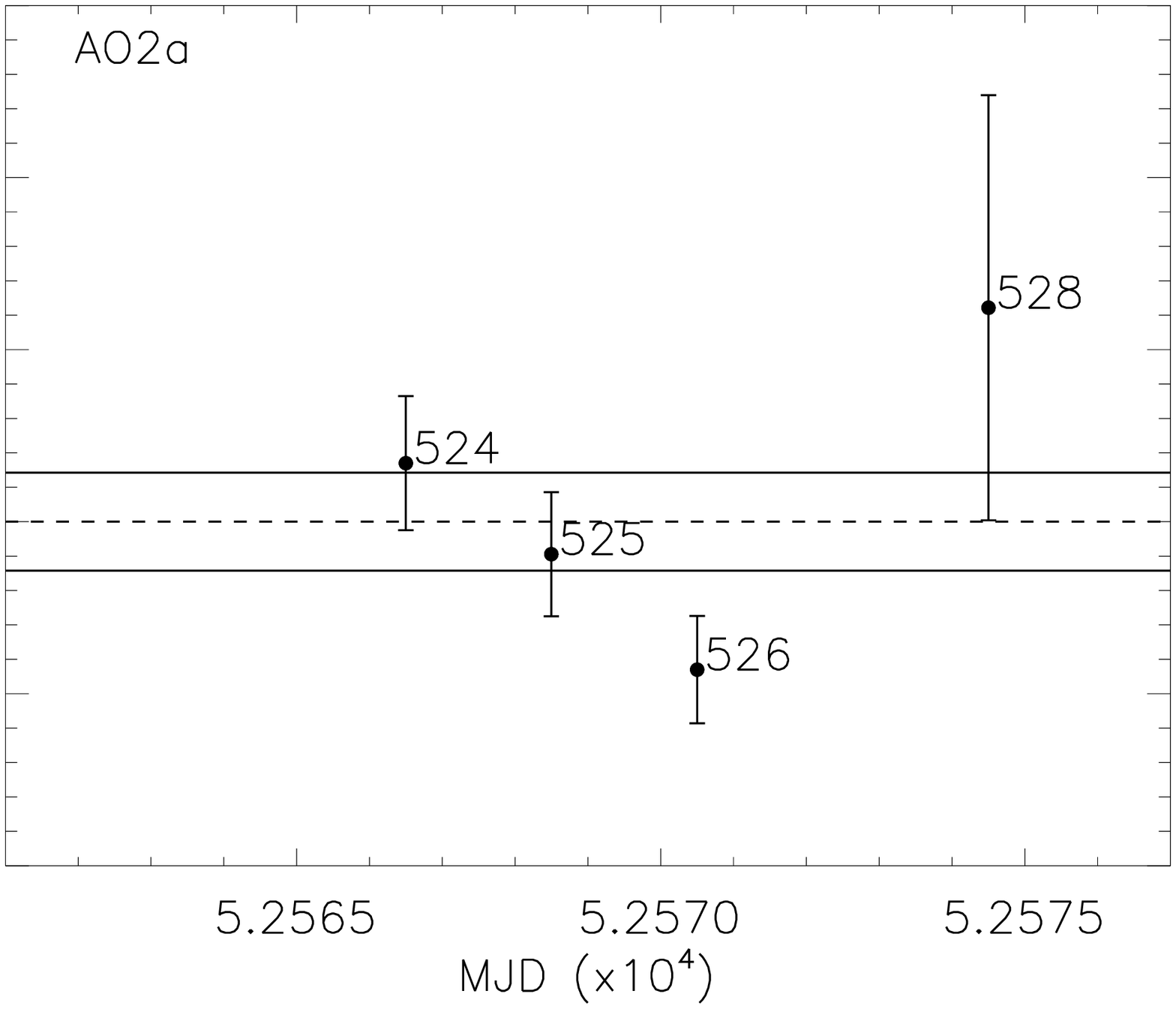}
    \includegraphics[angle=90,width=0.245\textwidth]{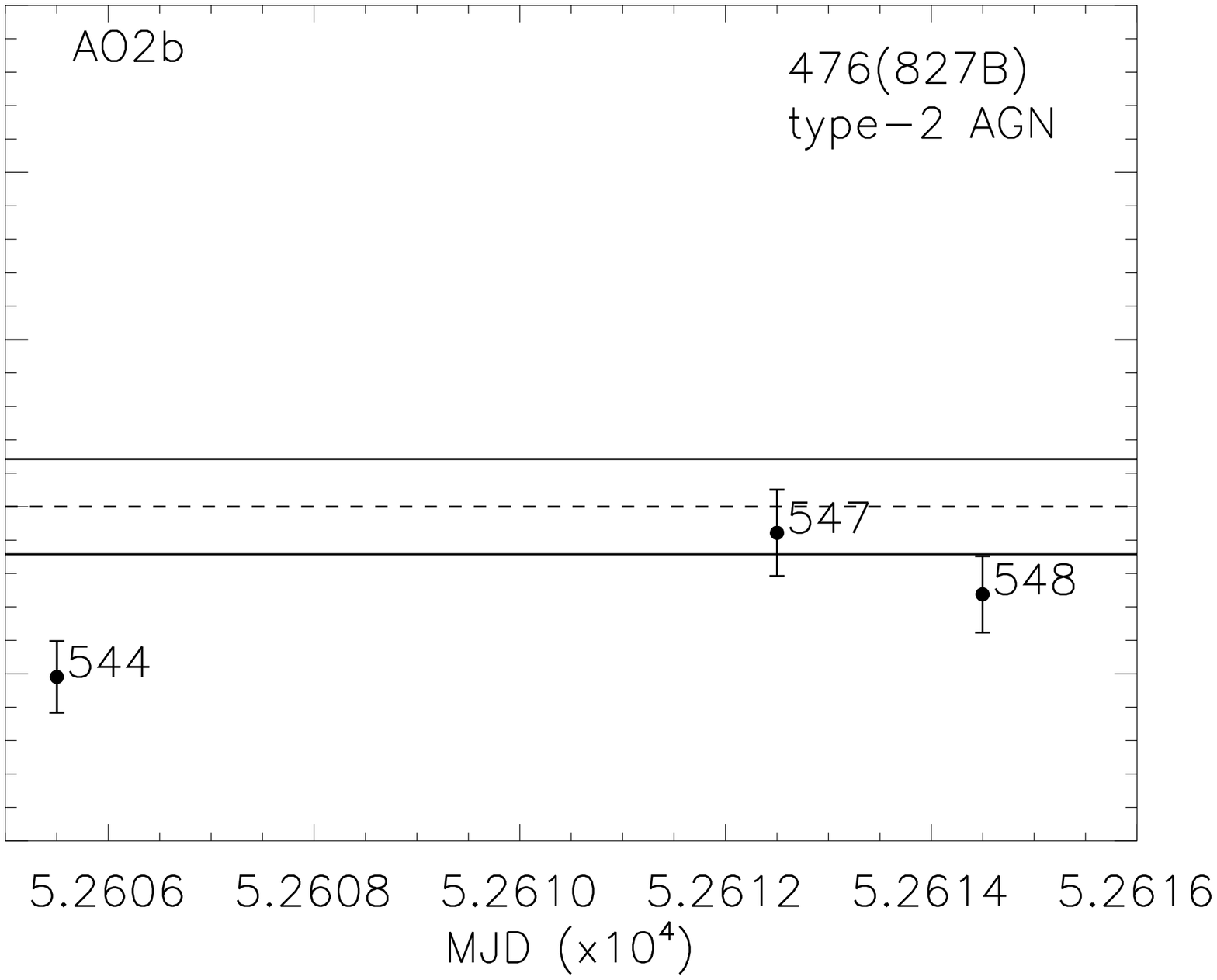}}
    \caption{Flux variability properties for 4 of the sources in our sample spectroscopically  
    classified as type-2 AGN but for which no absorption signatures were found in their co-added spectra (Mateos et al.~\cite{Mateos05b}). 
    Errors correspond to the 1$\sigma$ confidence interval.
    Horizontal lines indicate the mean {\it CR} over all revolutions (dashed lines) and the corresponding 
    1$\sigma$ confidence interval (solid lines).}
    \label{flux_var_unabsorbed_agn}
\end{figure*}

\begin{figure*}[!tb]
    \hbox{
    \hspace{0.40cm}\vspace{0.25cm}\includegraphics[angle=90,width=0.245\textwidth]{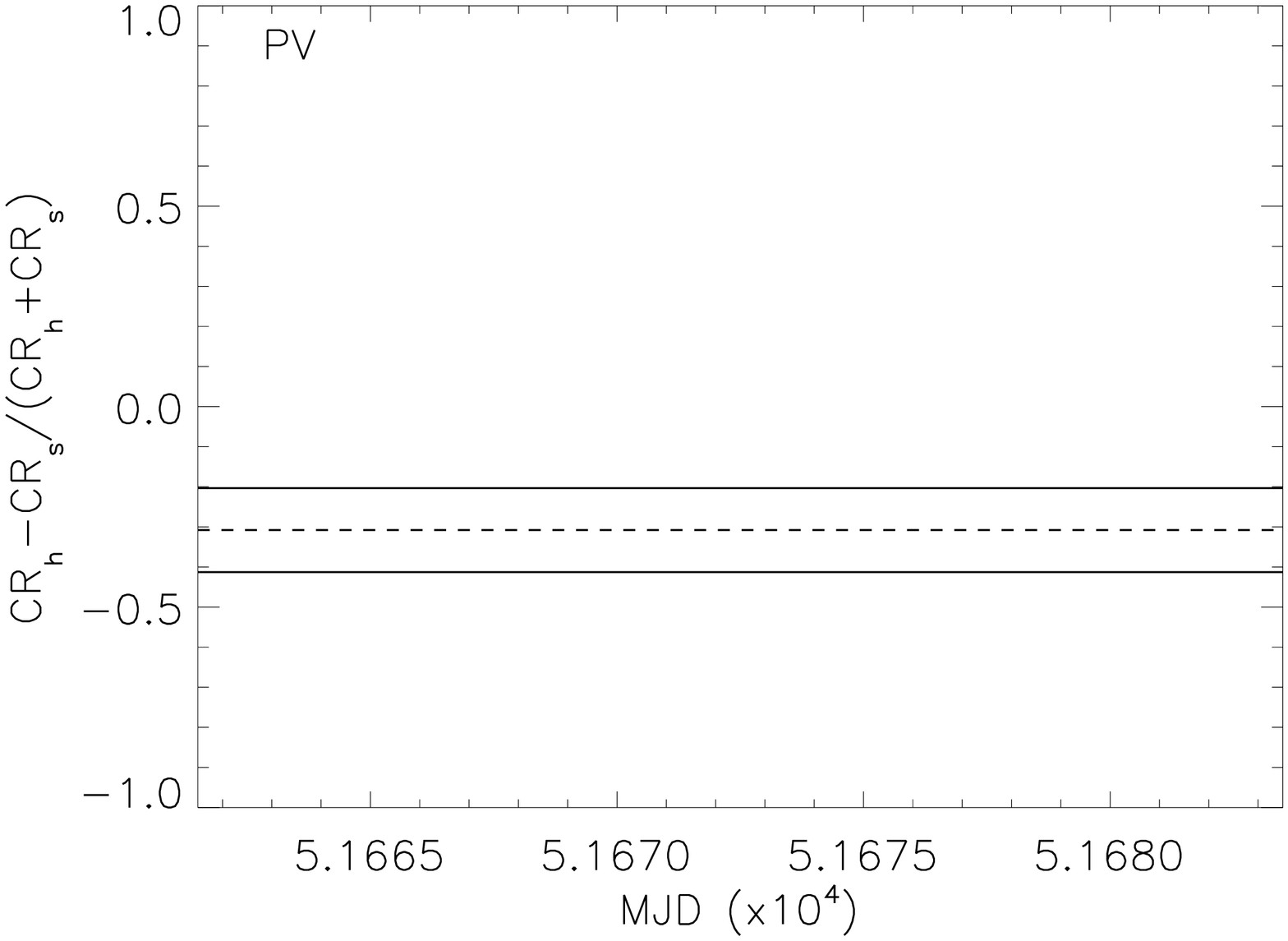}
    \includegraphics[angle=90,width=0.245\textwidth]{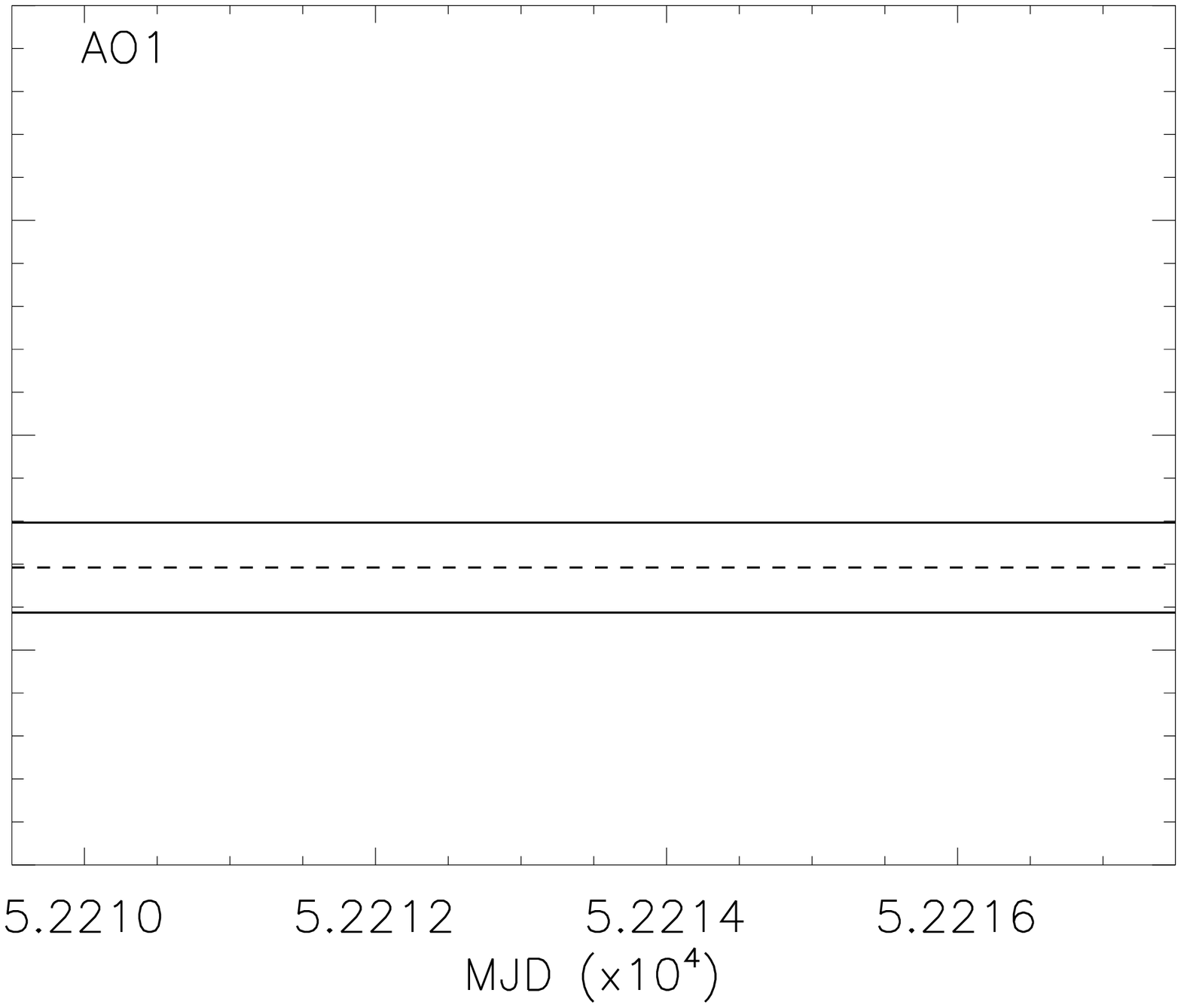}
    \includegraphics[angle=90,width=0.245\textwidth]{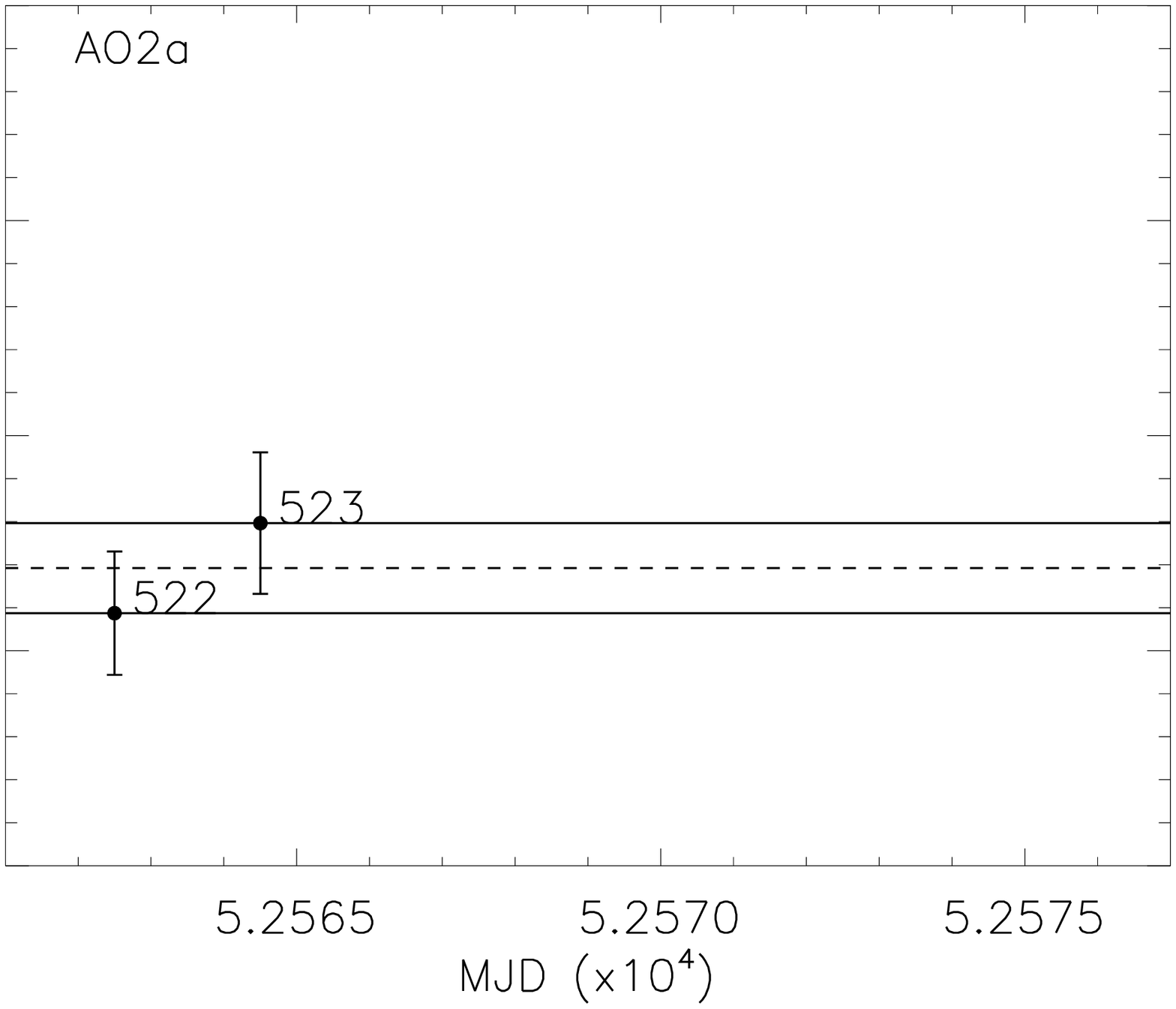}
    \includegraphics[angle=90,width=0.245\textwidth]{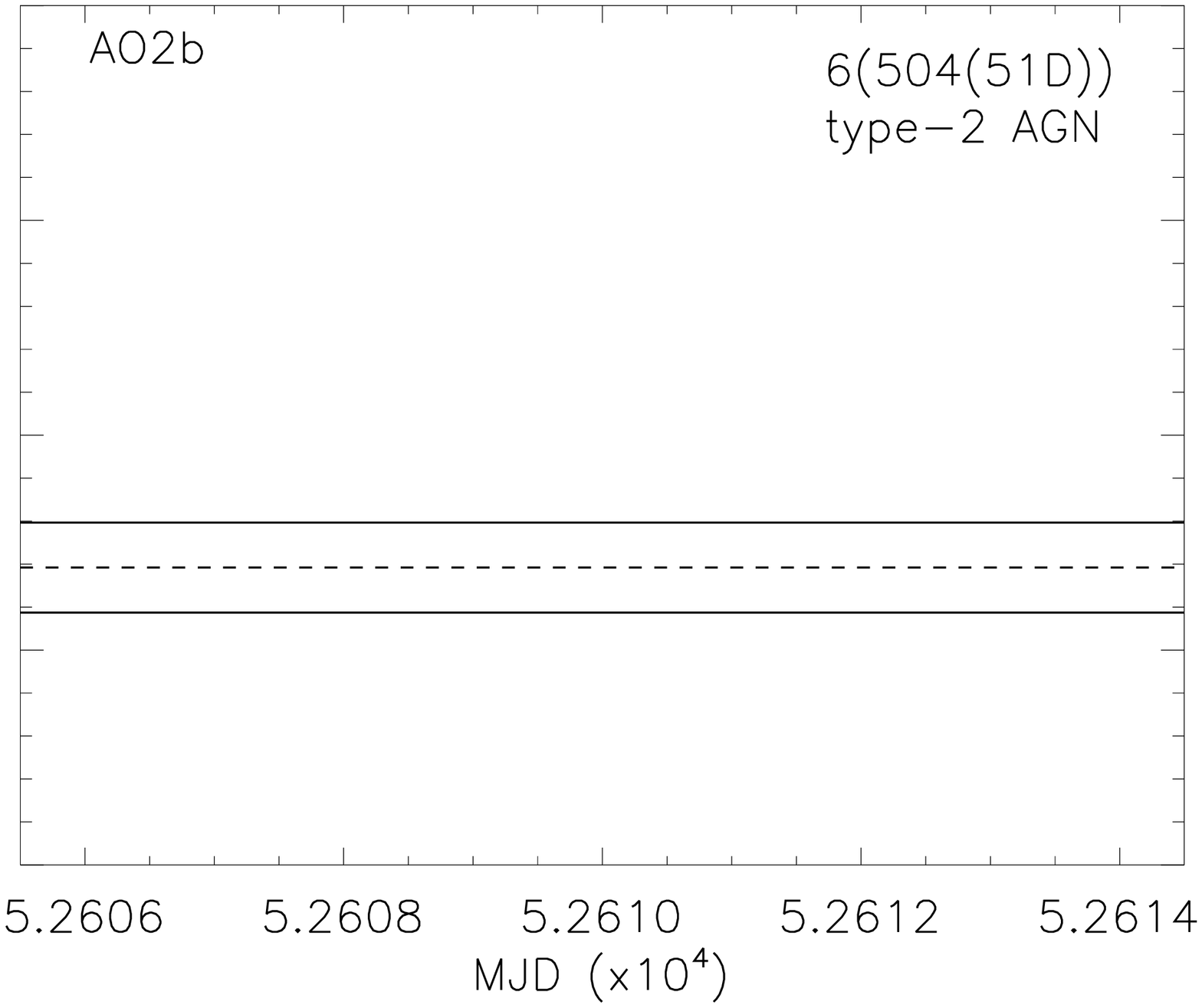}}
    \hbox{
    \hspace{0.40cm}\vspace{0.25cm}\includegraphics[angle=90,width=0.245\textwidth]{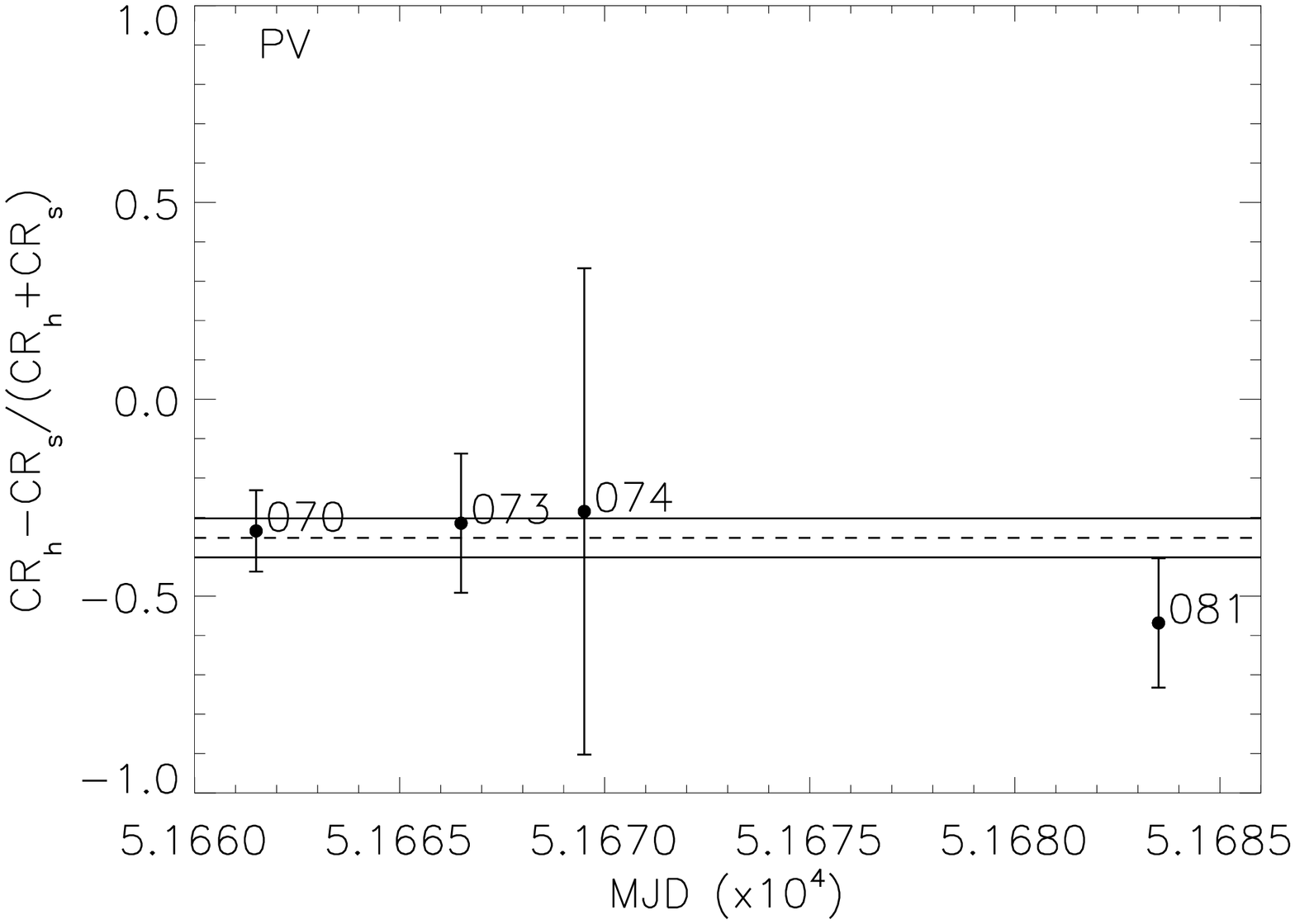}
    \includegraphics[angle=90,width=0.245\textwidth]{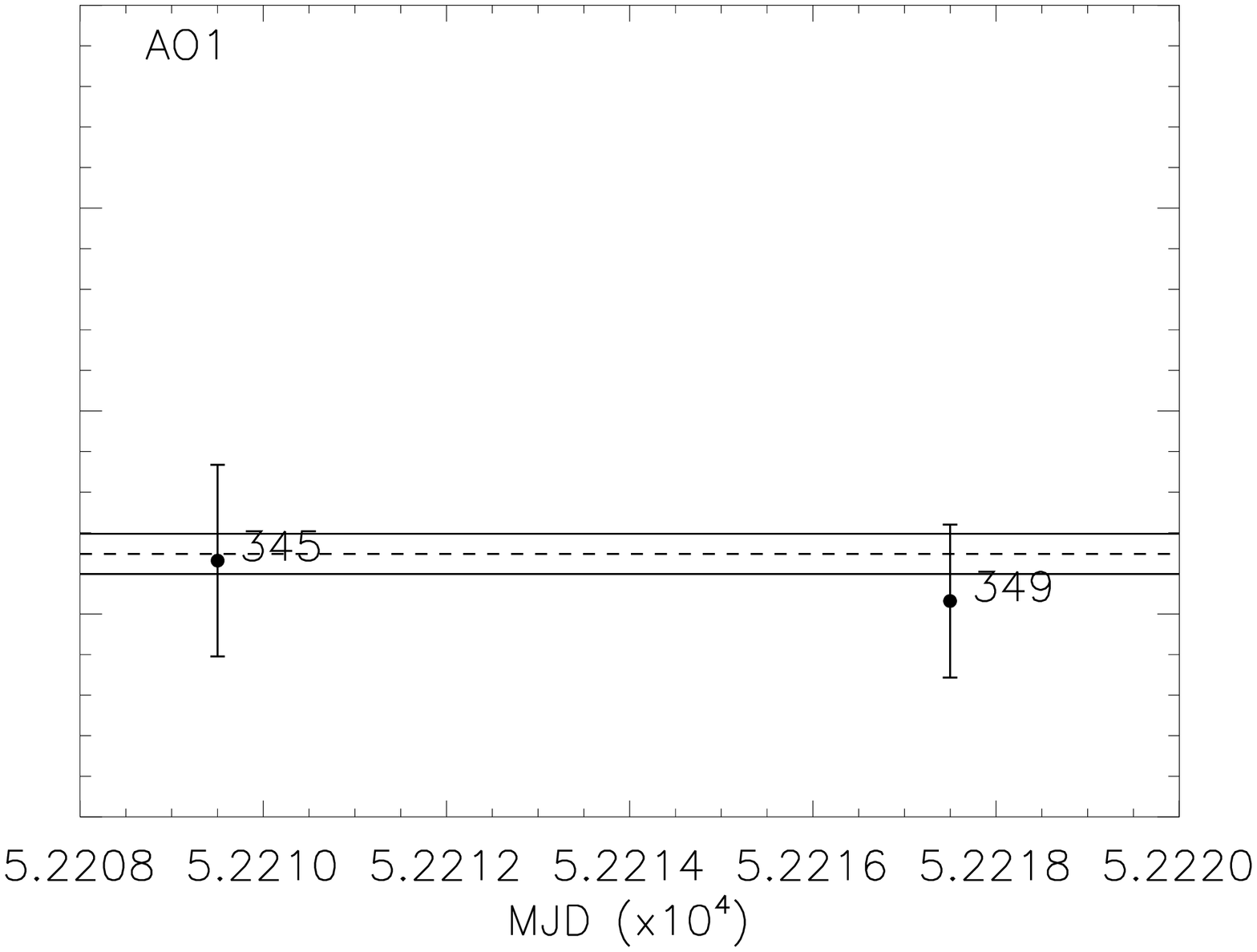}
    \includegraphics[angle=90,width=0.245\textwidth]{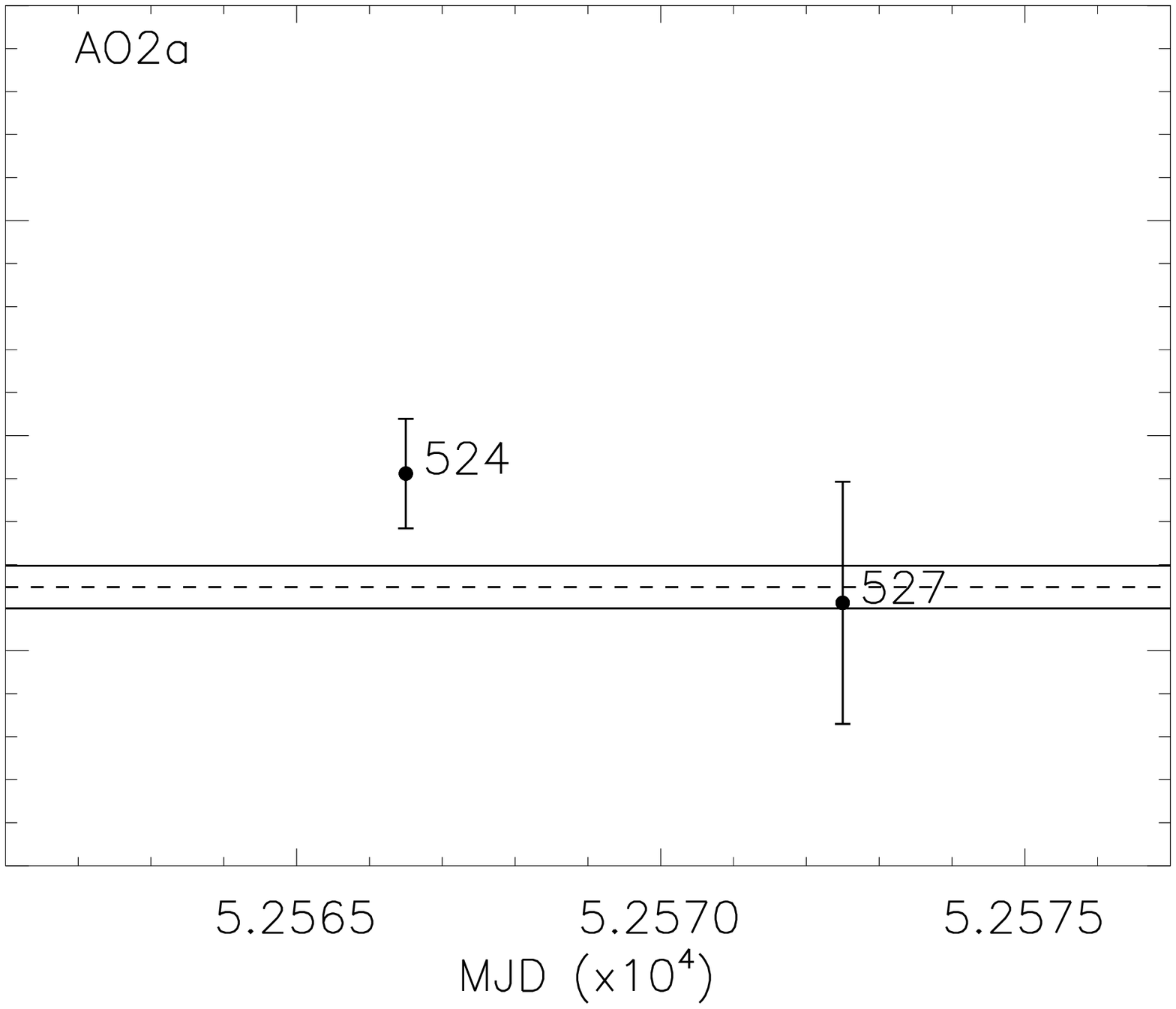}
    \includegraphics[angle=90,width=0.245\textwidth]{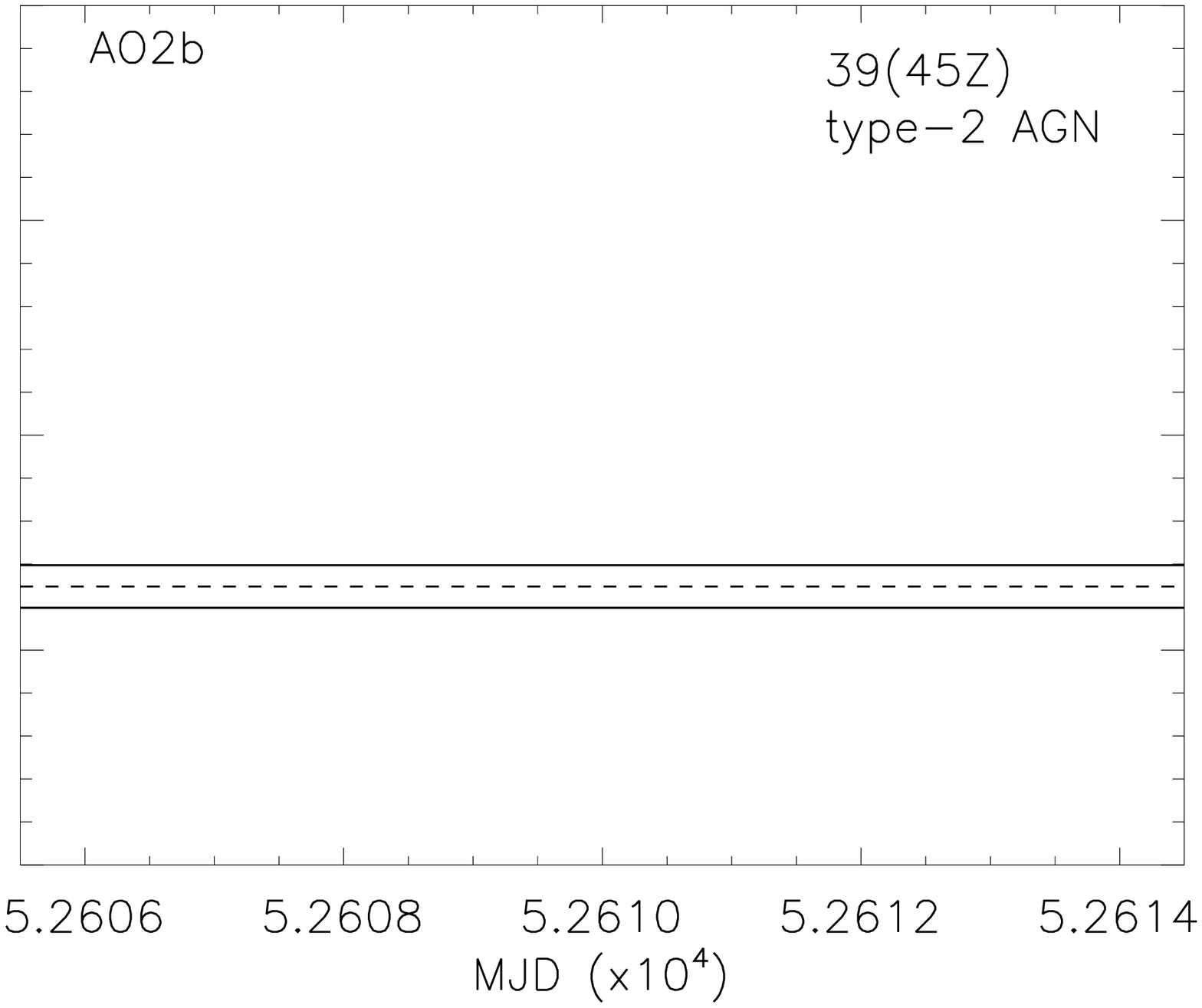}}
    \hbox{
    \hspace{0.40cm}\vspace{0.25cm}\includegraphics[angle=90,width=0.245\textwidth]{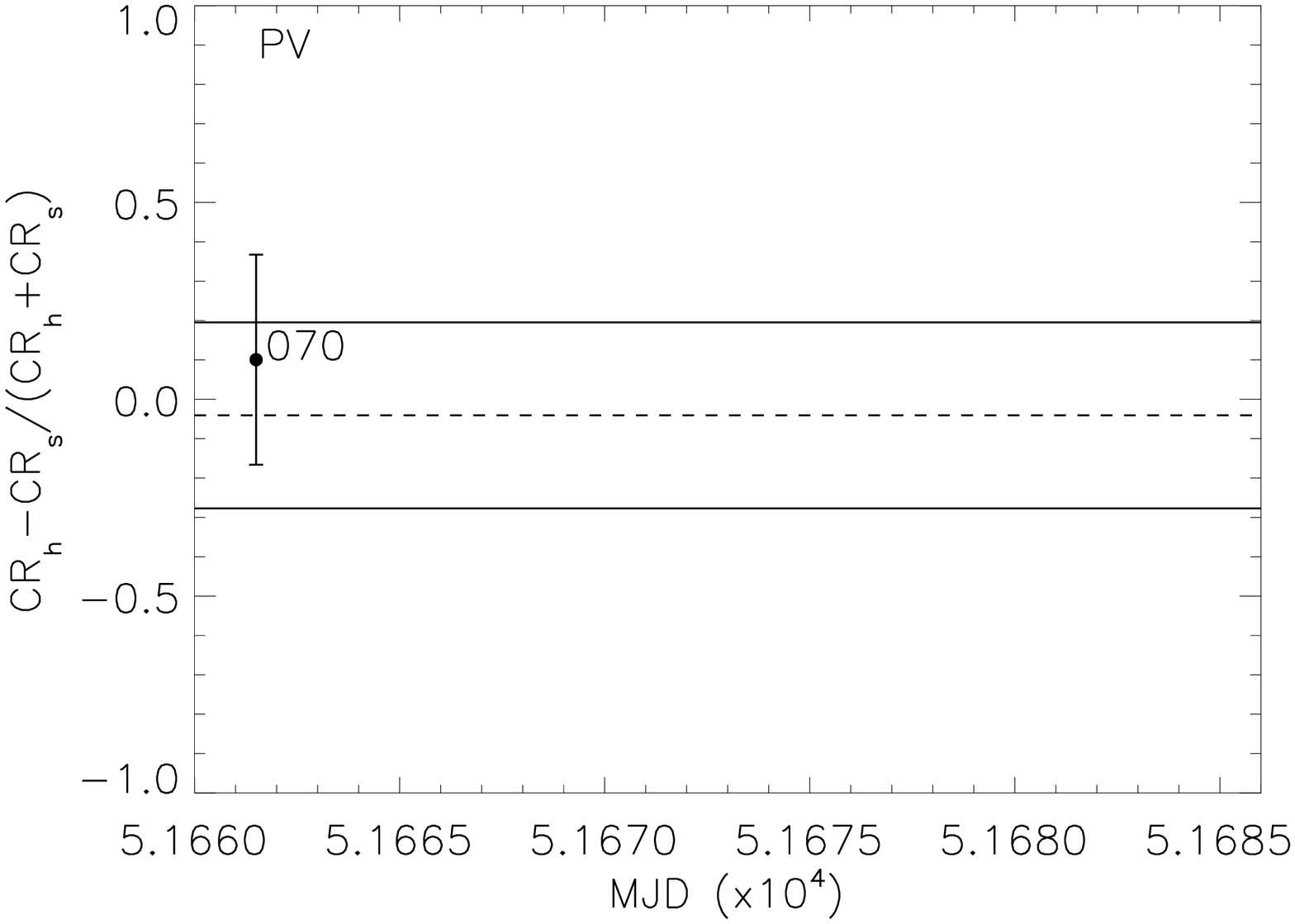}
    \includegraphics[angle=90,width=0.245\textwidth]{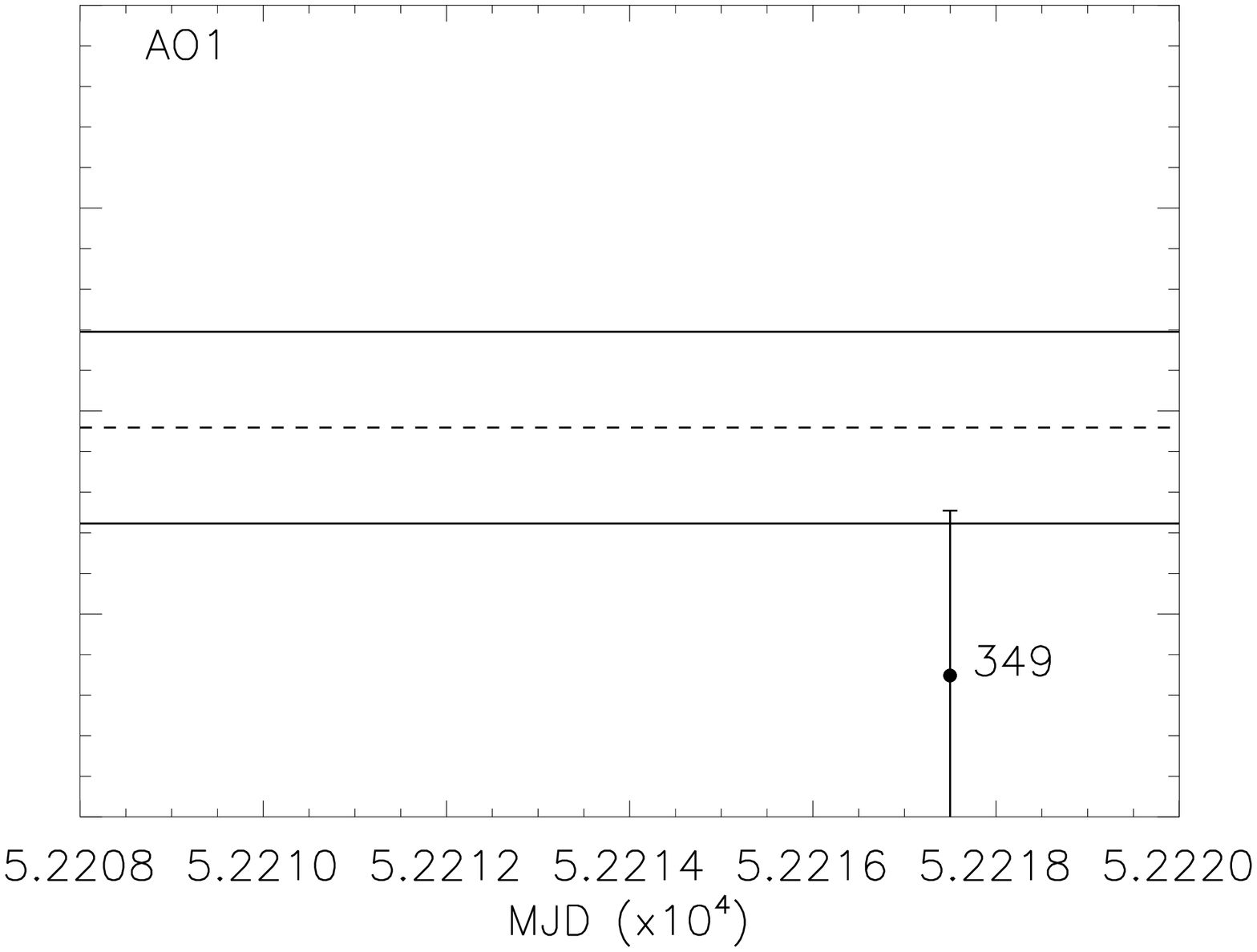}
    \includegraphics[angle=90,width=0.245\textwidth]{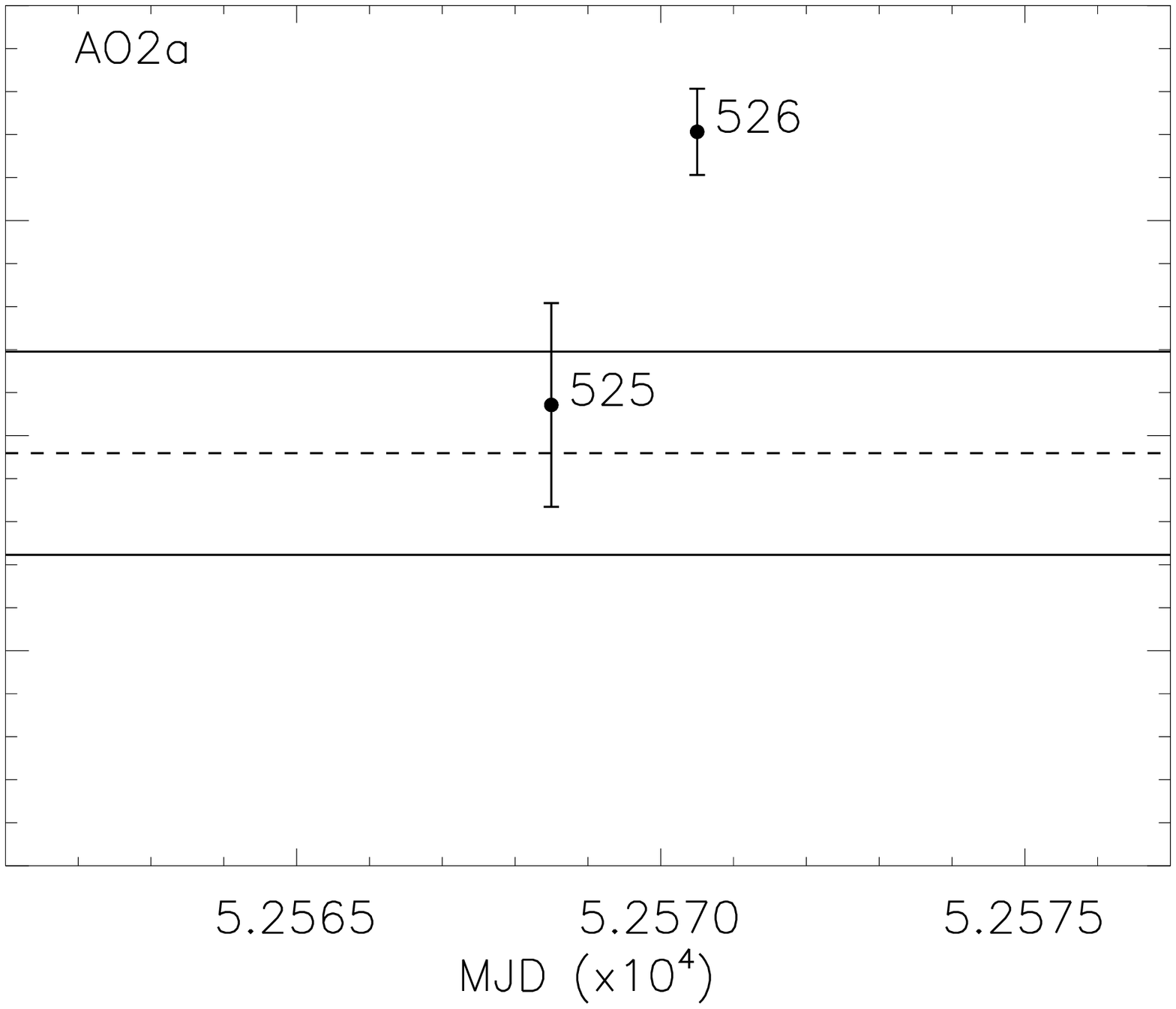}
    \includegraphics[angle=90,width=0.245\textwidth]{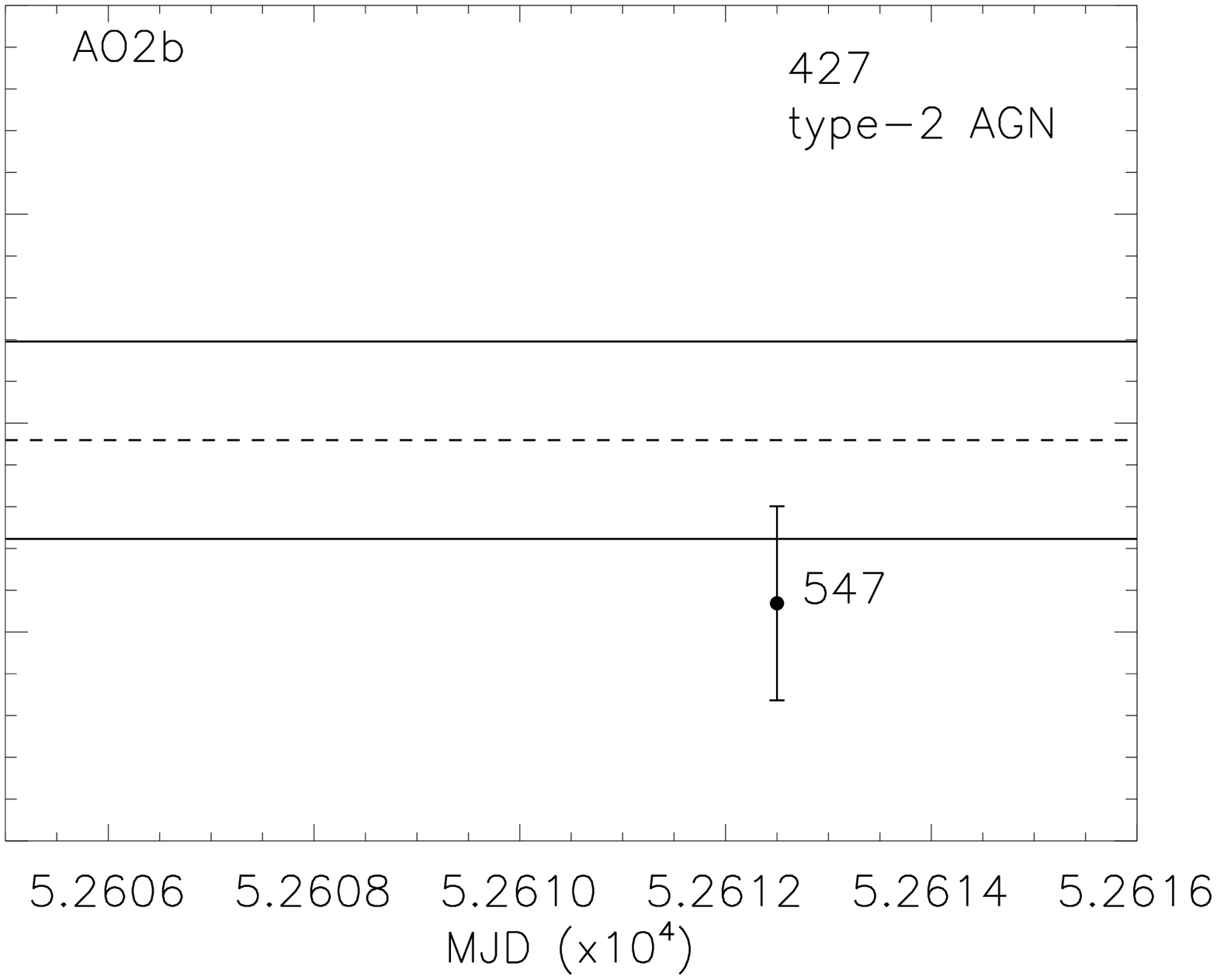}}
    \hbox{
    \hspace{0.40cm}\vspace{0.25cm}\includegraphics[angle=90,width=0.245\textwidth]{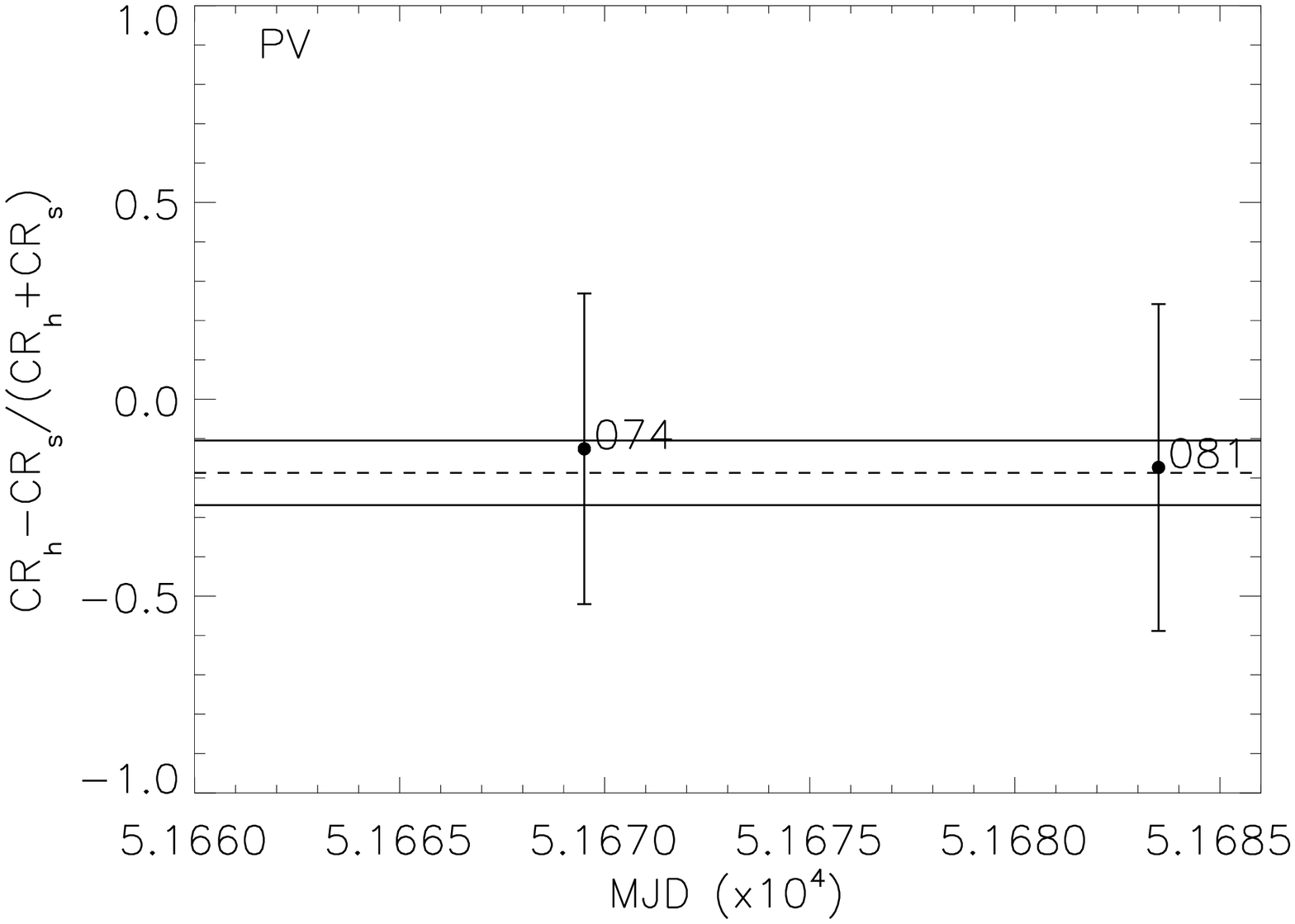}
    \includegraphics[angle=90,width=0.245\textwidth]{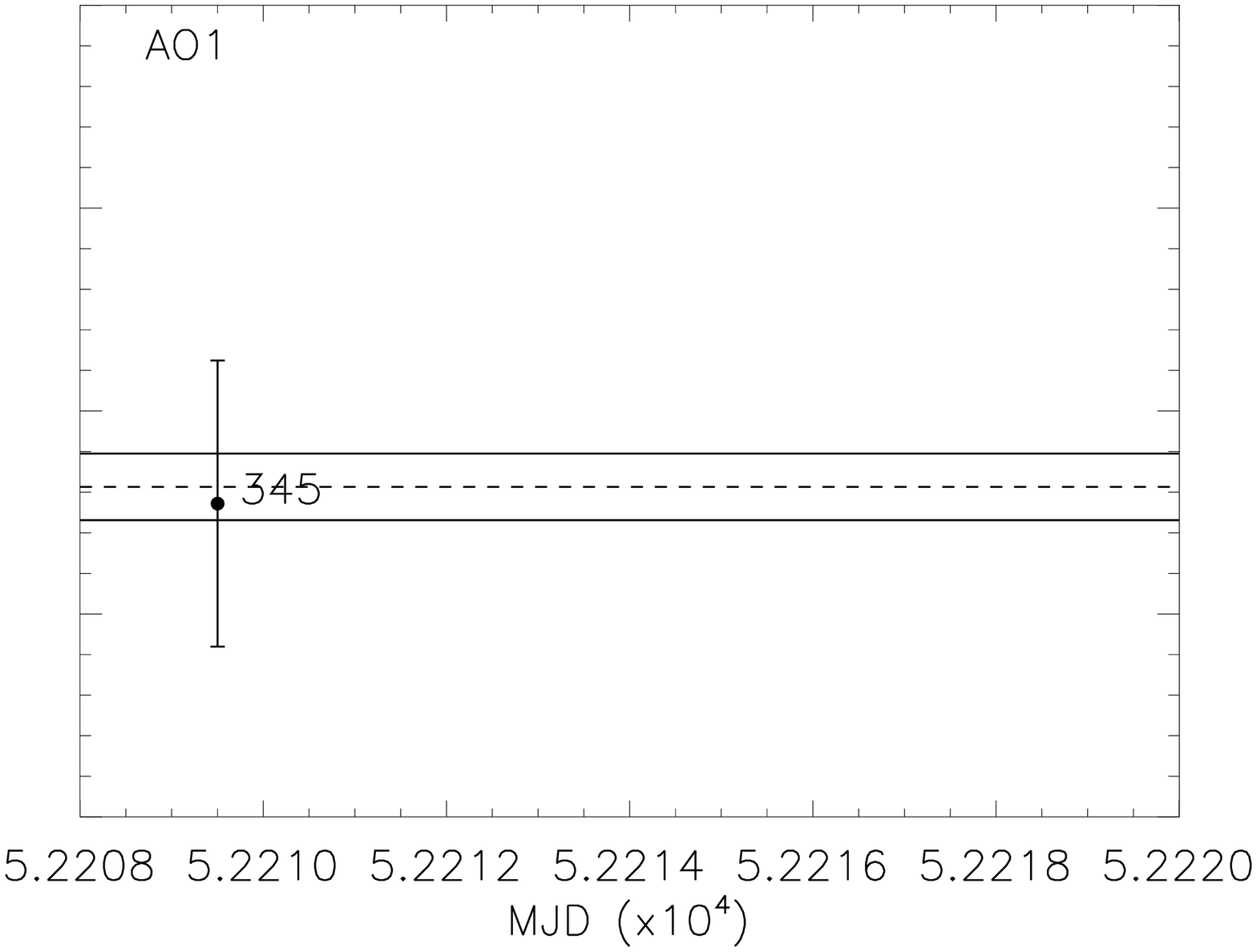}
    \includegraphics[angle=90,width=0.245\textwidth]{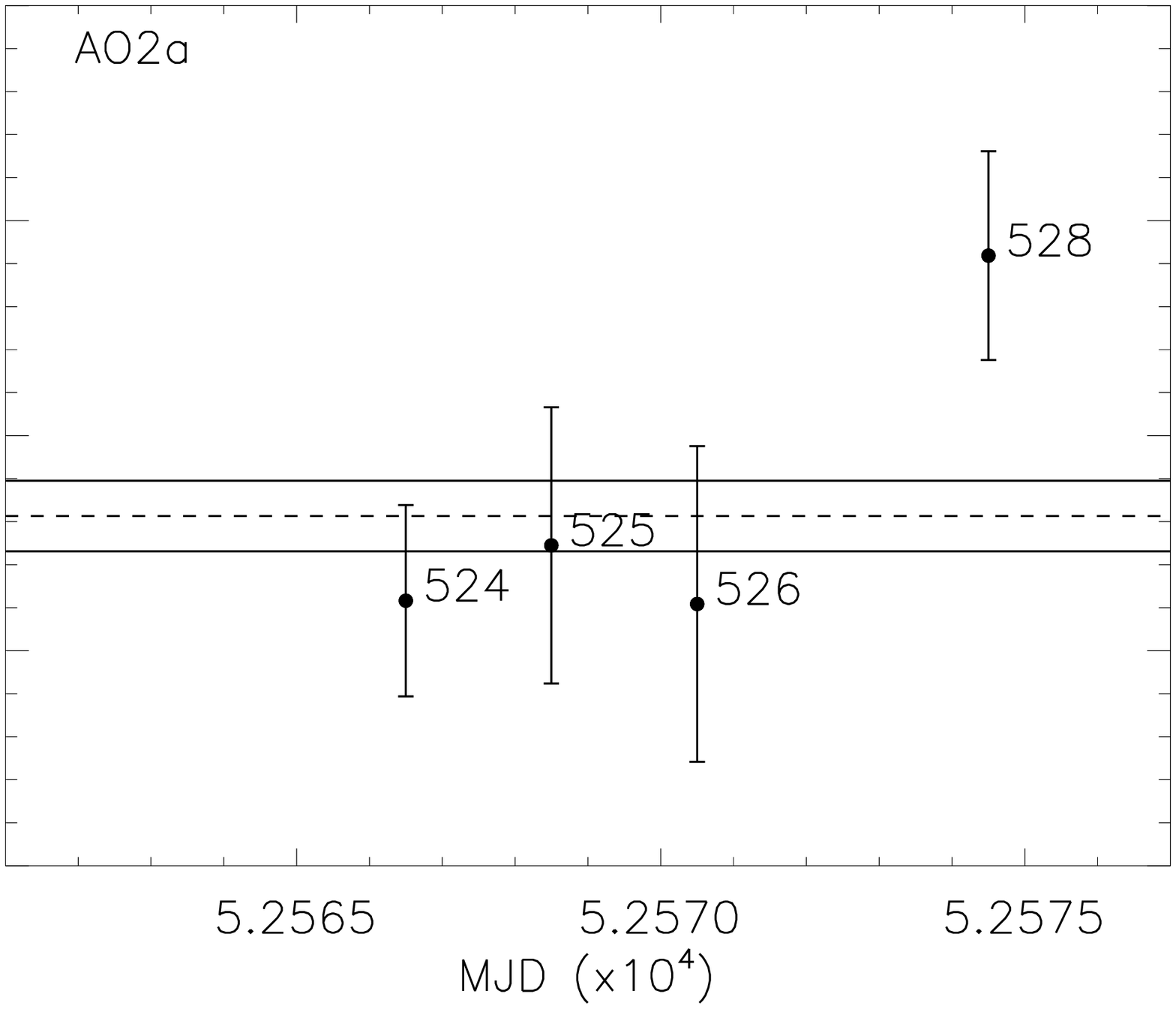}
    \includegraphics[angle=90,width=0.245\textwidth]{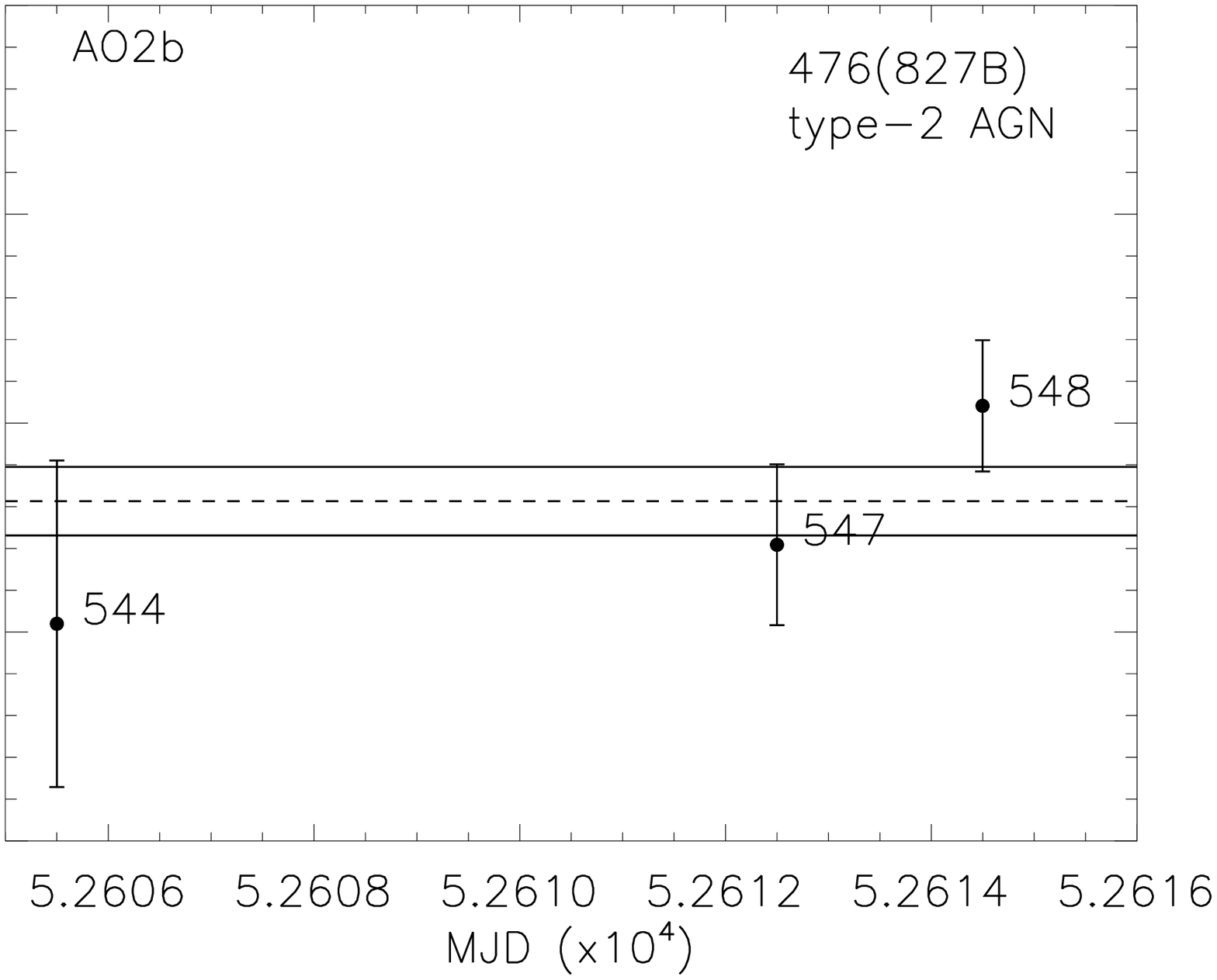}}
    \caption{Spectral variability properties for the sources in our sample spectroscopically  
    classified as type-2 AGN but for which no absorption signatures were found in their co-added 
    X-ray spectra. 
    Errors correspond to the 1$\sigma$ confidence interval.
    Horizontal lines indicate the mean {\it HR} over all revolutions (dashed lines) and the corresponding 
    1$\sigma$ confidence interval (solid lines).}
    \label{sp_var_unabsorbed_agn}
\end{figure*}

\section{On the long-term X-ray variability}
\label{origin_var}
The results of our variability analysis suggest that flux and spectral variability are not correlated for a 
significant fraction of the objects in our sample. Indeed 38 sources in our sample exhibit  
flux variability but not spectral variability, while 9 show spectral but not flux variability (see 
Table.~\ref{summary_var_det}).

In addition, no evident correlation between flux and spectral variability was found for the 15 objects for which 
both were detected. The {\it HR}-{\it CR} plots for these objects are shown in 
Fig.~\ref{sp_var_fl_var}. We see in Fig.~\ref{sp_var_fl_var} that significant changes in flux without evident 
changes in the X-ray colour are common in our objects. On the other hand important changes in the X-ray colour without significant flux variability 
have been detected in many of our sources.

The {\it two-component spectral model} and the {\it spectral pivoting model} have been frequently used to describe the 
flux and spectral variability properties in local Seyfert 1 galaxies.
We have studied whether these phenomenological models can accommodate the variability properties of our objects, i.e. whether 
changes in the continuum shape alone can explain the observed spectral and flux 
variability detected in our objects. To do that we have assumed that the spectra of our sources can be well described 
with a power law (absorbed by the Galaxy). Then we have calculated the changes 
in {\it HR} and {\it CR} associated with variations of $\Gamma$ alone for different rest-frame 
pivoting energies and for a source at a redshift of 1\footnote{If the pivoting energy is in the soft 
end of the {\it XMM-Newton} band pass then the amplitude 
of flux variability associated with changes in $\Gamma$ alone increases with the redshift and therefore a correlation 
between {\it HR}-{\it CR} is more significant at higher redshifts.}.
The results are shown in Fig.~\ref{pivoting} (top) for pivoting energies of 1, 10, 30, 100 and 1000 keV.  

We see that typical changes in $\Gamma$$\sim$0.2 
correspond to small changes in the X-ray colour $\sim$0.1-0.2, which are undetectable in most of our sources. 
This could easily explain the lack of detection of spectral variability in a significant fraction of 
our sources. In addition, if the pivoting energy falls in the observed energy band (in the example 
of Fig.~\ref{pivoting} this corresponds to pivoting energies of 1 keV and $\sim$10 keV), then 
spectral variability without significant changes in flux could be observed.
However, in the cases where both flux and spectral variability are detected a correlation 
between the X-ray colour and flux should be evident, but as we see in Fig.~\ref{sp_var_fl_var} in the majority 
of the objects X-ray colour-flux trends are not observed. 
Furthermore, in order to explain with the spectral pivoting model 
the observed dispersion of X-ray colours 
changes in $\Gamma$ above 1 are required. If we take into account that in the majority of the Seyferts for which 
detailed variability studies are available, the detected pivoting energies are $\gg$10 keV (for example 
see Table 1 in Zdziarski et al.~\cite{Zdziarski03}), then in most of our sources the 
pivoting energies are expected to be outside the observed energy band, and hence it is expected that changes 
in $\Gamma$$\gtrsim$1 will produce changes in flux much higher than those detected: these flux and spectral changes 
are not observed simultaneously in any of our sources.
Finally, changes in $\Gamma$ of at least 1 are unlikely based on the predictions of thermal Comptonization 
models and the results shown in Sec.~\ref{sp_var_sp}. 

We conclude that it is unlikely that spectral pivoting alone can explain the observed trends in 
flux and spectral variability in our sources. The same result applies also to the {\it two-component spectral model}, as in this case as for the {\it spectral pivoting model} similar X-ray colour-flux trends are expected.

We have repeated our analysis to study the expected changes in X-ray colour and flux associated with changes 
in the amount of absorption alone. The results are shown in Fig.~\ref{pivoting} (bottom), for different amounts of absorption 
and at different redshifts. Changes in absorption can be produced, for example, due to variations in the 
ionisation state of the absorber following changes in the ionising radiation, making the absorber more transparent. 
On the other hand, if the absorber is not homogeneous but clumpy (for example if it is made of clouds of 
gas moving around the central source), then variations in the column density along the line of sight will be 
expected due to clouds passing through the line of sight. 

In this case both spectral and flux variability should be detected (see  Fig.~\ref{pivoting} bottom), but again, this is not the 
case for many objects in our sample. To produce changes in the X-ray colour that would be detected in our data, 
changes in absorption of ${\rm N_H\ge5\times10^{22}\,cm^{-2}}$ are required, especially if the sources are at 
redshifts greater than 1. Changes in ${\rm N_H}$ above this value could explain the observed amplitude of spectral 
variability seen in a fraction of our objects. However in this case, as we see in Fig.~\ref{pivoting} (bottom), flux and 
spectral variability should be correlated: significant variations in absorption would be associated with 
significant changes in flux. It is very unlikely that variations in X-ray absorption alone can explain the variability properties of our sources. 

Hence, from our analysis we conclude that 
\begin{enumerate}
\item{the lack of correlation between the observed flux and spectral variability properties of our objects 
indicates that the observed spectral variability is not triggered by changes in the X-ray luminosity of the sources.}
\item{the amplitude of the observed flux variability cannot be explained due to changes in the X-ray continuum shape or the 
amount of X-ray absorption alone.}
\item{ The {\it two-component spectral model} and the {\it spectral pivoting model} do not describe 
  properly the variability of our sources.}
\end{enumerate}

In order to explain both the flux and spectral variability properties of our sources, at least two parameters of the model 
must vary independently, such as, for example, the shape of the continuum or the 
amount of X-ray absorption, but also the continuum normalisation.

\section{Variability properties of unabsorbed type-2 AGN}
\label{sp_var_unabs_type2}
In Mateos et al. (\cite{Mateos05b}) we showed that among the spectroscopically identified type-2 AGN 
in the sample of {\it Lockman Hole} sources selected, five did not show 
any signs of X-ray absorption in their co-added X-ray spectra (objects with identification number 6, 21, 39, 427 and 476). 
We have studied the flux and spectral variability properties of all these sources except 
source 21, for which only one data point was available. One of the hypotheses that could explain the 
disagreement between observed optical and X-ray properties in these sources is 
spectral variability: because the optical and X-ray observations were not obtained 
simultaneously, there might have been changes in the absorber
during the time interval between the observations. Table~\ref{summary_var_det} lists the flux 
and spectral variability properties of these sources, while their flux and X-ray colour curves 
are shown in Fig~\ref{flux_var_unabsorbed_agn} and Fig.~\ref{sp_var_unabsorbed_agn} respectively.
We detected flux variability in 3 of the sources (39, 427 and 476); however spectral variability was 
only detected in one of the objects, source 427. Fig.~\ref{sp_var_fl_var} shows the flux vs spectral 
variability 
properties for source 427. We see that in four of the available observations flux variability was not seen, 
however changes in the {\it HR} are evident. 

These sources were detected in a narrow redshift interval, from 0.5-0.7. In the previous section we have seen that 
at redshifts below 1, changes in the column density of the order of ${\rm N_H\sim10^{21}\,cm^{-2}}$, affect the observed flux by 
$\sim$20\%. However the corresponding change in the X-ray colour is only {\it $\Delta$HR}$\sim$0.05, and therefore 
impossible to detect in our data. 
If the changes in absorption were to be of the same magnitude as  
the typical absorption detected in our absorbed type-2 AGN 
(${\rm \sim10^{22}\,cm^{-2}}$; see Figure 12 in Mateos et al.~\cite{Mateos05b}) or larger, then 
changes in flux and in 
the X-ray colour will be significant ({\it $\Delta$CR}$\ge30\%$ and {\it $\Delta$HR}$\ge0.3$) and therefore 
we should have been able to detect them. Therefore, the lack of correlation 
between flux and spectral variability in these objects indicates that, for our 
unabsorbed type-2 AGN, variations in absorption of at least ${\rm N_H\ge10^{22}\,cm^{-2}}$ 
are not observed in these objects. Moreover, the lack of detection of significant spectral variability 
in most of the sources makes unlikely that spectral variability alone can explain the observed mismatch of their optical-X-ray 
spectral properties as in these sources variations in the amount of absorption of at least ${\rm 10^{22}\,cm^{-2}}$ are expected.

Finally, the non detection of X-ray absorption in these objects cannot be explained by these sources 
being Compton-thick, because in this case, the X-ray emission should be dominated by scattered 
radiation (the varying nuclear component is largely undetected), and therefore we would not expect to detect flux or 
spectral variability in the sources on a time scale of $\le$2 years.

\section{Conclusions}
\label{conclusions}
We have carried out a detailed study of the X-ray spectral and flux variability properties  
of a sample of 123 sources detected with {\sl XMM-Newton} in a deep observation in the 
{\it Lockman Hole} field. To study flux variability, we built for each object a light curve using 
the EPIC-pn 0.2-12 keV count rates of each individual observation. We obtained light curves for 120 out of 
123 sources (the other three sources only had one bin in their light curves). We have searched for variability 
using the $\chi^2$ test. We detected flux variability with a confidence level of at least $3\sigma$ in 62 
sources ($\sim$52\%). 
However the efficiency of detection of variability depends on the amplitude of 
the variability. The observed strong decrease in the detection of variability in our sources at 
amplitudes $\le$0.20 indicates that amplitudes of variability lower than this value cannot be detected in 
the majority of our sources.
We also found that the fraction of sources with detected flux variability depends substantially on the quality 
of the data. The observed dependence suggests that the fraction of flux-variable sources in our sample could be 
80\% or higher, and therefore 
that the great majority of the X-ray source population (which at these fluxes is dominated by AGN) vary in 
flux on time scales of months to years, perhaps responding to changes in the accretion rate.
Our results are consistent with the variability studies of Bauer et al. (\cite{Bauer03}) in the 
Chandra Deep Field-North and Paolillo et al. (\cite{Paolillo04}) in the Chandra Deep Field-South which 
found that the fraction of variable sources could be higher than $\sim$90\%. 

To measure the amplitude of the detected flux variability on each light curve above the statistical 
fluctuations of the data (excess variance), we used the maximum likelihood method described 
in detail in Almaini et al. (\cite{Almaini00}).

From this analysis we found the mean value of the relative excess variance to be $\sim$0.15 (68\% upper limit being $\le$0.36)
for the whole sample of objects 
($\sim0.34\pm0.01$ considering only sources with detected flux variability),
although with a large dispersion in observed values (from $\sim$0.1 to $\sim$0.65). Values 
of excess variance larger than $\sim$50-60\% were found not to be very common in our sources.

Similar values of the mean flux variability amplitude were found for the sub-samples of sources 
identified as type-1 and type-2 AGN, the fraction of variable sources being 
68$\pm$11\% for type-1 AGN and 48$\pm$15\% for type-2 AGN. 

We carried out extensive simulations in order to quantify the intrinsic true distribution of excess 
relative variance for our sources corrected for 
any selection biases. The most probable value of the true intrinsic  
amplitude of flux variability for our sample of objects 
is $\sim$0.2, as before correcting for selection 
biases. In addition, the simulations 
showed that the probability of detection of large variability amplitudes ($\ge$0.5) in our sample of sources is severely hampered by the poor quality of light curves.

Short time scale X-ray variability of Seyfert 1 galaxies indicate that the majority of these objects 
soften as they brighten. Our fainter and more distant sources do not generally follow this pattern.
This should not be too surprising, as this effect is observed in light curves that sample a uniform time interval 
and on time scales much smaller than in our study, where the strength of the effect seems to be higher.  

Because of the low signal-to-noise ratio of some of our data, we could not study spectral variability using the X-ray spectra of the sources from each observation, as the uncertainties on the fitted parameters would be too large.
Hence, we used a broad band hardness ratio or X-ray colour to quantify the spectral shape of the emission of the objects 
on each observation. 
A $\chi^2$ test shows spectral (``colour'') variability with a confidence 
of more than 3$\sigma$ in 24 objects ($\sim$20$\pm$6\% of the sample). 
However, we found the fraction of sources with detected spectral variability to increase with the quality of 
the data, hence this fraction 
could be as high as $\sim$40\%. This fraction is still significantly lower than the fraction of sources with 
flux variability.
However, we have shown that spectral variability is more 
difficult to detect than flux variability: a change in spectral slope 
of $\Delta\Gamma$$\sim$0.2-0.3 corresponds to a change in the observed X-ray Hardness Ratio of just 
$\Delta${\it HR}$\sim$0.1. The typical dispersion in {\it HR} values in our sources with detected spectral variability 
is $\sim$0.1-0.2, i.e. of the same magnitude as most errors in the {\it HR} in a single observation.
Therefore changes in the spectral index $\le$0.3 will be only detectable in a small number of objects in our 
sample. 

The fractions of AGN with detected spectral variability were found to be $\sim$14$\pm$8\% for type-1 AGN 
and $\sim$34$\pm$14\% for type-2 AGN with a significance of the fractions being different of $\sim$99\%, i.e. 
0.2-12 keV spectral variability on long time scales might be more common in type-2 AGN than in type-1 AGN. 
Although of only marginal significance, this result could be explained if part of the emission in type-2 AGN has a strong scattered radiation component.
This component is expected to be much less variable
and therefore changes in the intensity of the hard X-ray component alone would result in larger changes in the 
observed X-ray colour. 

Our samples of type-1 and type-2 AGN cover a broad range in 
redshifts, especially the former. Therefore we have studied variability properties on 
significantly different rest-frame energy bands. 
We did not find a strong dependence on redshift of the observed 
fraction of sources with detected spectral and flux variability. On the other hand
our analysis suggests that the amplitude of flux variability and the redshift might be correlated 
but the strength of the effect seems to be rather weak. Therefore we conclude 
that the overall flux and spectral X-ray variability properties of our AGN 
are very similar over the rest frame energy band from $\sim$0.2 keV to $\sim$54 keV.

We also grouped the spectra of our 
objects in four different observational phases, and compared the spectral fits to the grouped spectra.
With this study we detected variations in spectral fit parameters 
in 8 objects, including 6 type-1 AGN and 1 type-2 AGN. A detailed analysis of their emission 
properties showed that the origin of the detected spectral variability is due to a change 
in the shape of their broad band continuum alone. In none of the sources we detected strong or 
significant changes in other spectral components, such as soft excess emission 
and/or X-ray absorption. 

This result is confirmed by the spectral variability analysis of the 4 out of the 5 spectroscopically 
identified type-2 AGN for which we did not detect absorption in their co-added spectra. Flux variability 
was detected in 3 of the objects while only for one of them we detected spectral variability.
The lack of detection of significant spectral variability in 3 out of 4 of the sources makes unlikely that 
variations in the absorber 
can explain the mismatch between their optical and X-ray spectral properties. 
In addition, we have shown that, if the change in absorption is of the same magnitude as 
the typical absorption detected in absorbed type-2 AGN, ${\rm N_H\sim10^{22}\,cm^{-2}}$, then, changes in flux and in 
the X-ray colour will be significant ({\it $\Delta$CR}$\ge30\%$ and {\it $\Delta$HR}$\ge0.3$) and therefore 
we should have detected them.
Therefore, the lack of correlation between flux and spectral variations in unabsorbed type-2 AGN 
indicates that, this is an intrinsic property and not just bad luck in non-simultaneous optical and X-ray observations.
Finally, a Compton-thick origin for these sources is also very unlikely, as they are seen (3 out of 4) to vary in flux.

X-ray spectral variability in AGN on long time scales appears less common than flux variability. The same result seems to hold for both samples of type-1 and type-2 AGN.
Therefore, flux and spectral variability might not be correlated in a significant fraction of our sources. 
Indeed in 38 objects we detected flux variability but not spectral variability, while spectral variability but not 
flux variability was detected in 9 objects. 
A correlation analysis of the flux-spectral variability properties on the 15 objects were both spectral and flux 
variability were detected only showed no obvious links.

Due to the apparent lack of correlation of flux and spectral variability, the 
{\it two-component spectral model} and the {\it spectral pivoting model} frequently used to explain the X-ray variability 
properties in local Seyfert 1 galaxies, cannot accommodate the variability properties of our sources, as both 
require flux and spectral variability to be correlated. 
For a source with a varying spectral index we have seen that in order to explain the detected amplitudes of flux 
variability a $\Delta{\it HR}\ge0.3$ is required (which corresponds to $\Delta\Gamma\sim0.6-1.2$). However such variability of HR is not observed.
Furthermore, the amplitude of the observed flux variability cannot be explained due to changes in the X-ray continuum 
shape alone. This is confirmed by the fact that even in the objects with both spectral and flux variability detected, 
significant changes in {\it HR} are seen for almost constant values of the {\it CR}. 
Because of the apparent lack of correlation between flux and spectral variability, it is very unlikely that variations 
in X-ray absorption alone can explain the variability properties of our sources. Changes in absorption can be produced, but
then both spectral and flux variability should be detected,and they are not.
Therefore, the lack of correlation between the observed flux and spectral variability properties of our objects 
indicates that the observed spectral variability does not simply reflect 
changes in the X-ray luminosity of the sources.
Furthermore, our variability study supports the idea that the driver for
spectral variability on month-years
scales in our AGN cannot be simply a change in the mass accretion rate. Clearly, more complex models are called for. 

\begin{acknowledgements}
SM acknowledges financial support for this research from PPARC.
XB, FJC and MTC acknowledge support from the Spanish Ministry of Education and Science, under project ESP 2006-13608-CO2-O1.
\end{acknowledgements}

\appendix
\section{Correction of count rates for calibration drifts}
\label{appendix_A}
\begin{figure}
    \hbox{
    \includegraphics[angle=90,width=8.5cm]{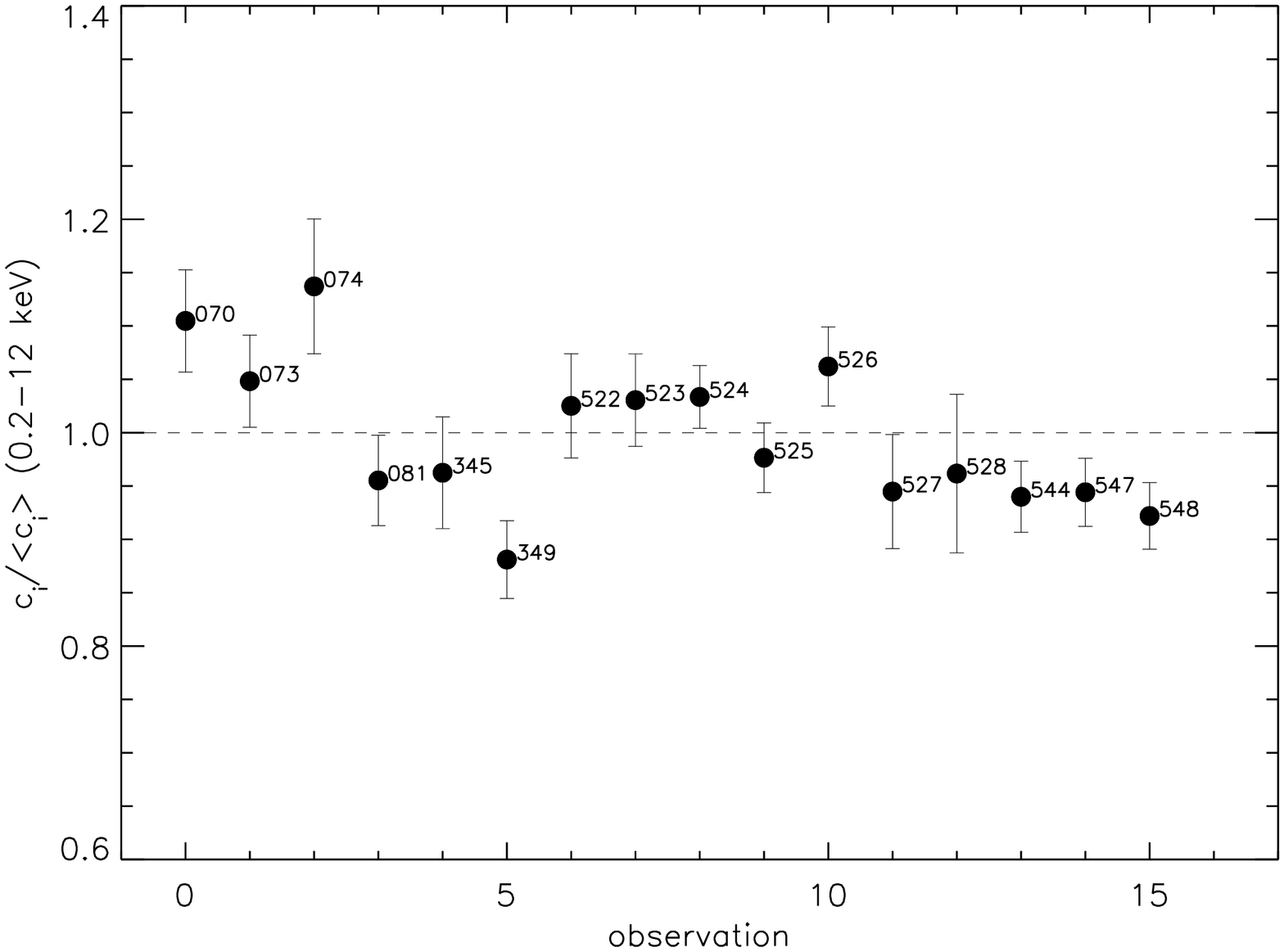}}
    \hbox{
    \includegraphics[angle=90,width=8.5cm]{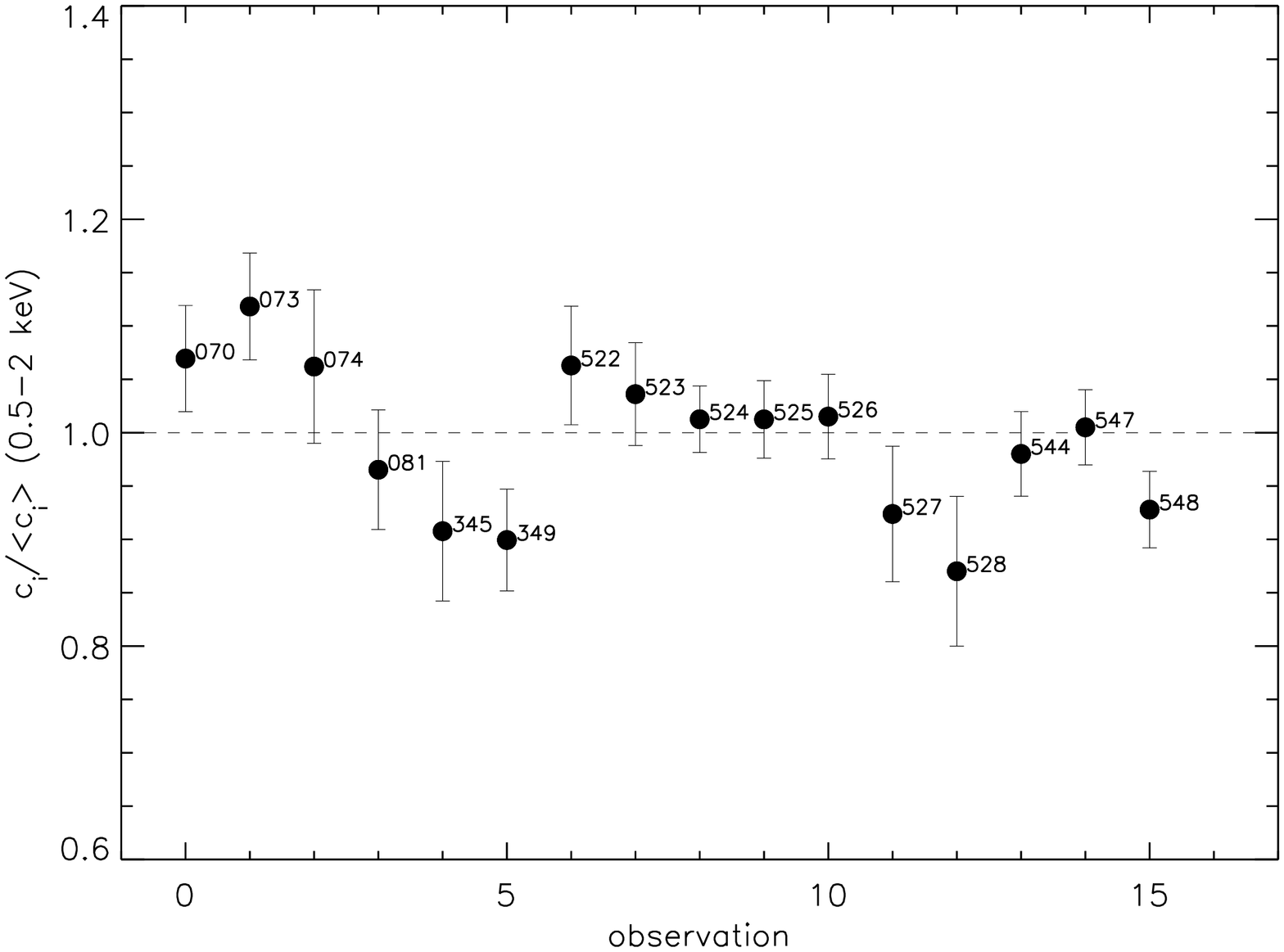}}
    \hbox{
    \includegraphics[angle=90,width=8.5cm]{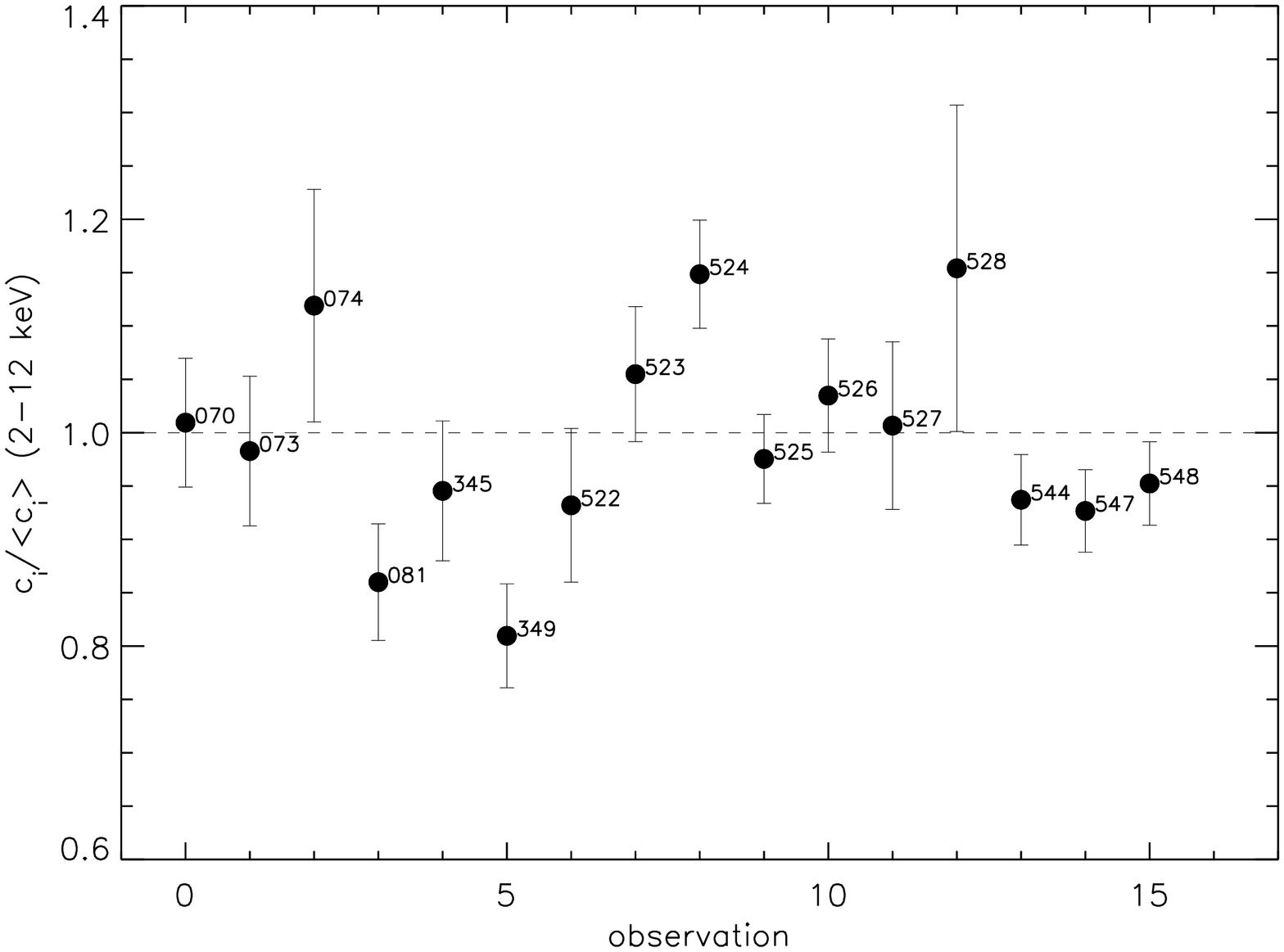}}
    \caption{Corrections of the count rates for each revolution 
      from systematic variations during the life time of the mission and from variations from using 
      different blocking filters for the observations. From top to bottom corrections for the 
      0.2-12, 0.5-2 and 2-12 keV count rates. Errors correspond to the 1$\sigma$ confidence interval.
      }
    \label{cal_ress}
\end{figure}

The {\sl XMM-Newton} observations in the {\it Lockman Hole} 
span a broad range of the mission lifetime, and therefore, 
changes might be expected in the calibration of the data. Particularly in the instrumental 
spectral response and systematic variations in the background modelling 
with time. That could introduce systematic differences in the measured count rates 
of the same source between observations.  In addition, pn observations were 
obtained with different blocking filters. A change in the blocking filter 
affects more the soft X-rays and therefore the measured 0.2-12 and 0.5-2 keV 
count rates.  

In order to remove from the light curves the effects listed above,
we have calculated for each revolution {\it r} 
the average deviation of the measured count rates in that revolution 
from their mean flux over all revolutions, for the subset of sources detected in that revolution. If there are no variations in the calibration, we would 
expect this value to be very close to unity, since 
variability is not expected to be correlated with time for different sources.
We have called this value $\Omega^r$ and it is defined as,

\begin{equation} \Omega^r={1 \over N(r)}{\sum_{i=1}^{N(r)}{c_i^r \over \langle c_i \rangle}}\end{equation}

The sum is over the number of sources detected on each revolution {\it r} ({\it N(r)}), $c_i^r$ is the 
count rate of source {\it i} in revolution {\it r} 
and $\langle c_i \rangle$ is the mean unweighted count rate of the source over all revolutions 
where it was detected. We found that the errors in the count rates were dominated by the quality 
of the 
observations (the errors in the background modelling and fitting of the source 
parameters in the {\tt SAS} source detection task {\tt emldetect}), and do not depend strongly 
on the values of the count rates themselves, in the sense that sources with larger count rates 
do not tend to have smaller errors. Because of this, the use 
of the weighted mean is not necessary in this case,  
and therefore we used the unweighted mean to calculate the average flux of each source 
and the mean dispersion of the count rates of the sources from their mean values 
for each revolution. However our mean values could be significantly affected by outliers 
due to flux variability of the sources. 
In order to avoid this, we applied a sigma-clipping (3$\sigma$) before calculating average 
values\footnote{Sigma-clipping only removed a small ($\sim$a few) number of points from each 
revolution, therefore we still have enough points to calculate reliable mean values for 
each revolution and energy band.}.

The 1$\sigma$ errors in $\Omega^r$ will be $\sigma(\Omega^r)/\sqrt{n(r)}$ where $n(r)$ is 
the number of points in revolution {\it r} after 
sigma-clipping i.e. $N(r) \rightarrow n(r)$, and 
$\sigma(\Omega^r)$ is 
\begin{equation} [\sigma(\Omega^r)]^2={1 \over n(r)-1}{\sum_{i=1}^{n(r)}\left ({c_i^r \over \langle c_i \rangle}-\Omega^r \right )}^2 \end{equation}

We calculated values of $\Omega^r$ for the measured 0.2-12 keV count rates which we used to 
study flux variability, and for 0.5-2 keV and 2-12 keV count rates, which we used to 
carry out the study of spectral variability. The values of $\Omega^r$ and their 
corresponding 1$\sigma$ errors for each revolution and energy band are shown in Fig.~\ref{cal_ress}.  

The corrected count rates and their corresponding 1$\sigma$ errors for each revolution were obtained as
\begin{equation} (c_i^r)^{corr}={c_i^r \over \Omega^r} \end{equation}
\begin{equation} \Delta (c_i^r)^{corr}=((c_i^r)^{corr})^2 \times \sqrt{\left ({\Delta c_i^r \over c_i^r} \right)^2+\left ({\Delta \Omega^r \over \Omega^r}\right)^2} \end{equation} 

It is evident from Fig.~\ref{cal_ress} that all the values are very close to unity, 
which means that we did not find calibration drifts 
between revolutions in any of the three energy bands, and therefore these corrections 
are very small in all cases.

\section{Simulations of variability in {\it Lockman Hole} sources}
\label{appendix_B}
\begin{figure}[!htb]
    \hbox{
    \hspace{-0.5cm}\includegraphics[angle=0,width=0.5\textwidth]{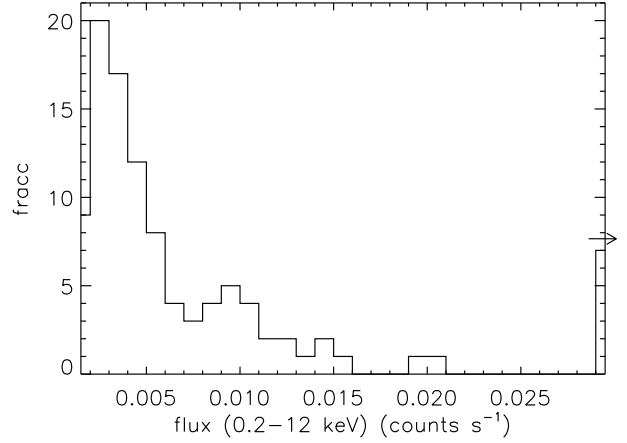}}
    \caption{
    Distribution of 0.2-12 keV count rates (corrected for calibration uncertainties) 
    in our sample of sources for which we studied flux variability.
      }
    \label{rate_dist}
\end{figure}

In Sec.~\ref{var_amplitude} we showed the distribution of variability amplitude that we 
found in our sample of objects. However measured values might differ from intrinsic 
values due to a number of effects, such as not enough quality in the light curves. In order 
to take into account all possible effects, we carried out simulations to calculate the 
distribution of intrinsic variability amplitude in our sample of objects 
after correcting for these effects. 

In order to simplify the notation we will refer to the intrinsic amplitude of flux 
variability, $\sigma_{intr}^i$, as {\it $S_i$}, while ${\it S_o}$ will be used to refer to the 
corresponding measured amplitude, $\sigma_{intr}^o$.
We defined a 3D parameter space ({\it $S_i$}, {\it $S_o$}, {\it CR}), where {\it $S_i$} is the 
intrinsic amplitude of flux variability, {\it $S_o$} is the observed amplitude as measured 
following the procedure described in Sec.~\ref{var_amplitude}, and {\it CR} is 0.2-12 keV count rate. 

We defined the grid in count rates such that the full range of count rates of our data is covered 
by simulations, from 0.001 to 0.481 ${\rm counts\,s^{-1}}$, while $S_i$ values (fraction of variability over the total count rate) covered the range from 0 to 4. 
As we will see later, the maximum simulated value of $S_i$ 
had to be chosen high enough to ensure that we obtain the correct distribution of intrinsic 
amplitudes within the covered range by our sources (below $S_o^{max}\sim0.7$).

\begin{figure}[!tb]
    \hbox{
     \hspace{-1.5cm}\includegraphics[angle=90,width=0.6\textwidth]{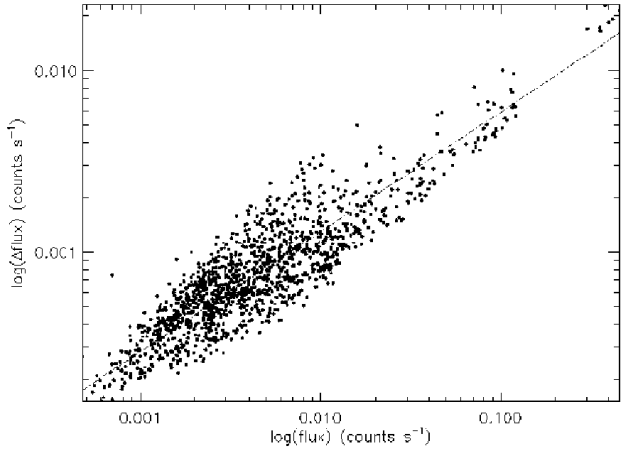}}
    \caption{
    Correlation between $\Delta CR$ and $CR$ for the sources in
    our sample. The values were obtained from the 0.2-12 keV energy band. The dot-dashed line 
    represents the power law best fit to the points.
      }
    \label{arate_rate}
\end{figure}

For each grid point ({\it $S_i$}, {\it CR}), and for each of the 120 sources in our sample for which 
we could study flux variability, we simulated a light curve. Simulated light curves have 
the same number of points and mean count rate, {\it CR}, as the real data. 
Count rates and their corresponding statistical errors are correlated as we show 
in Fig.~\ref{arate_rate} where we plotted the values obtained on each revolution 
for all sources (we did not see variations in the correlation between different revolutions). 
The correlation in logarithmic scale is linear, i.e., $\ln(\Delta CR)=\alpha \times \ln(CR)+ \beta$.
From this expression the analytical form of the correlation between $\Delta CR$ and {\it CR} 
is a power law with 

\begin{equation} \Delta CR=CR^\alpha \times e^\beta\label{eq:corr}\end{equation}

The values that we obtained from the best fit 
were $\alpha$=0.66 and $\beta$=-3.61. Using this correlation we obtained for each simulated 
count rate $CR$ the corresponding statistical error, $\sigma_{stat}$. 
The values (count rates) of each point in the light curves, $cr$, were assumed to follow a 
Gaussian distribution of mean equal to the mean count rate, {\it CR}, 
and dispersion $\sigma^2=\sigma_{stat}^2+(S_i\times CR)^2$, where $\sigma_{stat}$ is the 
statistical count rate error that corresponds to a count rate value $CR$. The second term 
is the intrinsic variability added to the simulated light curves.

\begin{figure}[!htb]
    \hbox{
    \hspace{-0.5cm}\includegraphics[angle=0,width=0.5\textwidth]{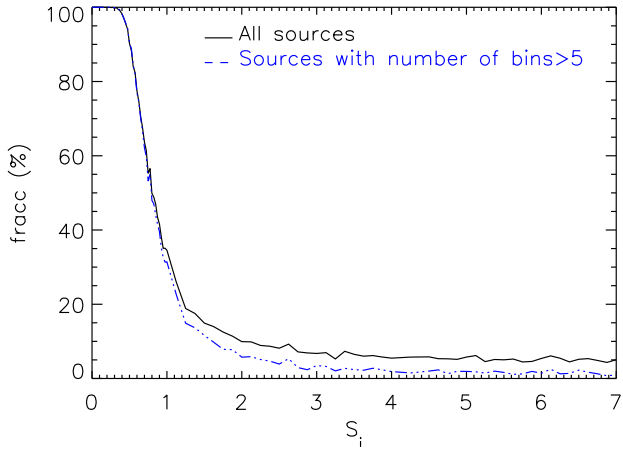}}
    \caption{
    Fraction of simulations that give $S_o$ values below $S_o^{max}$ as a function of 
    simulated $S_i$ for all $CR$, when all sources are used in the simulations (solid line) and when only sources 
    with at least 5 points in their light curves are used (dotted-dashed line). 
      }
    \label{bad_srcs}
\end{figure}

Finally, for each $cr$ value we obtained the corresponding statistical error from Eq.~\ref{eq:corr}. 
We then obtained for each simulated light curve the 
measured values of the intrinsic variability amplitude, $S_o$, following the same procedure as for the real data. 

For each set of values ($S_i$, $CR$) we computed the function $N(S_o|S_i,CR)$ that gives 
the number of sources with observed amplitude $S_o$. Then we calculated the function 
$N(S_o|S_i)$, that gives for each $S_i$ the number of sources in the simulations with 
value $S_o$, weighted with the distribution of $CR$, as

\begin{equation} N(S_o|S_i)=\sum_j P(CR_j) \times N(S_o|S_i,CR_j) \end{equation}
where $P(CR)$ is the distribution of count rates in our sources 
which is shown in Fig.~\ref{rate_dist}. The count rates were obtained 
as the arithmetic mean of the corrected count rates of all the points in the light curves. 
The total number of sources in the simulations with observed value $S_o$ is 

\begin{equation} N(S_o)=\sum_{S_i} N(S_o|S_i) \end{equation}

$N(S_o)$ can be used to obtain 
the probability distribution of $S_i$ values for a given $S_o$ as 

\begin{equation} P(S_i|S_o)={N(S_o|S_i) \over N(S_o)} \end{equation}

Then the probability distribution of $S_i$ will be 

\begin{equation} P(S_i)=\sum_{S_o} P(S_i|S_o) \times P(S_o) \end{equation}
where $P(S_o)$ is the distribution of $S_o$ values observed in our sources 
(see Fig~\ref{hist_excess_var}). 

\begin{figure}[!tb]
    \hbox{
    \includegraphics[angle=90,width=0.5\textwidth]{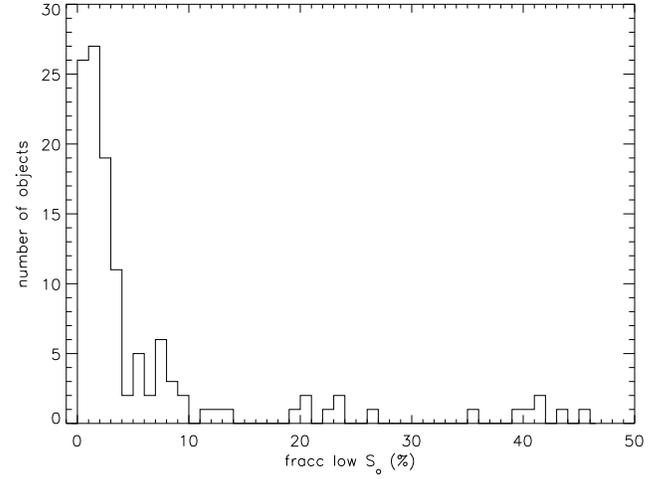}}
    \hbox{
    \includegraphics[angle=0,width=0.5\textwidth]{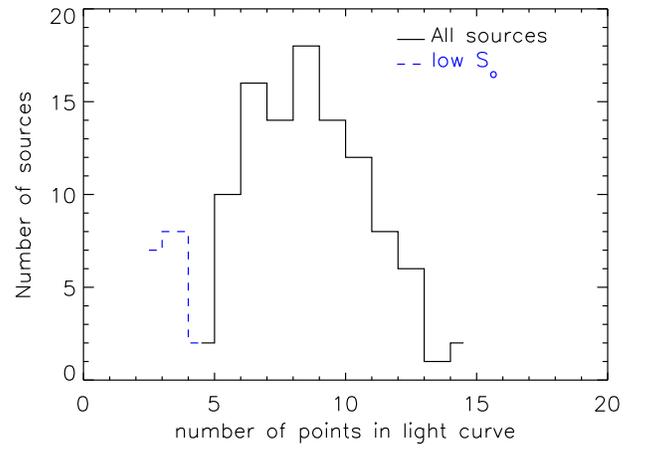}}
    \caption{Left: Distribution of the fraction of simulations for each source (for all simulated 
	values of ($S_i, CR$)) for which $S_o\le S_o^{max}$ ($\sim$0.7). 
	Right: Distribution of number of bins in the light curves of our 
	sources. The corresponding distribution for the objects 
	where the fraction of simulations with $S_o\le S_o^{max}$ was 
	higher than 10\% is indicated with a dashed-line.}
    \label{fracc_bad_sims}
\end{figure}

In order to get the correct distribution of $P(S_i)$ we need to simulate up to a value of $S_i$ for which all
values of $S_o$ are above the maximum observed variance 
$\sigma_{intr}=0.7\equiv S_o^{max}$, otherwise 
the normalisation of $P(S_i)$ will be wrong. 
However during the simulations we found a number of sources for which we obtained values of $S_o$ 
below 0.7 independently of how large was the simulated $S_i$. This is shown in Fig.~\ref{bad_srcs}
where the fraction of these cases is plotted as a function of simulated $S_i$ (solid line). Although this fraction 
decreases rapidly for low $S_i$ (below $\sim$2) it remains $\sim$constant for the highest simulated values.

We identified these anomalous cases with sources with a small number of bins in the light curves 
(see dotted-dashed line in Fig.~\ref{bad_srcs}). 

Fig.~\ref{fracc_bad_sims} (left) shows the distribution of 
the fraction of simulations for each source (for all values of ($S_i, CR$)) for which 
$S_o\le S_o^{max}$ ($\sim$0.7). We see that only a small number of 
sources have a fraction of $S_o\le S_o^{max}$ ($\sim$0.7) above 10\%.
Fig.~\ref{fracc_bad_sims} (right) shows the distribution of 
the number of bins in the light curves of our objects (solid line). 
The corresponding distribution for the sources for which the fraction 
of simulations with $S_o\le S_o^{max}$$\ge$10\% is indicated with a 
dashed-line. Our results show that, selecting a value of the fraction of 
10\% we pick up most of the ``problematic'' light curves, withough reducing 
significantly the number of light curves.
In addition we found that these ``problematic'' cases 
correspond to the sources with lower number of bins in the light curves. 

As we need to simulate values of $S_i$ where the fraction of simulations with $S_o \le S_o^{max}$
is low, we have not used sources with less than 5 points in their light curves 
in the calculation of the distributions $P(S_i)$ and $P(S_o)$. 

This implies that the number of sources for which 
we can obtain the distribution of $S_i$ is 103, instead of the original sample of 120 sources having at least 
2 points in their light curves. By doing this selection we can be confident that we can represent the expected 
distribution of $S_i$, $P(S_i)$, using simulations of $S_i$ values up to $\sim4$. 

\begin{figure}[!tb]
    \hbox{
    \includegraphics[angle=0,width=0.5\textwidth]{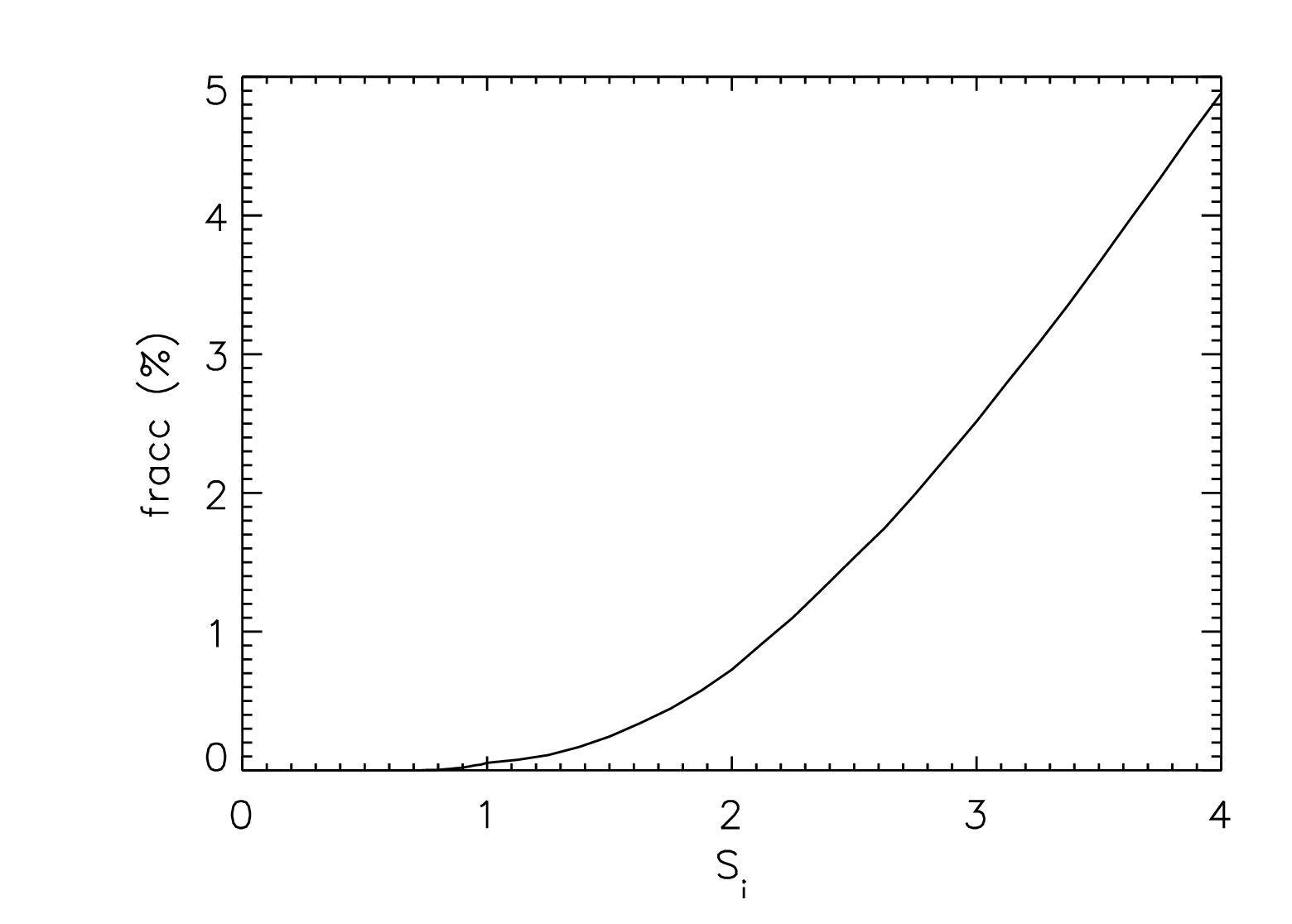}}
    \hbox{
    \includegraphics[angle=0,width=0.5\textwidth]{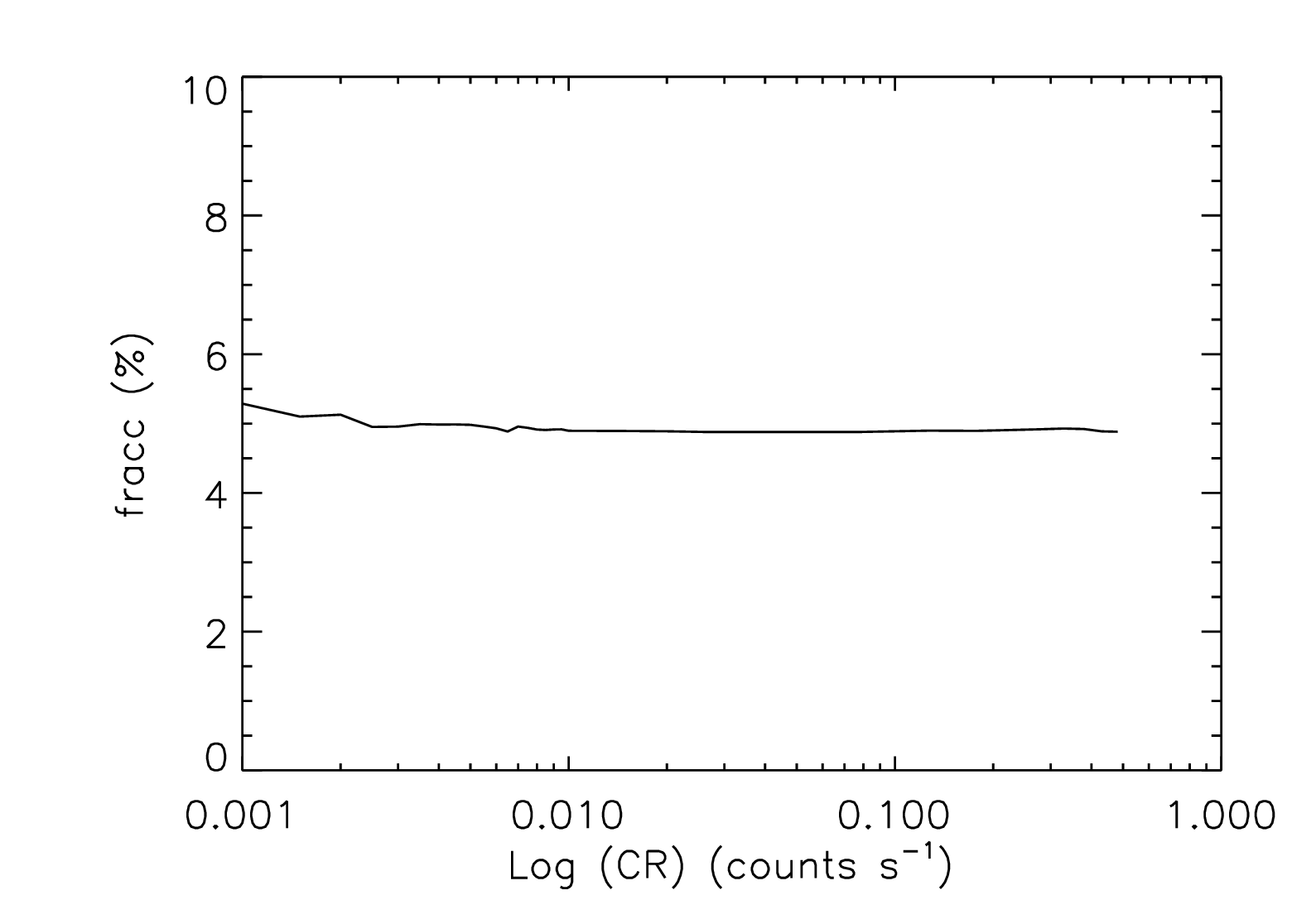}}
    \caption{Fraction of simulated light curves with negative mean count rates as a function 
    of $S_i$ (intrinsic amplitude; top) and $CR$ (0.2-12 keV count rate; bottom). 
      }
    \label{fracc_negative}
\end{figure}

A fraction of the simulated light curves were found to have negative mean count rates, and therefore 
in these cases the measured amplitudes of variability, $S_o$, were also negative as 
$S_o=\sigma_Q/\langle CR \rangle$($\equiv\sigma_{intr}$; see Eq.~\ref{eq:sintr}) 
with $\langle CR \rangle$ being the mean unweighted count rate of the simulated 
light curve. Our simulations showed that the fraction of simulated light curves with negative mean count 
rates did not vary significantly with the simulated count rates, $CR$, but it was a strong function of 
$S_i$. This can be seen in Fig.~\ref{fracc_negative} where we plotted this fraction 
as a function of $S_i$ (left) and $CR$ (right). We see that using only simulations with $S_i \le 4$ the maximum fraction of 
simulated light curves with negative values of $S_o$ is $\sim$5\% for $S_i$=4.

This result is expected as for larger values of 
$S_i$ the Gaussian distribution that we are using to obtain simulated count rates becomes 
broader, while the mean does not move significantly from zero, and 
hence the probability of obtaining negative average values increases. In all these cases 
we used the absolute values of the mean count rates to calculate the values of $S_o$. 

\clearpage
\end{document}